\documentclass[aps,prb,twocolumn,superscriptaddress,floatfix]{revtex4-2}
\usepackage[utf8]{inputenc}
\usepackage{amsmath,amssymb,bm,graphicx}

\usepackage{comment}
\usepackage{color}
\usepackage{xcolor}
\usepackage{physics}
\usepackage{graphicx}
\usepackage{bm}
\usepackage{amsfonts}
\usepackage{braket}
\usepackage[normalem]{ulem}

\def\manuscriptmode{clean}   

\usepackage{ifthen}
\definecolor{BLUE}{rgb}{0,0,1}

\ifthenelse{\equal{\manuscriptmode}{clean}}{%
  \newcommand{\tred}[1]{#1}
  \newcommand{\tb}[1]{#1}
  
  \newcommand{\tmag}[1]{#1}
  \newcommand{\tor}[1]{#1}
  \newcommand{\tro}[1]{}
  \newcommand{\tbo}[1]{}
  \newcommand{\tgo}[1]{}
}{%
\ifthenelse{\equal{\manuscriptmode}{submission}}{%
  \newcommand{\tred}[1]{\textcolor{red}{#1}}
  \newcommand{\tb}[1]{\textcolor{red}{#1}}
  
  \newcommand{\tmag}[1]{\textcolor{red}{#1}}
  \newcommand{\tor}[1]{\textcolor{red}{#1}}
  \newcommand{\tro}[1]{\textcolor{red}{\sout{#1}}}
  \newcommand{\tbo}[1]{\textcolor{red}{\sout{#1}}}
  \newcommand{\tgo}[1]{\textcolor{red}{\sout{#1}}}
}{%
  \newcommand{\tred}[1]{\textcolor{red}{#1}}
  \newcommand{\tb}[1]{\textcolor{blue}{#1}}
  
  \newcommand{\tmag}[1]{\textcolor{magenta}{#1}}
  \newcommand{\tor}[1]{\textcolor{orange}{#1}}
  \newcommand{\tro}[1]{\textcolor{red}{\sout{#1}}}
  \newcommand{\tbo}[1]{\textcolor{blue}{\sout{#1}}}
  \newcommand{\tgo}[1]{\textcolor{green}{\sout{#1}}}
}}
\DeclareRobustCommand{\rev}[1]{#1}
\newcommand{\revdel}[1]{}

\newcommand{\gl}{\mathrm{gl}}   
\newcommand{\gf}{\mathrm{gf}}   

\usepackage{slashed}
\usepackage{wasysym}
\newcommand{\be}{\begin{equation}}
\newcommand{\en}{\end{equation}}

\providecommand{\U}[1]{\protect\rule{.1in}{.1in}}

\def\XXint#1#2#3{{\setbox0=\hbox{$#1{#2#3}{\int}$}
     \vcenter{\hbox{$#2#3$}}\kern-.5\wd0}}

\makeatletter
\DeclareRobustCommand{\cev}[1]{{\mathpalette\do@cev{#1}}}
\newcommand{\do@cev}[2]{%
  \vbox{\offinterlineskip
    \sbox\z@{$\m@th#1 x$}%
    \ialign{##\cr
      \hidewidth\reflectbox{$\m@th#1\vec{}\mkern4mu$}\hidewidth\cr
      \noalign{\kern-\ht\z@}
      $\m@th#1#2$\cr
    }%
  }%
}
\makeatother

\begin{document}

\title{Inverse Chiral Phonon Zeeman Effect in Noncentrosymmetric Crystals}

\author{Jun-ichiro~Kishine}
\affiliation{Division of Natural and Environmental Sciences, The Open University of Japan, Chiba 261-8586, Japan}
\affiliation{Quantum Research Center for Chirality, Institute for Molecular Science, Okazaki, Aichi 444-8585, Japan}

\author{A.~S.~Ovchinnikov}
\affiliation{M.N. Mikheev Institute of Metal Physics, Ural Division, Russian Academy of Sciences, Ekaterinburg 620219, Russia}
\affiliation{Institute of Natural Science and Mathematics, Ural Federal University, Ekaterinburg 620002, Russia}

\author{Masahiro~Sato}
\affiliation{Department of Physics, Chiba University, Chiba 263-8522, Japan}

\author{G.~N.~Makarov}
\affiliation{Landau Institute for Theoretical Physics, Chernogolovka 142432, Russia}
\affiliation{Moscow Institute of Physics and Technology, Dolgoprudnyi 141700, Russia}

\author{A.~D.~Lyakhov}
\affiliation{Institute of Natural Science and Mathematics, Ural Federal University, Ekaterinburg 620002, Russia}

\date{\today}

\begin{abstract}
We present \rev{a quantized effective theory of} the inverse chiral phonon Zeeman 
effect in noncentrosymmetric crystals. Within micropolar elasticity, coupled 
translational displacements and microrotations give rise to intrinsically chiral 
phonons, which generate an elliptically polarized internal magnetic field through 
dynamical piezoelectricity. \tmag{Provided that electronic screening is incomplete, this 
field acts as an effective longitudinal Zeeman field on electronic spins, driving a 
Floquet response for localized states and a stationary adiabatic 
band splitting for itinerant electrons, leading to macroscopic spin polarization.} 
\rev{In contrast to earlier discussions based on nearly degenerate axial optical modes 
or externally driven circular lattice motion, the present mechanism relies on 
finite-wave-number truly chiral phonons whose counter-rotating branches are 
intrinsically split by crystal chirality. Quantization is introduced to define 
the single-phonon magnetic-field scale, coherent-state enhancement, and 
magnetic chirality in a controlled manner.} The results establish a purely 
lattice-driven mechanism for the inverse chiral phonon Zeeman effect in 
noncentrosymmetric crystals.
\end{abstract}

\maketitle

\section{Introduction}

The magnetic activity of lattice vibrations was already recognized in the 1970s by pioneering Raman studies of rare-earth compounds~\cite{Schaack1975,Schaack1976}.  
In recent years, this long-standing concept has been revitalized by modern spectroscopic techniques, which revealed that circularly polarized (axial) phonons can carry angular momentum and induce magnetization even in nonmagnetic solids~\cite{Juraschek2017,Juraschek2019,Juraschek2020,Geilhufe2021,Chaudhary2024}.  
The associated phonon magnetic moments are now known to reach values comparable to several Bohr magnetons, establishing a rapidly growing research field on phono-magnetic effects~\cite{Shabala2025}.

This family of phenomena is collectively referred to as the \textit{phonon Zeeman effect}, where an external magnetic field lifts the degeneracy of two orthogonal optical phonon modes and produces right- and left-handed branches.  
Giant phonon magnetic moments \tmag{of order $0.1 - 1.0$ $\mu_{\mathrm{B}}$} have been experimentally identified in some compounds~\cite{Cheng2020,Baydin2022,Wu2023,Hernandez2023}.  
These observations provide quantitative evidence that lattice dynamics, assisted by electron--phonon coupling, can generate effective magnetic responses far beyond the nuclear-magneton scale.

\tb{The inverse process---in which lattice motion itself generates magnetization rather than responding to it---is known as the \textit{inverse phonon Zeeman effect}.  
Two phenomenological pictures have been put forward for it. 
In \textit{dynamical multiferroicity}~\cite{Juraschek2017,Juraschek2019}, the circular ionic motion of a driven polar phonon carries a time-dependent polarization $\boldsymbol{P}(t)$ and hence a magnetization of the lattice itself, $\boldsymbol{M}\!\sim\!\boldsymbol{P}\!\times\!\dot{\boldsymbol{P}}$, which with Born effective charges is of nuclear-magneton order~\cite{Shabala2025}. 
In the \textit{phonon inverse Faraday effect}~\cite{Juraschek2020,Shabala2024,Shabala2025}, by contrast, circularly polarized phonons act on the electrons as an effective magnetic field, and the induced electronic magnetization is proportional to the helicity imbalance of the phonon amplitudes, $|u_R|^2-|u_L|^2$. 
The Bohr-magneton-scale moments observed experimentally are attributed to this electronic channel, for which second-order electron--phonon coupling~\cite{Shabala2024}, orbit--lattice coupling of localized electrons~\cite{Chaudhary2024}, electron--nuclear quantum geometry~\cite{Klebl2025}, and inertial spin--rotation coupling~\cite{Geilhufe2022} have been proposed as microscopic mechanisms; Merlin has argued that the resulting field is non-Maxwellian, a time-reversal-odd perturbation $\propto\boldsymbol{Q}\times\boldsymbol{Q}^{\ast}$ that mimics a magnetic field inside the sample but is undetectable outside it~\cite{Merlin2024,Merlin2025}.}

\tb{Both pictures are, in essence, macroscopic statements dictated by symmetry: $\boldsymbol{P}\!\times\!\dot{\boldsymbol{P}}$ and $|u_R|^2-|u_L|^2$ are the simplest time-reversal-odd, inversion-even combinations that can be formed from a polar mode and its time derivative. 
Which microscopic degrees of freedom carry the resulting moment, and how it is amplified from the nuclear- to the Bohr-magneton scale, remain open questions~\cite{Shabala2024,Shabala2025}. 
Noncentrosymmetric crystals offer a route to an internal magnetic response whose origin can be traced explicitly. 
Crystals with helical or screw-axis symmetries lift the degeneracy between phonons of opposite helicities from the outset, giving rise to truly chiral phonons~\cite{Ishito2023}, in which rotational and translational lattice dynamics are intrinsically intertwined, and their deformations couple to the electric polarization through the piezoelectric effect---mediated by the elastic waves of a micropolar medium, comprising both gapless and gapful branches, rather than by the internal optical ion displacements of dynamical multiferroicity. 
Motivated by this situation, the present work explores a possible alternative to descriptions based on merely rotating (axial) phonon modes, which exist also in centrosymmetric crystals, by considering the piezoelectric polarization current carried by a finite-wave-number chiral mode. 
Through the Maxwell--Amp\`ere law, such a current may give rise to a Maxwellian magnetic field, and we suggest that this provides one tractable scenario in which, within an effective continuum theory, the steps from lattice chirality to a field acting on the electron spins can be followed in a transparent manner.}

\rev{The specific goal of the present work is therefore threefold. First, we focus on finite-$q$ chiral phonons propagating along a screw axis, rather than on zone-center optical modes whose circular motion is created or selected by an external pump. Second, the two counter-rotating branches are already inequivalent at the level of the lattice eigenmodes because inversion symmetry is broken by the non-centrosymmetric crystal itself. Third, the dynamical piezoelectric \tmag{response - governed by the Maxwell-Amp\`ere relation} - converts this intrinsic phonon chirality into an internal elliptically polarized magnetic field, which can then be read out either as an effective Floquet Zeeman field for a localized spin or as a spin-dependent spectral response of itinerant electrons. In this sense, the electronic sector discussed below is not an independent driving mechanism but a physically transparent probe of the field generated by the chiral lattice motion.}

To substantiate this idea, we present \rev{a quantized effective theory of} the inverse chiral phonon Zeeman effect based on dynamical piezoelectricity within the framework of micropolar elasticity~\cite{Eringen1999,Nowacki1986}.
Focusing on chiral phonons propagating with non-zero wave number along a screw axis, we show that intrinsic lattice dynamics generates an internal elliptically polarized magnetic field, establishing a  \tmag{lattice-mediated route} to magnetoelectric coupling.
In particular, we show that chiral lattice dynamics alone generates an effective longitudinal Zeeman field acting on electronic spins, without invoking external light or static magnetic fields.

\rev{\tmag{To clear up the key physical assumptions of our approach, three points deserve to be stated explicitly at the outset.} (i) The parameter set used below is a representative one taken from the continuum description introduced in Ref.~\cite{Kishine2020}; it is meant to provide order-of-magnitude estimates rather than a first-principles prediction for a specific compound. (ii) Quantization is introduced not as a formal embellishment but in order to define \tmag{the magnetic-field scale per single phonon}, the coherent-state enhancement, and the vacuum expectation value of magnetic chirality. (iii) The Floquet and conduction-electron discussions address complementary questions: the former isolates the emergent static Zeeman field generated by the oscillating magnetic texture, while the latter shows how that field appears in an electronic spectrum and in the induced magnetization.}

\tmag{The paper is organized as follows. In Sec.~II we introduce the micropolar 
elastic model and quantize its chiral transverse phonon modes. In Sec.~III we 
derive the internal elliptically polarized magnetic field generated by dynamical 
piezoelectricity, its coherent-state enhancement, and the associated magnetic 
chirality (Zilch). Section~IV describes the resulting electron spin polarization, 
both as an effective Floquet Zeeman field for localized spins and as a spin-dependent 
band splitting for itinerant electrons. Sections~V and VI discuss the validity of 
the piezoelectric scenario with respect to electronic screening and the 
wave-vector-selective coherent excitation of chiral phonons, respectively. 
Section~VII summarizes the results and discusses the scope of the theory and its 
relation to previous work. Technical details are collected in the Appendices.}

\section{Model}

We consider a noncentrosymmetric crystal whose elastic properties are described by the theory of micropolar elasticity.
In this framework, translational polar displacements $\boldsymbol{u}(\boldsymbol{r})$ and axial microrotations $\boldsymbol{\varphi}(\boldsymbol{r})$ at position $\boldsymbol{r}$ are treated as independent but dynamically coupled fields~\cite{Eringen1999,Nowacki1986,Kishine2020}.  The strain (Cauchy) and wryness (Cosserat) tensors are defined as
$\varepsilon_{kl}=\partial_l u_k-\epsilon_{klm}\varphi_m$
and $\gamma_{kl}=\partial_l \varphi_k$, respectively. Here, $\epsilon_{klm}$ is the Levi-Civita symbol.
\begin{figure}[t]
\centering
\includegraphics[width=1.0\linewidth]{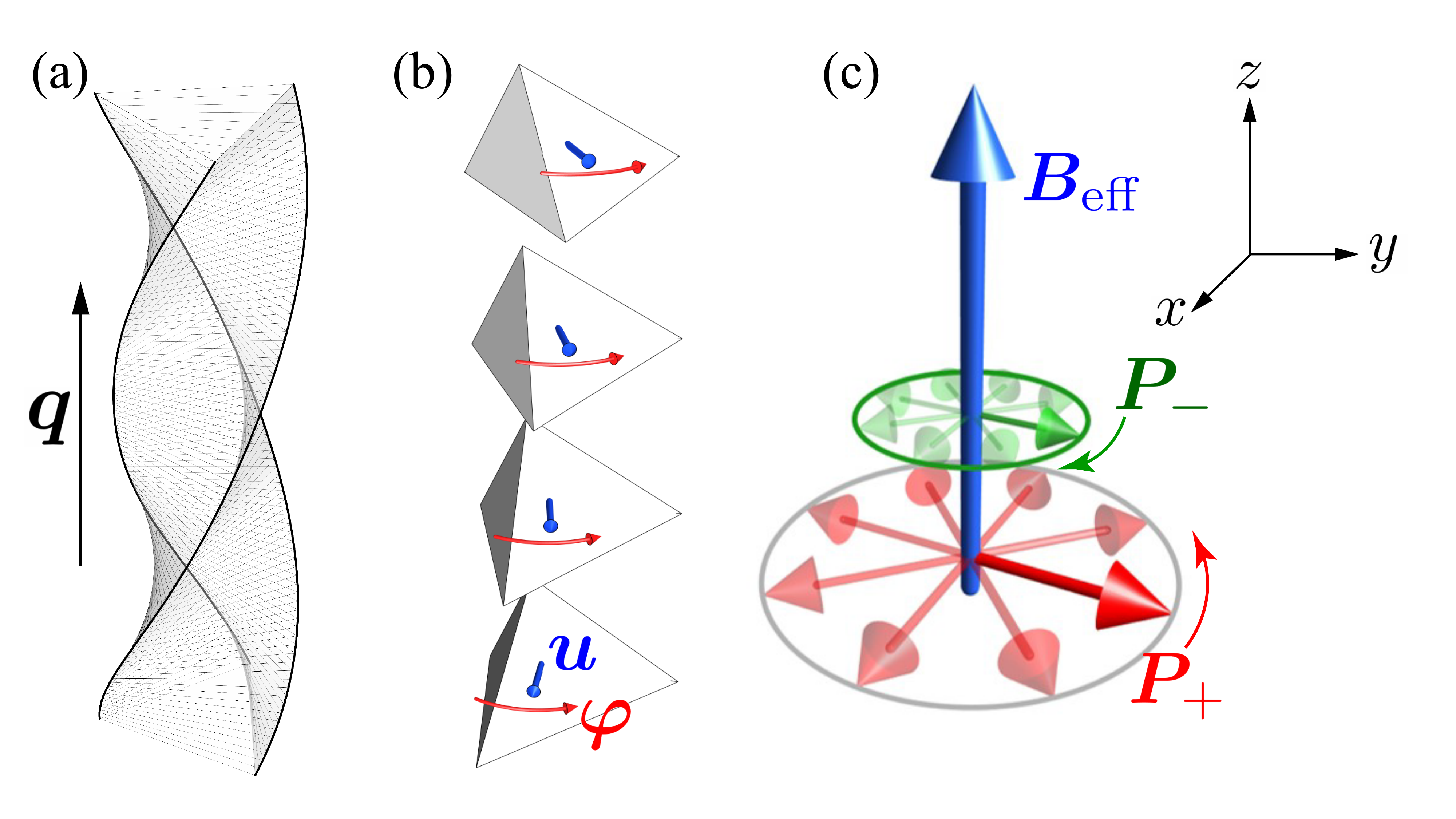}
\caption{Schematic illustration of (a) a chiral elastic wave in a continuum description,
(b) independent translational and rotational degrees of freedom of rigid blocks, which constitute the basic viewpoint of micropolar elasticity,
and (c) an effective longitudinal \tmag{magnetic field \(\boldsymbol{B}_{\text{eff}}\)} induced by lifting the degeneracy of counter-rotating \tmag{piezoelectric waves \(P_{\pm}\)}.
}
\label{fig:model}
\end{figure}
For definiteness and following our previous work~\cite{Kishine2020}, we consider a crystal with a hexagonal space group $P6_322$ and analyze plane-wave solutions propagating along the chiral $z$ axis [see Fig.~\ref{fig:model}(a) and (b)].

The Lagrangian density for the transverse  components $u_i$ and $\varphi_i$ ($i=1$ and $2$ correspond to $x$ and $y$, respectively) reads as
\begin{align}
\mathcal{L}_T & =
\frac{\rho}{2} \dot{u}^2_i
- \frac{A_{55}}{2}  \left( \partial_z u_i \right)^2
+ \frac{\rho j_1}{2} \dot{\varphi}^2_i
- \frac{B_{44}}{2} \left( \partial_z \varphi_i \right)^2
\nonumber \\
&\quad
-\frac12 \left( A_{44} - 2 A_{47} + A_{55} \right) \varphi^2_i
+ \left( A_{47} - A_{55} \right) \epsilon_{ij} u_i \partial_z \varphi_j
\nonumber \\
&\quad
- C_{74} \partial_z \varphi_i \partial_z u_i
+ \left( C_{44} - C_{74} \right) \epsilon_{ij} \varphi_i \partial_z \varphi_j ,
\end{align}
where $\rho$ is the mass density and $j_1$ is the microinertia associated with lattice rotations, $\epsilon_{ij}$ is the 2-dimensional Levi-Civita symbol. The components $A_{ij}$, $B_{ij}$ and $C_{ij}$ of the even-rank polar tensors and the pseudotensor, respectively, are written in the generalized Voigt (Sirotin) notation~\cite{Sirotin1982}.  The chiral coupling terms with $C_{ij}$ lift the degeneracy between right- and left-handed phonon modes even in the absence of any magnetic or optical pumping field~\cite{Kishine2020}.

\begin{figure}[t]
\centering
\includegraphics[width=1.0\linewidth]{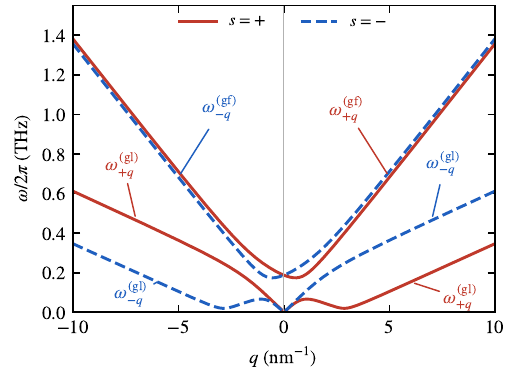}
\caption{\tb{Dispersion relations $\omega^{(\gl/\gf)}_{\pm q}$ of the transverse micropolar modes, Eqs.~(\ref{wtran1}) and (\ref{wtran2}), for the parameter set given in Sec.~III. Solid red (dashed blue) lines correspond to $s=+$ ($s=-$). The gapful (gf) branches retain a finite cutoff frequency at $q=0$, whereas the gapless (gl) branches vanish as $q\to0$. For $q\neq0$ the counter-rotating branches are split by the chiral couplings $C_{44}$ and $C_{74}$, with $\omega^{(\gl/\gf)}_{-,-q}=\omega^{(\gl/\gf)}_{+,q}$.}}
\label{fig:disp}
\end{figure}

\tmag{Since, to the best of our knowledge, the canonical quantization of \tred{the gapless 
and gapful modes} within the framework of a micropolar continuum (Cosserat 
elasticity) has not been previously addressed in the literature, we present 
the corresponding quantum mechanical formalism in detail in Appendix~\ref{app:quant}.}
Canonical quantization of the transverse modes and diagonalization using the Bogoliubov $uv$-transformation yields the Hamiltonian
\begin{equation}
\hat{H}_{T}
=
\Delta E^{(T)}_0
+
\sum_{q,s}
\left[
\hbar \tred{\omega^{(\gl)}_{s q}} \hat{\alpha}^{\dagger}_{s,q} \hat{\alpha}_{s,q}
+
\hbar \tred{\omega^{(\gf)}_{s q}} \hat{\beta}^{\dagger}_{s,q} \hat{\beta}_{s,q}
\right],
\label{HTrans}
\end{equation}
where $\Delta E^{(T)}_0$ is a correction term of the vacuum energy, $s=\pm$ labels the circular polarization (right- or left-handed),
and $\hat{\alpha}_{s,q}$ and $\hat{\beta}_{s,q}$ are the annihilation operators for the \tred{gapless (gl) and gapful (gf) branches}, respectively, with the wave-vector $\boldsymbol{q}=(0,0,q)$ [see Fig.~\ref{fig:model}(a)]. 
\tred{The two branches are labeled by their behavior in the long-wavelength limit: the gapless branch continuously connects to the bare translational mode, $\omega^{(\gl)}_{sq}\to0$ as $q\to0$, whereas the gapful branch connects to the bare microrotational mode and retains a finite cutoff frequency at $q=0$, set by the microinertia $j_1$ and the rotational stiffness $A_{44}-2A_{47}+A_{55}$ (see Appendix~\ref{app:quant}). At finite $q$ both branches are translation--rotation mixed, so the labels refer to the $q\to0$ character only; in particular, the gapful branch should not be identified with an optical phonon of a diatomic lattice, although it plays an analogous role below.} 
The corresponding dispersion relations $\tred{\omega^{(\gl/\gf)}_{s q}}$ [\tor{see Eqs.~(\ref{wtran1}) and (\ref{wtran2})}] coincide with those of the classical theory~\cite{Kishine2020}\tb{, and are shown in Fig.~\ref{fig:disp} for the parameter set used throughout this work (see Sec.~III)}. 

\rev{At this stage, the passage from the classical normal-mode problem to Eq.~(\ref{HTrans}) deserves a brief remark. The classical theory determines the dispersion relations and eigenvectors of the coupled translational-rotational modes, whereas quantization is essential for three reasons. It enables us to introduce the magnetic field generated by a \emph{single} phonon quantum, provides a natural foundation for coherent-state enhancement in nonequilibrium cases, and allows one to evaluate a vacuum measure of magnetic chirality such as the Zilch introduced below. The quantum treatment is therefore not merely formal; it is the step that connects the continuum elastic description to field amplitudes, occupations, and electronic response.}

A similar analysis can be carried out for longitudinal micropolar elastic waves propagating along the $z$ direction. Since the inverse chiral phonon Zeeman mechanism discussed below is governed by circularly polarized transverse modes, the longitudinal modes play no essential role in the following and will be ignored.

\section{Magnetic Fields Induced by Dynamical Piezoelectric Waves}

In a piezoelectric micropolar medium, elastic waves are accompanied by a time-dependent electric polarization $\boldsymbol{P}$.
In centrosymmetric crystals, such polarization dynamics induces \tmag{a lattice} magnetization in the form
$\boldsymbol{M}\sim \boldsymbol{P}\times \partial_t \boldsymbol{P}$,
which reflects the collective circular motion of ions~\cite{Juraschek2017}.
\tmag{In contrast, as we will show below, in the absence of inversion symmetry,} the dynamical piezoelectric coupling generates internal circularly polarized \textit{magnetic fields} that interact directly with electronic spins. 

From the constitutive relation of the micropolar medium~\cite{Nowacki1986} with the point group $622$ (see Appendix~\ref{app:em}), the transverse components  $u_{\pm}=u_1\pm iu_2$ and $\varphi_{\pm}=\varphi_1\pm i\varphi_2$ determine the decoupled circularly polarized modes of the electric polarization, $P_{\pm}=P_1\pm i P_2$, in the form  
\begin{align}
P_{\pm}(q,t)
&=
\pm q\!\left(d_{14}+\tfrac12 X_{11}\right)u_{\pm}(q,t)\nonumber\\
&+\left(X_{11}\mp q\Gamma^{(1)}_{11}\right)\varphi_{\pm}(q,t).
\end{align}
Here, $d_{14}$, and $X_{11}$, $\Gamma^{(1)}_{11}$ are the components of the conventional piezoelectric tensor and two additional tensors related to the rotational degrees of freedom, respectively.  The longitudinal polarization $P_3$ does not contribute to the mechanism discussed below.

The magnetic field associated with the polarization dynamics follows from the Maxwell--Amp\`ere equation for nonmagnetic media,
$\boldsymbol{\nabla}\times\boldsymbol{B}=\mu_0\partial_t\boldsymbol{P}$.
For the circularly polarized magnetic components it takes the form [see Fig.~\ref{fig:model}(c)]
\begin{align}
B_{\pm}(q,t)=\mp\frac{\mu_0}{q}\,\partial_t P_{\pm}(q,t).
\end{align}

Upon canonical quantization of the displacement and rotation fields, one obtains the magnetic-field operators
$\hat B_{\pm,q}(t)$, satisfying
$\hat B^{\dagger}_{\pm,q}(t)=\hat B_{\mp,-q}(t)$ (Appendix~\ref{app:em}).
The single-phonon magnetic-field states
$|B_{\pm,q}\rangle=\hat B^{\dagger}_{\pm,q}|0\rangle$
define the intrinsic magnetic-field scale associated with a \textit{single quantum} of a chiral phonon mode.

The inner product of the single phonon magnetic-field states takes the form
\begin{equation}
\left\langle B_{\pm,q}\big|B_{\pm,q}\right\rangle
=
\frac{\hbar\mu_0^2}{2\rho V_0 q^2}
\sum_{\tred{s=\gl,\gf}}
\left[\omega^{(s)}_{\pm q}\right]^2
\mathcal{F}^{(s)}_{\pm}(q),
\label{Bpm}
\end{equation}
where the explicit expression $\mathcal{F}^{(s)}_{\pm}(q)$ is given in Appendix~\ref{app:em}.
Importantly, Eq.~(\ref{Bpm}) is time independent and defines the \textit{single-phonon rms magnetic-field amplitude}
$B^{(1)}_{\pm}(q)\equiv\sqrt{\langle B_{\pm,q}|B_{\pm,q}\rangle}$.

\begin{figure}[t]
\centering
\includegraphics[width=0.9\linewidth]{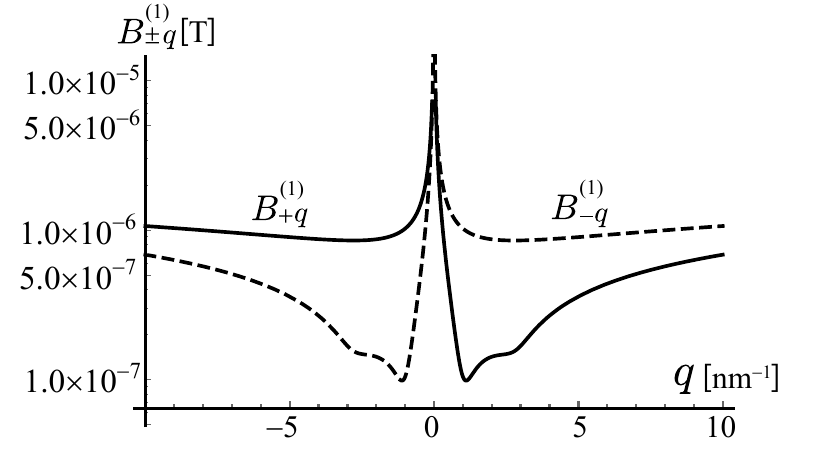}
\caption{Calculated single-phonon magnetic-field amplitudes $B_{\pm}^{(1)}(q)$ as a function of phonon wave number $q$.
Material parameters are taken from Ref.~\cite{Kishine2020}; see text for details. \rev{More precisely, the adopted values constitute a representative continuum parameter set used here for order-of-magnitude estimates rather than for a material-specific first-principles prediction.}}
\label{fig:Bpm}
\end{figure}

\rev{For Fig.~\ref{fig:Bpm}, the following values were taken: $\rho = 5029~\mathrm{kg\,m^{-3}}$, $V_0=3.988\times10^{-28}~\mathrm{m^{3}}$, $j_1 = 0.5\times10^{-19}~\mathrm{m^{2}}$, $A_{55}=0.215\times10^{10}~\mathrm{N\,m^{-2}}$, $A_{44}=0.21\times10^{10}~\mathrm{N\,m^{-2}}$, $A_{47}=0.195\times10^{10}~\mathrm{N\,m^{-2}}$, $B_{44} = 1.0\times10^{-10}~\mathrm{N}$, $C_{44} = 0.44~\mathrm{N\,m^{-1}}$, and $C_{74} = 0.36~\mathrm{N\,m^{-1}}$, i.e., the same as in Ref.~\cite{Kishine2020}. For illustrative purposes, we choose $d_{14} =X_{11} = 1.0\times10^{-2}~\mathrm{C\,m^{-2}}$ and $\Gamma^{(1)}_{11} = 1.0\times10^{-11}~\mathrm{C\,m^{-1}}$ to {\it{balance}} the different contributions from $u_{\pm}$ and $\varphi_{\pm}$ to $P_{\pm}$.}
Figure~\ref{fig:Bpm} shows the resulting wave-number dependence of $B^{(1)}_{\pm}(q)$.
The asymmetry between the two circular polarizations,
$B^{(1)}_{+}(q)\neq B^{(1)}_{-}(q)$,
is a direct consequence of broken inversion symmetry and implies that electrons experience an elliptically polarized internal magnetic field.
In the long-wavelength limit $q\to0$, the dominant contribution originates from \tred{the gapful (microrotation-dominated) branches} with frequencies $\tred{\omega^{(\gf)}_{\pm q}}$, which play a role analogous to IR-active optical \tred{phonons} in centrosymmetric crystals~\cite{Schaack1975}, but without requiring any external perturbation to lift their degeneracy.

\tmag{Physically, the sharp increase and formal divergence of the single-phonon 
magnetic field $\boldsymbol{B}_{\pm}^{(1)}(\boldsymbol{q})$ in the long-wavelength 
limit $\boldsymbol{q} \to 0$ originates from the rotational nature of the micropolar 
continuum. In this regime, the gapful microrotational modes correspond to coherent, 
in-phase microrotations of structural units over macroscopically large volumes. 
Since these dynamic microrotations are directly coupled to the piezoelectric 
response, they generate a highly collective, non-canceling effective polarization 
current that scales up with the wavelength. In any realistic experimental setup, 
this continuum-model divergence is naturally regularized by the finite sample 
dimension $L$. The physical boundaries impose a strict lower bound on the allowed 
wave vector, $q_{\text{min}} \sim 2\pi/L$, thereby ensuring that the induced 
magnetic field remains strictly finite while retaining its enhanced value near the 
zone center.}

\rev{Accordingly, the numerical scale of the field should be interpreted as representative of a symmetry-allowed mechanism in a realistic parameter window, not as a fit to a particular compound.}

\subsection{Many-Phonon Enhancement}

Once the magnetic field associated with a single phonon quantum is established, the enhancement in coherent phonon states follows naturally and can be discussed in a controlled and quantitative manner.

Equation~(\ref{Bpm}) and Fig.~\ref{fig:Bpm} quantify the magnetic-field scale associated with a \textit{single} phonon quantum.
In nonequilibrium situations, however, the relevant chiral phonon mode may acquire a macroscopic occupation.
For a mode with occupation number $N_{\pm}(q)$, the cycle-averaged field intensity scales as
$\overline{\langle B_{\pm}^2(q)\rangle}\propto[2N_{\pm}(q)+1]\,[B^{(1)}_{\pm}(q)]^2$,
so that the rms magnetic-field amplitude grows as
$B_{\pm}^{\rm rms}(q)\sim\sqrt{2N_{\pm}(q)+1}\,B^{(1)}_{\pm}(q)$.
In particular, a coherent phonon state yields a classical magnetic-field component with amplitude $\propto\sqrt{N_{\pm}}$,
providing a natural route to amplify the internal magnetic field by orders of magnitude relative to the single-phonon scale.

\subsection{Zilch and Phonon-Induced Magnetic Chirality}

The elliptically polarized magnetic field generated by a dynamical piezoelectric wave
carries an intrinsic measure of magnetic chirality.
The latter can be characterized by Lipkin's Zilch density~\cite{Lipkin1964},
\begin{equation}
Z(\boldsymbol{r},t)=\boldsymbol{B}(\boldsymbol{r},t)\cdot\left[\boldsymbol{\nabla}\times\boldsymbol{B}(\boldsymbol{r},t)\right],
\end{equation}
which provides a gauge-invariant measure of optical and magnetic chirality.

Making Fourier transformation,
$Z(z,t)=\int (dq/2\pi)\,Z_q(t)\,e^{iqz}$,
and using quantization of the magnetic field and polarization in terms of phonon operators, the expectation value of the Zilch density in the phonon vacuum becomes time independent (see Appendix~\ref{app:zilch}),
\begin{align} \label{Zqint}
\langle\hat{Z}_q\rangle
=
-\frac{\delta_{q,0}}{4\pi\hbar}
\int dk\,k
\left(
\langle B_{+,k}|B_{+,k}\rangle
-
\langle B_{-,k}|B_{-,k}\rangle
\right).
\end{align}
\tmag{The appearance of $\delta_{q,0}$ in the right-hand side is a 
rigorous consequence of the translational invariance of the phonon 
vacuum. Since the ground state $\lvert 0 \rangle$ is spatially homogeneous, 
the expectation value of any local observable must be position independent, 
restricting its spatial Fourier transform to the zero momentum. Thus, Eq.~(\ref{Zqint}) represents the total, integrated zilch density of the quantum ground state, accumulating the chirality of 
zero-point fluctuations across all wave numbers $k$.}

\tmag{A nonzero} Zilch directly reflects the intrinsic asymmetry between the
counter-rotating magnetic components,
$\langle B_{+,k}|B_{+,k}\rangle\neq\langle B_{-,k}|B_{-,k}\rangle$,
which originates from the chiral splitting of the underlying phonon modes.
This result establishes that the dynamical piezoelectric waves generate not only
an internal magnetic field, but also a finite magnetic chirality rooted in broken \tor{inversion} symmetry.

\section{Phonon-Induced Electron Spin Polarization}

The elliptically polarized magnetic field $\boldsymbol{B}(t)$ generated by the
dynamical piezoelectric wave acts on electronic spins as an effective ac Zeeman field.
In this work, electrons are treated as passive probes of the phonon-induced magnetic field.
\rev{This separation of roles is important. The Floquet problem below answers the minimal question ``what static Zeeman field is generated by the oscillating internal magnetic texture?'' The itinerant-electron model then addresses the complementary question ``how does that field manifest itself in an electronic spectrum and in the net magnetization of a Fermi sea?'' The latter therefore complements, rather than replaces, the Floquet discussion.}

\subsection{Effective Floquet Zeeman Field}

We consider the Hamiltonian of a single-electron spin $\boldsymbol{\sigma}$ coupled to the
counter-rotating magnetic field components,
\begin{align}
\hat{H}(t)
=
\mu_B\,\boldsymbol{\sigma}\cdot\boldsymbol{B}_{+}(t)
+
\mu_B\,\boldsymbol{\sigma}\cdot\boldsymbol{B}_{-}(t),
\label{FloquetH}
\end{align}
where $\mu_B$ is the Bohr magneton and
$\boldsymbol{B}_{\pm}(t)$ represent clockwise and anticlockwise rotating magnetic fields
with the frequencies $\omega_{\pm}=\tred{\omega^{(\gf)}_{\pm q}}$.
The two frequencies are nearly degenerate,
$|\omega_{+}-\omega_{-}|\ll\omega=(\omega_{+}+\omega_{-})/2$,
while the amplitudes generally differ due to lattice chirality,
$B_{+}\neq B_{-}$.
\tb{Here $B_{\pm}\equiv|B_{\pm}(Q)|$ denote the classical amplitudes of the counter-rotating field components of the coherently populated mode with wave number $q=Q$; for a coherent state with occupation number $N_{\pm}(Q)$ they are related to the single-phonon amplitudes of Sec.~III by $B_{\pm}\simeq\sqrt{N_{\pm}(Q)}\,B^{(1)}_{\pm}(Q)$ (see Appendix~\ref{app:floquet}).}

When the typical electronic energy scale \tmag{is much smaller} than 
$\hbar\omega_{\pm}$, the time-periodic Hamiltonian~(\ref{FloquetH})
can be treated within the Floquet high-frequency expansion~\cite{Bukov2014,Eckardt2015,Eckardt2017,Oka2019,Sato2025}. To leading order, this yields the effective static Floquet Hamiltonian~\cite{Takayoshi2014b,Sato2016}
(see Appendix~\ref{app:floquet}),
\begin{align}
\hat{H}_{\mathrm{eff}}
\simeq
\tb{-\frac{\mu_B^2}{\hbar\omega}}
\left(
B_{+}^2-B_{-}^2
\right)
\sigma_z,
\end{align}
which describes an emergent Zeeman field along the $z$ axis.
The effective field is proportional to the \textit{intensity difference}
of the counter-rotating magnetic components, reflecting the phonon-induced magnetic chirality.

Unlike optically driven magnetization~\cite{Juraschek2020},
the present effect originates intrinsically from the internal lattice dynamics,
without any external light or static magnetic field.
While this statement concerns the microscopic mechanism,
the magnitude of the emergent Zeeman field is controlled by the occupation
of the underlying chiral phonon modes.

Since the Floquet-induced static Zeeman term depends on the intensity difference,
one finds
\begin{align} \label{Beff}
B_{\rm eff}
\propto
\frac{N_{+}(\tb{Q})}{\omega}\,[B_{+}^{(1)}(\tb{Q})]^2
-
\frac{N_{-}(\tb{Q})}{\omega}\,[B_{-}^{(1)}(\tb{Q})]^2,
\end{align}
\tb{where $N_{\pm}(Q)$ are the occupation numbers of the two counter-rotating components of the mode $Q$.}
Hence, in a coherent phonon state or a strongly populated nonequilibrium regime,
the effective Zeeman field increases \textit{linearly} with the phonon occupation.

\subsection{Inverse Zeeman Phonon Splitting Effect}

\tmag{The localized, single-electron Floquet framework, which assumes that individual 
electrons respond independently to the elliptically polarized magnetic field, 
cannot be directly applied  to itinerant electron systems where a well-defined electronic band structure is formed.}
\tmag{Nevertheless,} the phonon-induced elliptically polarized magnetic field gives rise to a
spin-dependent modification of the electronic spectrum, enabling the emergence of a net magnetization along the \(z\)-direction. 
At this stage, the role of electrons is purely passive:
the electronic system responds to the effective Zeeman field generated
by chiral phonons, without feeding back onto the lattice dynamics.
\tmag{Such an assumption is well justified by the fact that the typical electronic 
energy scale (on the order of~eV) corresponds to a femtosecond timescale, 
which is much faster than the picosecond dynamics of terahertz micropolar 
phonons ($\hbar\omega \ll \varepsilon_{\text{F}}$); hence, the electronic 
states follow the periodic driving adiabatically.}

\tb{The coupling between conduction electrons and the phonon-induced field is the Zeeman interaction of the electron magnetic moment $\boldsymbol{m}=\tfrac12 g_e\mu_B\boldsymbol{\sigma}$ with the internal field,
\begin{equation} \label{HintReal}
\mathcal{H}_{\mathrm{int}}
=
-\tfrac12 g_e\mu_B
\int d^3r\,
\left[\hat{\psi}^{\dagger}(\boldsymbol{r})\,\boldsymbol{\sigma}\,\hat{\psi}(\boldsymbol{r})\right]
\cdot\boldsymbol{B}(\boldsymbol{r}),
\end{equation}
where $\hat{\psi}=(\hat{\psi}_{\uparrow},\hat{\psi}_{\downarrow})^{T}$ is the electron field operator, $g_e$ is the electron $g$ factor, and the slow time dependence of $\boldsymbol{B}$ is treated adiabatically as discussed above. 
For the transverse field one has $\boldsymbol{\sigma}\cdot\boldsymbol{B}=\tfrac12\left(\sigma_{+}B_{-}+\sigma_{-}B_{+}\right)$ with $\sigma_{\pm}=\sigma_x\pm i\sigma_y$ and $B_{\pm}=B_x\pm iB_y$, so that each circular component of the field flips the electron spin. 
Expanding $\hat{\psi}_{\sigma}(\boldsymbol{r})=V^{-1/2}\sum_{\boldsymbol{k}}c_{\boldsymbol{k}\sigma}e^{i\boldsymbol{k}\cdot\boldsymbol{r}}$ and $\boldsymbol{B}(\boldsymbol{r})=\sum_{\boldsymbol{q}}\boldsymbol{B}(\boldsymbol{q})e^{i\boldsymbol{q}\cdot\boldsymbol{r}}$, Eq.~(\ref{HintReal}) takes the second-quantized form}
\begin{align} \label{HintFM}
\mathcal{H}_{\mathrm{int}}
=
-\tfrac{1}{2} g_e \mu_B
\sum_{\boldsymbol{k},\boldsymbol{q}}
\left[
B_{-}(\boldsymbol{q})\,
c^{\dagger}_{\boldsymbol{k}+\boldsymbol{q}\uparrow}
c_{\boldsymbol{k}\downarrow}
+
B_{+}(\boldsymbol{q})\,
c^{\dagger}_{\boldsymbol{k}+\boldsymbol{q}\downarrow}
c_{\boldsymbol{k}\uparrow}
\right],
\end{align}
where $c^{\dagger}_{\boldsymbol{k}\sigma}$ ($c_{\boldsymbol{k}\sigma}$) are the electron creation (annihilation) operators in the state ${\boldsymbol{k}\sigma}$ ($\sigma=\uparrow,\downarrow$)\tb{, and $B_{\pm}(\boldsymbol{q})$ are the Fourier components of the classical counter-rotating field}.
This interaction has the same structure as a Zeeman coupling to a
circularly polarized magnetic field, with the crucial difference that
the field here is generated internally by lattice dynamics.  

\tmag{Note that  the interaction (\ref{HintFM}) is written in the general momentum representation 
where chiral phonons transmit a wave vector $\boldsymbol{q}$ to the electronic subsystem. 
To obtain the concomitant net magnetization, $\langle m_z\rangle$, we focus on a single phonon wave vector $\boldsymbol{Q}=(0,0,Q)$ near the center of the Brillouin zone, which provides a solid ground for employing the long-wavelength micropolar continuum theory.  In this limit, the induced fields $B_{\pm}(\boldsymbol{Q})$\tb{, with $|B_{\pm}(\boldsymbol{Q})|=B_{\pm}$ the amplitudes introduced in Sec.~IV.A,} become macroscopically homogeneous, and produce a Zeeman-like spin-dependent band splitting (see Appendix~\ref{app:split}).}

\tmag{Physically, the characteristic timescale for a conduction electron moving at the 
Fermi velocity to encounter spatial inhomogeneities of the field (with a typical 
length scale of $1/Q$) is given by $\tau_{\text{col}} \sim 1 / (Q v_{\text{F}})$, while 
the rotation of the field occurs on a timescale of $\tau_{\text{rot}} \sim 1 / \omega$. 
Therefore, selecting a small wavenumber near the zone center ensures the kinematic 
criterion $Q \ll \omega/v_{\text{F}}$, which is physically equivalent to the condition 
$\tau_{\text{rot}} \ll \tau_{\text{col}}$. This implies that the temporal dynamics of the 
moving electron are governed by a continuous adiabatic adjustment to the rotating 
field direction, closely resembling 
the "atomic" Floquet scenario, rather than by scattering processes on the spatially inhomogeneous background.}

\tmag{For the case of weak splitting,
$\tor{(g_e\mu_B)^2\left(|B_{+}(\boldsymbol{Q})|^2-|B_{-}(\boldsymbol{Q})|^2\right)\ll\varepsilon_{\text{F}}^2}$,
$\langle m_z\rangle$ can be obtained perturbatively.
Assuming that only the lower quasiparticle branches are occupied, one gets}
\begin{align}  \label{NetMag}
\langle m_z\rangle
&\approx
\frac{1}{8}
\left(
g_e \mu_B
\right)^3
S(\boldsymbol{Q})
\left(
|B_{+}(\boldsymbol{Q})|^2
-
|B_{-}(\boldsymbol{Q})|^2
\right),
\end{align}
where the factor $S(\boldsymbol{Q})$, \tor{which has the dimension of inverse energy squared and is} defined in
Appendix~\ref{app:split}, encodes the electronic density of states and
kinematic details.
\tor{Equation}~(\ref{NetMag}) explicitly shows that the phonon chirality controls the
induced electronic magnetization.
\rev{Equation~(\ref{NetMag}) also makes clear why the conduction-electron problem carries information beyond the single-spin Floquet reduction: the induced magnetization depends on the electronic phase space through $S(\boldsymbol{Q})$ and therefore on the structure of the partially filled band.}

Here we note that Eq.~(\ref{NetMag}) does not take into account the breaking of inversion 
symmetry in the electronic system. In noncentrosymmetric crystals, however, the bulk 
Rashba effect is symmetry-allowed. It can be shown that weak Rashba splitting leads 
only to second-order corrections with respect to the Rashba coupling constant 
($\propto \alpha_{\text{R}}^2$) to the net electronic magnetization (see Appendix~\ref{app:rashba}). 
\tmag{\rev{We intentionally keep the Rashba contribution subleading here to isolate as clearly 
as possible the lattice-generated Zeeman response itself, confirming that the primary, 
linear spin-polarization effect is driven entirely by the phonon-induced piezoelectric 
field and cannot be obscured by intrinsic spin-orbit couplings.}
}

\section{Validity of the Piezoelectric Scenario}

A key requirement for the dynamical piezoelectric mechanism is that the induced
polarization is not efficiently screened by conduction electrons.
This imposes the condition that the characteristic phonon frequency $\omega$
should be comparable to or exceed the electronic plasma frequency $\omega_{\mathrm{p}}$.
For a free-electron gas,
$\omega_{\mathrm{p}} \simeq 56.4\sqrt{n}~\mathrm{Hz}$,
so that for \tred{gapful-branch frequencies} in the range
$6$--$12~\mathrm{THz}$ the condition $\omega\gtrsim\omega_{\mathrm{p}}$ is satisfied for
carrier densities below $n\sim 10^{23}~\mathrm{m^{-3}}$.
\tmag{The present theory therefore naturally applies to piezoelectric insulators and 
low-carrier-density degenerate semiconductors, where electronic screening remains 
incomplete but a well-defined Fermi surface is established.}

\tmag{This frequency hierarchy successfully reconciles the two complementary electronic 
regimes discussed in our analysis. \tb{For localized states in the insulating matrix, the Floquet high-frequency 
expansion requires only that the spin-dependent low-energy scales of the localized 
electron---the Zeeman energy $\mu_B B_\pm$ and any exchange or crystal-field 
splitting of the relevant level---be small compared with $\hbar\omega$, so that 
the time-periodic Zeeman coupling generated by the lattice dynamics can be 
replaced by an effective static Hamiltonian.} Concurrently, for the itinerant electron subsystem in the low-density 
degenerate regime, the condition $\omega \gtrsim \omega_{\text{p}}$ ensures that 
electronic screening remains incomplete. Since the typical Fermi energy in this 
conducting framework significantly exceeds the single-phonon energy 
($\hbar\omega \ll \varepsilon_{\text{F}}$), the carrier dynamics are governed by 
a stationary adiabatic response rather than high-frequency transitions. The condition, $\hbar\omega_{\rm{p}}
\lesssim\hbar\omega
\ll
\varepsilon_{\rm{F}}$, that suppresses screening simultaneously guarantees the 
internal consistency of both the Floquet description and the collective band framework.
}

\rev{The same argument clarifies the material platform  to realize the present theory. 
In chiral piezoelectric 
insulators and \tmag{low-carrier-density degenerate semiconductors with screw-axis 
symmetry}, where the finite-$\boldsymbol{q}$ chiral mode remains weakly screened 
over several oscillation cycles, the mechanism is expected to be relevant. In a good metal, by contrast, strong electronic 
screening would substantially reduce the dynamical piezoelectric field; \tmag{hence}, 
the present scenario \tor{is not expected to apply}.
}

\section{Coherent-Phonon Excitation and Wave-Vector Selectivity}

The inverse chiral phonon Zeeman effect discussed in this work is governed
predominantly by the occupation of a \textit{specific} chiral phonon mode with a
well-defined \tb{wave number $q = Q$} along the rotational axis.
\revdel{Therefore, the ability to selectively {exciting} and coherently {controlling} a target
phonon wavenumber provides a direct experimental handle to tune both the
magnitude and the sign of the induced Zeeman field.}
\rev{Therefore, the ability to selectively excite and coherently control a target phonon wave number provides a direct experimental handle for tuning both the magnitude and the sign of the induced Zeeman field.}
In practice, such mode selectivity is most naturally achieved by preparing the
phonon mode in a coherent state, which simultaneously allows one to address a
single wave vector and to enhance the magnetic-field amplitude.

For the single circularly polarized phonon mode, we consider a coherent phonon state
\begin{equation}
\lvert \alpha_{\pm}(\tb{Q}) \rangle = \exp\left( -\frac{1}{2} \lvert \alpha_{\pm} \rvert^2 \right) \sum_{n=0}^{\infty} \frac{\alpha_{\pm}^n}{\sqrt{n!}} \lvert n_{\pm}(\tb{Q}) \rangle,
\label{eq:13}
\end{equation}
which satisfies $\langle \alpha_{\pm} \rvert \hat{a}_{\pm,\tb{Q}}^{\dagger} \hat{a}_{\pm,\tb{Q}} \lvert \alpha_{\pm} \rangle = N_{\pm}(\tb{Q}) = \lvert \alpha_{\pm} \rvert^2$. In such a state, the expectation value of the lattice displacement acquires a classical amplitude $u_0$, providing a direct link between the coherent excitation and the phonon occupation number.

Using the material parameters adopted in Fig.~\ref{fig:Bpm}, the phonon occupation 
associated with a coherently excited elastic wave can be estimated directly 
from the lattice displacement amplitude. Specifically, the occupation number 
is related to $u_0$ as $N_{\pm}(\tb{Q}) \simeq (u_0/2u_{\text{zp}})^2$, 
with $u_{\text{zp}} = \sqrt{\hbar/2\rho V_0 \omega}$, where $u_{\text{zp}}$ 
is the zero-point displacement, $\rho$ is the mass density, $V_0$ is the unit 
cell volume, and $\omega$ is the phonon frequency. With the parameters used in 
Fig.~\ref{fig:Bpm} and a typical \tred{gapful-branch} frequency $\omega/2\pi \sim 10$~THz, 
one finds $u_{\text{zp}} \sim 6 \times 10^{-13}$~m $\approx 0.006$~\AA. 
Consequently, an \AA-scale coherent lattice displacement, which may correspond 
to the upper bound of physically reasonable amplitudes, leads to an occupation 
number at most of the order of $N_{\pm} \sim (u_0/2u_{\text{zp}})^2 \sim 10^4$.

\tmag{Since the phonon-driven effective Zeeman field scales linearly with the 
phonon occupation number\tb{, $B_{\rm eff} \propto N_{+}(Q)[B^{(1)}_{+}(Q)]^2 - N_{-}(Q)[B^{(1)}_{-}(Q)]^2$ [Eq.~(\ref{Beff})]}, 
preparing the target lattice mode in a coherent state enables a collective 
enhancement of the spin-splitting effect across both electronic regimes. 
For the localized states in the insulating matrix, this macroscopic population 
directly amplifies the Floquet-induced Zeeman splitting. Concurrently, for the 
itinerant electronic subsystem, the steady-state \tb{helicity imbalance of the coherently populated mode} 
amplifies the rigid adiabatic shift of the spin subbands by orders of magnitude 
compared to the vacuum-fluctuation baseline. Thus, the high photon-to-phonon 
conversion efficiency provided by modern intense terahertz pulses directly 
translates into an experimentally detectable spin-dependent energy splitting at 
the macroscopic scale, offering \tor{robust signatures} in both insulating and low-density 
conducting channels.}

In addition, to facilitate comparison with experimentally reported phonon magnetic moments, 
which are often quoted in units of the Bohr magneton $\mu_{\text{B}}$, it is 
convenient to convert the internal magnetic-field scale into an equivalent 
magnetic moment per unit cell. Identifying the magnetization as 
$M = B/\mu_0$, we define $\mu_{\text{cell}} \equiv M V_0 = (V_0/\mu_0)B$, 
and $\mu_{\text{cell}}/\mu_{\text{B}} = (V_0/\mu_0\mu_{\text{B}})B$\tb{, where $B$ stands for the internal field amplitude under consideration, e.g., $B^{(1)}_{\pm}(Q)$ for a single phonon or $|B_{\pm}(Q)|$ for a coherent state}. 
For the unit-cell volume $V_0 = 3.988 \times 10^{-28}$~m$^3$ used throughout 
this work, this relation yields $\mu_{\text{cell}}/\mu_{\text{B}} \simeq 34.2 \times B\,[\text{T}]$, 
providing a direct correspondence between the internal magnetic-field amplitude 
and the effective phonon magnetic moment.

Finally, we note that selectively addressing a target wave \tb{number $Q$} 
can be realized by a momentum-selective drive---for example, a transient-grating 
setup that defines \tb{$Q$} or a patterned electromechanical transducer 
that launches a piezoelectric wave with \tb{$q = Q$} along the 
chiral axis. In this language, the driven chiral mode is prepared in a coherent 
state $\lvert \alpha_{\pm}(\tb{Q}) \rangle$ with 
$\langle \hat{a}_{\pm,\tb{Q}} \rangle = \alpha_{\pm}$ and 
$N_{\pm}\tb{(Q)} = \lvert \alpha_{\pm} \rvert^2$, so that \tmag{the phonon-induced effective} 
Zeeman field is directly tunable through the helicity imbalance of the coherently 
populated mode [see Eq.~(\ref{Beff}) for $B_{\rm eff}$ above]. 
\rev{From the experimental side, the essential control knob is therefore a coherent population imbalance between 
the two chiral branches at a well-defined finite $\boldsymbol{Q}$. Once such a state 
is realized, the resulting spin splitting may be probed through magneto-optical signals, 
spin-sensitive transport, or pump-probe spectroscopy that resolves the helicity 
dependence of the transient response.}

\section{Summary and Concluding Remarks}

We have demonstrated that chiral lattice dynamics in noncentrosymmetric 
crystals generate an internal magnetic field through dynamical piezoelectricity, 
even in the absence of external light or static magnetic fields. Within a 
microscopic description based on micropolar elasticity, this field acts as an 
effective longitudinal Zeeman field on electronic spins, \tmag{driving a 
 Floquet response for localized states and a stationary adiabatic 
band splitting for itinerant electrons}, provided that electronic screening is 
incomplete. The resulting inverse chiral phonon Zeeman effect establishes a 
lattice-driven mechanism for spin polarization and spin-dependent band splitting, 
providing a route to controlling magnetoelectric responses via chiral phonons.

\rev{The present theory is intentionally minimal. It neglects disorder, impurity 
scattering, and explicit anharmonic mode coupling. These effects broaden the 
phonon linewidth, shorten the coherence time, and limit the attainable 
nonequilibrium occupation number; \tmag{however}, they do not alter the symmetry 
principle by which a chiral piezoelectric mode generates an internal elliptically 
polarized magnetic field. The present work is therefore not intended as a 
material-specific quantitative prediction, but as a reduced effective framework 
that identifies the minimal ingredients required for the effect: a 
noncentrosymmetric piezoelectric crystal with finite-$\boldsymbol{q}$ truly 
chiral phonons, incomplete electronic screening, and a coherent or nonequilibrium 
protocol capable of establishing a helicity imbalance.}

\rev{This viewpoint also clarifies the relation to the inverse phonon Zeeman 
effect and dynamical multiferroicity discussed previously. Earlier formulations 
start from nearly degenerate optical modes and ask how a circular drive or 
circular ionic motion generates magnetization. Here, the crystal chirality itself 
pre-splits the counter-rotating branches, and the finite-momentum piezoelectric 
wave converts this intrinsic chirality into a propagating internal magnetic field. 
The present mechanism is thus not a rephrasing of 
$\boldsymbol{P}\times\dot{\boldsymbol{P}}$ at $\boldsymbol{q}=0$, but the 
magnetic manifestation of a symmetry-split dynamical piezoelectric mode.}

\begin{acknowledgments}
J.~K. acknowledges JSPS KAKENHI Grant Nos.~JP25K00962, JP25H02149, JP23H00091, and JP26H02234, and JST ERATO Grant No.~JPMJER2503.
J.~K. also acknowledges the JSPS International Joint Research Program (JRP-LEAD with UKRI, Grant No.~JPJSJRP20241710) and the CNRS International Research Project ``Chiral Field--Matter Interaction'' (IRP ChiFiMI) for the opportunities of discussion.
M.S.\ acknowledges support from JSPS KAKENHI Grant Nos.~JP25K07198, JP25H02112, JP22H05131, JP25H01609, and JP25H01251, JST CREST (Grant No.~JPMJCR24R5), and the JSPS Program for Forming J-PEAKS (Grant No.~JPJS00420230002).
\end{acknowledgments}

\appendix

\section{Quantization of the micropolar elastic theory}
\label{app:quant}

\subsection{Longitudinal  modes}

  Below, we construct a quantum version of the classical theory\cite{Kishine2020} using the standard second quantization procedure\cite{Feynman1972}.  

Following the scheme of the classical treatment, we consider plane waves propagating along the chiral $z$-axis.  The dynamics of the longitudinal micropolar elastic waves is determined by the set of equations for the $u_3$ and $\varphi_3$ fields, 
\begin{equation}  \label{EOM1}
\rho \ddot{u}_3 = A_{33} \partial^2_z u_3 + C_{33} \partial^2_z \varphi_3, 
\end{equation}
\begin{equation} \label{EOM2}
\rho j_3 \ddot{\varphi}_3 = C_{33} \partial^2_z u_3 + B_{33} \partial^2_z \varphi_3 - 2 \left( A_{66} - A_{69} \right) \varphi_3,  
\end{equation}
where the nonzero tensor components of the point group $622$ are associated with couplings $A_{ij}$, $B_{ij}$ and $C_{ij}$ of the micropolar strain tensors $\varepsilon_{kl}$  and $\gamma_{kl}$.\cite{Eringen1999}  Here, the elements of the four-rank tensor are represented in the Sirotin (generalized Voigt) scheme \cite{Sirotin1982}, $\rho$, $j_3$  are the mass density and the longitudinal component of the microinertia tensor, respectively.  

Eqs. (\ref{EOM1},\ref{EOM2}) can be obtained by minimizing the  Lagrangian density 
$$
\mathcal{L}_L =  \frac{\rho}{2} \dot{u}^2_3 (z) + \frac{\rho j_3}{2} \dot{\varphi}^2_3 (z) - \frac{A_{33}}{2} \left(\partial_z u_3 \right)^2    
$$
\begin{equation}
 - \frac{B_{33}}{2}  \left( \partial_z \varphi_3 \right)^2
  - C_{33} \partial_z u_3 \partial_z \varphi_3 
-  \left( A_{66} - A_{69} \right) \varphi^2_3 (z). 
\end{equation}

The momentum densities conjugated to the modes $u_3(z)$ and $\varphi_3(z)$ for each coordinate $z$ are $p_3 (z)  = {\delta \mathcal{L}_L}/{\delta \dot{u}_3(z)} = \rho \dot{u}_3(z)$  and $\Pi_3 (z) = {\delta \mathcal{L}_L}/{\delta \dot{\varphi}_3(z)} = \rho j_3 \dot{\varphi}_3(z)$, respectively.

Then the Hamiltonian density in the first quantization is 
$$
\mathcal{H}_L =    
 \frac{\hat{p}^2_3}{2\rho}  + \frac{{\hat{\Pi}}^2_3}{2\rho j_3}  + \frac{A_{33}}{2} \left( \partial_z \hat{u}_3 \right)^2 
$$
\begin{equation} \label{HL}
 + \frac{B_{33}}{2} \left( \partial_z \hat{\varphi}_3 \right)^2  +  C_{33} \partial_z \hat{u}_3 \partial_z \hat{\varphi}_3  
+ \left( A_{66} - A_{69} \right) \hat{\varphi}^2_3 (z), 
\end{equation}
where $\hat{u}_3$  and $\hat{p}_3$, $\hat{\varphi}_3$ and $\hat{\Pi}_3$ are operators that satisfy the commutation relations $\left[ \hat{u}_3 (z), \hat{p}_3 (z^{'}) \right] = i\hbar \delta (z-z^{'})$ and $\left[ \hat{\varphi}_3 (z), \hat{\Pi}_3 (z^{'}) \right] = i\hbar \delta (z-z^{'})$.  

Performing the Fourier transformation for the fields $ \hat{u}_3$, $\hat{\varphi}_3$ and their associated momenta, $\hat{u}_3(z) = V^{-1/2} \sum_q e^{iqz} \hat{u}_{3q}$, etc., where $V$ is a volume of the crystal,  one can introduce the second-quantized representations
 \begin{equation} \label{uVK}
\hat{u}_{3q} = \sqrt{\frac{\hbar}{2 \rho  \Omega^{(u)}_{3q}}} \, \left( \hat{a}_{3,q} + \hat{a}^{\dagger}_{3,-q} \right), 
\end{equation}
\begin{equation}  \label{pVK}
\hat{p}_{3q} = -i \sqrt{\frac{\rho  \hbar \Omega^{(u)}_{3q}}{2}} \, \left( \hat{a}_{3,q} - \hat{a}^{\dagger}_{3,-q} \right),   
\end{equation} 
\begin{equation} 
\hat{\varphi}_{3q} =  \sqrt{\frac{\hbar}{2 \rho  j_3 \Omega^{(\varphi)}_{3q}}} \, \left( \hat{b}_{3,q} + \hat{b}^{\dagger}_{3,-q} \right),
\end{equation} 
\begin{equation}  
\hat{\Pi}_{3q} = -i \sqrt{\frac{\rho j_3 \hbar \Omega^{(\varphi)}_{3q}}{2}} \, \left( \hat{b}_{3,q} - \hat{b}^{\dagger}_{3,-q} \right). 
\end{equation} 
The Bose-operators satisfy the standard commutation relations $\left[ \hat{a}_{3,q}, \hat{a}^{\dagger}_{3,q'} \right] = \delta_{q,q'}$, $\left[ \hat{b}_{3,q}, \hat{b}^{\dagger}_{3,q'} \right] = \delta_{q,q'} $ with all other commutators being zero.  

Upon integration of the crystal volume,  provided $\int dV  e^{i(q-q')z}=V\delta_{q,q'}$, the total  Hamiltonian of the longitudinal modes becomes 
$$
H_L =  \sum_q  \left[
\hbar \Omega^{(u)}_{3q}  \left(   \hat{a}^{\dagger}_{3,q}  \hat{a}_{3,q} + \frac12 \right)  
+  \hbar \Omega^{(\varphi)}_{3q} \left(  \hat{b}^{\dagger}_{3,q}  \hat{b}_{3,q} + \frac12 \right) 
\right.
$$
\begin{equation}  \label{SQHam}
\left.
+  \hbar K_{3q}  \left(   \hat{a}_{3,q} \hat{b}_{3,-q} +  \hat{a}_{3,q} \hat{b}^{\dagger}_{3,q} +  \hat{a}^{\dagger}_{3,-q}  \hat{b}_{3,-q} +  \hat{a}^{\dagger}_{3,-q} \hat{b}^{\dagger}_{3,q}  \right) 
\right]. 
\end{equation}
 where  
\begin{equation} 
\Omega^{(u)}_{3q} = \sqrt{\frac{A_{33}}{\rho}} \, |q|, 
\end{equation} 
\begin{equation} 
\Omega^{(\varphi)}_{3q} = \sqrt{\frac{  B_{33} q^2 + 2 \left( A_{66} - A_{69} \right) }{\rho j_3}}, 
\end{equation} 
and 
\begin{equation} 
2 K_{3q}= \frac{C_{33}q^2}{\sqrt{\rho^2 j_3 \Omega^{(u)}_{3q}  \Omega^{(\varphi)}_{3q}}}. 
\end{equation} 

  To diagonalize the Hamiltonian (\ref{SQHam}) the Bogolyubov transformation is used
\begin{equation} \label{aDecomp-app}
\hat{a}_{3,q} = u_{11}(q) \hat{\alpha}_{3,q} + u_{12}(q) \hat{\beta}_{3,q} + v^{\ast}_{11}(-q) \hat{\alpha}^{\dagger}_{3,-q} + v^{\ast}_{12}(-q)  \hat{\beta}^{\dagger}_{3,-q}, 
\end{equation}
\begin{equation}  \label{bDecomp-app}
\hat{b}_{3,q} = u_{21}(q) \hat{\alpha}_{3,q} + u_{22}(q) \hat{\beta}_{3,q} + v^{\ast}_{21}(-q) \hat{\alpha}^{\dagger}_{3,-q} + v^{\ast}_{22}(-q)  \hat{\beta}^{\dagger}_{3,-q},  
\end{equation}
where the $u-v$ coefficients are listed in the Supplemental Material~\cite{SupplementalMaterialArXiv}.  The technique to obtain their explicit form is based on equations of motion for the double-time  Green functions  \cite{Krivoruchko1981}.

As a result, one has
\begin{equation}  \label{Hdiag-app}
\tilde{H}_L =  \Delta E_0 + \sum_q \left( \hbar \tred{\omega^{(\gl)}_{3q}}  \hat{\alpha}^{\dagger}_{3,q} \hat{\alpha}_{3,q}  + \hbar \tred{\omega^{(\gf)}_{3q}}  \hat{\beta}^{\dagger}_{3,q} \hat{\beta}_{3,q}  \right),     
\end{equation}
where $\Delta E_0$ is a correction term of the vacuum energy,  $\hat{\alpha}^{\dagger}_{3,q}$ ($\hat{\alpha}_{3,q}$)  and $\hat{\beta}^{\dagger}_{3,q}$  $(\hat{\beta}^{\dagger}_{3,q})$ are the pertinent creation (annihilation) Bose-operators for the \tred{gapless (gl) and gapful (gf) branches} of the excitation spectrum, respectively. 

\begin{widetext}
The latter are defined by the expressions
\begin{equation}
 \left[ \tred{\omega^{(\gf/\gl)}_{3q}} \right]^2 = \frac12 \left[   \left( \Omega^{(\varphi)}_{3q} \right)^2 + \left( \Omega^{(u)}_{3q} \right)^2 \right]  
\pm \frac12 \left\{  
\left[    \left( \Omega^{(\varphi)}_{3q} \right)^2 - \left( \Omega^{(u)}_{3q} \right)^2 \right]^2  + 16 K^2_{3q} \Omega^{(\varphi)}_{3q} \Omega^{(u)}_{3q} 
\right\}^{1/2},
\end{equation} 
and coincide with the result of the classical theory\cite{Kishine2020}
\begin{equation}
\left[ \tred{\omega^{(\gf/\gl)}_{3q}}\right]^2 = \frac{1}{2\rho j_3} \left[ b_3(q) + j_3 a_3(q)  
\pm \sqrt{\left[ b_3 (q) - j_3 a_3 (q) \right]^2  + 4 j_3 \Delta^2_3 (q)}  \right]
\end{equation} 
with $a_3 (q) = A_{33} q^2$, $b_3 (q) = B_{33} q^2 + 2 \left( A_{66} - A_{69} \right)$   and $\Delta_3 (q) = C_{33} q^2$.
\end{widetext}

\tmag{Henceforth, we assume a discrete atomic structure of the crystal with total volume $V=V_0 N$, where $V_0$ is the elementary cell volume and $N$ is the number of atomic sites, then the discrete Fourier transform reads $u_3 (\boldsymbol{r}_i)=1/\sqrt{N}\sum_{\boldsymbol{q}} u_{3\boldsymbol{q}} e^{i\boldsymbol{q} \boldsymbol{r}_i}$ etc.} 

\begin{widetext}
In this representation, the quantized field operators are determined as
\begin{equation}
\hat{u}_{3q} = \sqrt{\frac{\hbar}{2 \rho V_0 \Omega^{(u)}_{3q}}} \,  
\left\{
\left[ u_{11}(q) + v_{11}(q) \right] \hat{\alpha}_{3,q}
+ 
\left[ u_{12}(q) + v_{12}(q) \right] \hat{\beta}_{3,q}
\right.
\end{equation}
$$
\left.  
+
\left[ u^{\ast}_{11}(-q) + v^{\ast}_{11}(-q) \right] \hat{\alpha}^{\dagger}_{3,-q}
+ 
\left[ u^{\ast}_{12}(-q) + v^{\ast}_{12}(-q) \right] \hat{\beta}^{\dagger}_{3,-q}
\right\},
$$
\begin{equation} 
\hat{\varphi}_{3q} = \sqrt{\frac{\hbar}{2 \rho V_0  j_3 \Omega^{(\varphi)}_{3q}}} \, 
\left\{
\left[ u_{21}(q) + v_{21}(q) \right] \hat{\alpha}_{3,q}
+ \left[ u_{22}(q) + v_{22}(q) \right] \hat{\beta}_{3,q}
\right. 
\end{equation}
$$
\left.  
+
\left[ u^{\ast}_{21}(-q) + v^{\ast}_{21}(-q) \right] \hat{\alpha}^{\dagger}_{3,-q}
+ 
\left[ u^{\ast}_{22}(-q) + v^{\ast}_{22}(-q) \right] \hat{\beta}^{\dagger}_{3,-q}
\right\}. 
$$

\subsection{Transverse modes}

Below, we discuss the procedure for quantization of transverse modes. 

It is easy to see that the Lagrangian density 

$$
\mathcal{L}_T =   \sum_{i=1}^2 \left[ \frac{\rho}{2} \dot{u}^2_i  - \frac{A_{55}}{2}  \left( \partial_z u_i \right)^2   \right]
+ \sum_{i=1}^2 \left[  \frac{\rho j_1}{2} \dot{\varphi}^2_i -  \frac{B_{44}}{2} \left( \partial_z \varphi_i \right)^2   \right] 
 -\frac12 \left( A_{44} - 2 A_{47}  + A_{55}  \right) \sum_{i=1}^2  \varphi^2_i  
$$ 
\begin{equation}  \label{LT1}
- C_{74} \sum_{i=1}^2   \partial_z \varphi_i  \partial_z u_i     
+  \left( C_{44} - C_{74}  \right)  \sum_{i,j=1}^2   \epsilon_{ij} \varphi_i  \partial_z \varphi_j 
+  \left( A_{47} - A_{55} \right)  \sum_{i,j=1}^2   \epsilon_{ij} u_i  \partial_z \varphi_j,    
\end{equation} 
\end{widetext}
where $\epsilon_{ij}$ is the Levi-Civita symbol, leads to the equations of motion for the transverse components of \tor{microdisplacements} $u_{1,2}$ and microrotations $\varphi_{1,2}$
\begin{equation}
\rho \ddot{u}_1  = A_{55}  \partial^2_z u_1 + C_{74}  \partial^2_z \varphi_1 + \left( A_{47} - A_{55} \right) \partial_z \varphi_2, 
\end{equation} 
\begin{equation}
\rho \ddot{u}_2 = A_{55}  \partial^2_z u_2 + C_{74}  \partial^2_z \varphi_2 -  \left( A_{47} - A_{55} \right) \partial_z \varphi_1,  
\end{equation} 
$$
\rho j_1 \ddot{\varphi}_1  =  B_{44}  \partial^2_z \varphi_1  - \left( A_{44} - 2 A_{47} + A_{55} \right)  \varphi_1  
$$
\begin{equation}
+ C_{74}  \partial^2_z u_1 + 2 \left( C_{44} - C_{74} \right)  \partial_z \varphi_2
 +  \left( A_{47} - A_{55} \right)  \partial_z u_2, 
\end{equation} 
$$
\rho j_1 \ddot{\varphi}_2 =  B_{44}  \partial^2_z \varphi_2  - \left( A_{44} - 2 A_{47} + A_{55} \right)  \varphi_2  
$$
\begin{equation}
+ C_{74} \partial^2_z u_2 - 2 \left( C_{44} - C_{74} \right) \partial_z \varphi_1 
 -  \left( A_{47} - A_{55} \right)  \partial_z u_1.  
\end{equation} 
Note that the last two terms in (\ref{LT1}) have the form of Lifshitz invariants, i.e. they are proportional to the gradient of the \tor{microrotation} field. 

Introducing  the canonical momentum densities $p_i (z)  =  \rho \dot{u}_i(z)$,  $\Pi_i (z) =  \rho j_1 \dot{\varphi}_i(z)$ ($i=1,2$)  and replacing them, together with the fields $u_i(z)$,  $\varphi_i(z)$, by the corresponding operators with the commutation relations 
\begin{equation}
 \left[ \hat{u}_i (z), \hat{p}_j (z')  \right] = i\hbar \delta_{ij} \delta (z-z'), 
\end{equation} 
\begin{equation}
 \left[ \hat{\varphi}_i (z), \hat{\Pi}_j (z')  \right] = i\hbar \delta_{ij} \delta (z-z')
\end{equation} 
leads to the first quantization Hamiltonian density
\begin{widetext} 
$$
\mathcal{H}_T =   \frac{1}{2\rho} \sum_{i=1}^2 \hat{p}^2_i   +  \frac{1}{2\rho j_1}  \sum_{i=1}^2  \hat{\Pi}^2_i +   \frac{A_{55}}{2} \sum_{i=1}^2   \left( \partial_z \hat{u}_i \right)^2  + \frac{B_{44}}{2}  \sum_{i=1}^2  \left( {\partial_z \hat{\varphi}_i} \right)^2 + \frac12 \left( A_{44} - 2 A_{47}  + A_{55}  \right) \sum_{i=1}^2  \hat{\varphi}^2_i   
$$
\begin{equation}
+  C_{74} \sum_{i=1}^2  \partial_z \hat{\varphi}_i  \partial_z \hat{u}_i  -  \left( C_{44} - C_{74}  \right)  \sum_{i,j=1}^2   \epsilon_{ij} \hat{\varphi}_i  \partial_z \hat{\varphi}_j 
-  \left( A_{47} - A_{55} \right)  \sum_{i,j=1}^2   \epsilon_{ij} \hat{u}_i  \partial_z \hat{\varphi}_j. 
\end{equation} 
\end{widetext}

Using  the momentum representation for the fields and momenta, $\hat{u}_{iq}$, $\hat{\varphi}_{iq}$, $\hat{p}_{iq}$  and   $\hat{\Pi}_{iq}$,  and expressing them in terms of the creation and annihilation operators, $\hat{a}^{\dagger}_{i,q}$ ($\hat{a}_{i,q}$) and $\hat{b}^{\dagger}_{i,q}$ ($\hat{b}_{i,q}$), which obey the Bose commutation rules $\left[ \hat{a}_{i,q},  \hat{a}^{\dagger}_{j,q'} \right] = \left[ \hat{b}_{i,q}, \hat{b}^{\dagger}_{j,q'} \right] =\delta_{q,q'}\delta_{ij}$,
\begin{equation} \label{uVK12}
\hat{u}_{iq} = \sqrt{\frac{\hbar}{2\rho \Omega^{(u)}_{1q}}} \, \left( \hat{a}_{i,q} + \hat{a}^{\dagger}_{i,-q} \right), 
\end{equation}
\begin{equation}  \label{pVK12}
\hat{p}_{iq} = - i \sqrt{\frac{\rho \hbar \Omega^{(u)}_{1q}}{2}} \, \left( \hat{a}_{i,q} - \hat{a}^{\dagger}_{i,-q} \right),  
\end{equation} 
\begin{equation} 
\hat{\varphi}_{iq} =  \sqrt{\frac{\hbar}{2 \rho j_1 \Omega^{(\varphi)}_{1q}}} \, \left( \hat{b}_{i,q} + \hat{b}^{\dagger}_{i,-q} \right),
\end{equation} 
\begin{equation}  
\hat{\Pi}_{iq} = -i \sqrt{\frac{\rho j_1 \hbar \Omega^{(\varphi)}_{1q}}{2}} \, \left( \hat{b}_{i,q} - \hat{b}^{\dagger}_{i,-q} \right), 
\end{equation} 
 one gets  the second quantization Hamiltonian density
\begin{widetext} 
$$
\mathcal{H}_T =  \sum_q \left[
\hbar \Omega^{(u)}_{1q} \left(  \hat{a}^{\dagger}_{1,q}  \hat{a}_{1,q} + \frac12 \right)  + \hbar \Omega^{(u)}_{2q} \left(  \hat{a}^{\dagger}_{2,q}  \hat{a}_{2,q} + \frac12 \right)
+  \hbar \Omega^{(\varphi)}_{1q} \left(   \hat{b}^{\dagger}_{1,q}  \hat{b}_{1,q}  + \frac12 \right) +  \hbar \Omega^{(\varphi)}_{1q} \left(   \hat{b}^{\dagger}_{2,q}  \hat{b}_{2,q}  + \frac12 \right) 
\right.  
$$
$$
+\hbar K_{0q} \left(
 \hat{a}_{1,q} \hat{b}_{1,-q} + \hat{a}_{1,q} \hat{b}^{\dagger}_{1,q} +  \hat{a}^{\dagger}_{1,-q} \hat{b}_{1,-q} + \hat{a}^{\dagger}_{1,-q} \hat{b}^{\dagger}_{1,q}
\right)
+\hbar K_{0q} \left(
 \hat{a}_{2,q} \hat{b}_{2,-q} + \hat{a}_{2,q} \hat{b}^{\dagger}_{2,q} +  \hat{a}^{\dagger}_{2,-q} \hat{b}_{2,-q} + \hat{a}^{\dagger}_{2,-q} \hat{b}^{\dagger}_{2,q}
\right)
$$
$$
+ i \hbar K_{1q} \left(
 \hat{b}_{1,q} \hat{b}_{2,-q} + \hat{b}_{1,q} \hat{b}^{\dagger}_{2,q} +  \hat{b}^{\dagger}_{1,-q} \hat{b}_{2,-q} + \hat{b}^{\dagger}_{1,-q} \hat{b}^{\dagger}_{2,q}
\right)
$$
\begin{equation} \label{Ht}
\left.
+  i\hbar K_{2q}  \left(
\hat{a}_{1,q} \hat{b}_{2,-q} + \hat{a}_{1,q} \hat{b}^{\dagger}_{2,q} + \hat{a}^{\dagger}_{1,-q} \hat{b}_{2,-q} + \hat{a}^{\dagger}_{1,-q} \hat{b}^{\dagger}_{2,q}
- \hat{a}_{2,q} \hat{b}_{1,-q} -  \hat{a}_{2,q} \hat{b}^{\dagger}_{1,q} -  \hat{a}^{\dagger}_{2,-q} \hat{b}_{1,-q} -  \hat{a}^{\dagger}_{2,-q} \hat{b}^{\dagger}_{1,q}
\right)
\right], 
\end{equation}
\end{widetext}
where 
\begin{equation} 
\Omega^{(u)}_{1q} =\Omega^{(u)}_{2q} = \sqrt{\frac{A_{55}}{\rho}} \, |q|, 
\end{equation} 
\begin{equation}
\Omega^{(\varphi)}_{1q} = \Omega^{(\varphi)}_{2q} = \sqrt{\frac{A_{44} - 2 A_{47} + A_{55} +  B_{44} q^2  }{\rho j_1}}, 
\end{equation}  

and the coupling coefficients are defined as
\begin{equation}
 K_{0q} = \frac{C_{74}q^2}{2\sqrt{\rho^2 j_1 \Omega^{(u)}_{1q}\Omega^{(\varphi)}_{1q}}},
\end{equation}
\begin{equation}
K_{1q} = \frac{q \left( C_{44} - C_{74}\right)}{\rho j_1 \Omega^{(\varphi)}_{1q}},
\end{equation}
\begin{equation}
K_{2q} = \frac{q \left( A_{47} - A_{55} \right)}{2\sqrt{\rho^2 j_1 \Omega^{(u)}_{1q} \Omega^{(\varphi)}_{1q}}}. 
\end{equation}

The diagonal form of the Hamiltonian (\ref{Ht}) is obtained using the Bogolyubov transformation 
\begin{equation}
\left(
\begin{array}{c}
\hat{\boldsymbol{c}}_q \\
\hat{\boldsymbol{c}}^{\dagger}_{-q}
\end{array}
\right)
=\left(
\begin{array}{cc}
 \hat{U}(q)  &  \hat{V}^{\ast}(-q) \\
\hat{V}(q)  &  \hat{U}^{\ast}(-q)
\end{array}
\right)
\left(
\begin{array}{c}
\hat{\boldsymbol{\gamma}}_q \\
\hat{\boldsymbol{\gamma}}^{\dagger}_{-q}  
\end{array}
\right)
\end{equation} 
with  $\hat{\boldsymbol{c}}^{T}_q = \left( \hat{a}_{1,q},\hat{a}_{2,q},\hat{b}_{1,q},\hat{b}_{2,q} \right)$  and $\hat{\boldsymbol{\gamma}}^{T}_q = \left( \hat{\alpha}_{+,q}, \hat{\alpha}_{-,q}, \hat{\beta}_{+,q}, \hat{\beta}_{-,q} \right)$.  $\hat{U}(q)$, $\hat{V}(q)$ are $4 \times 4$ matrices whose elements are the coefficients of the Bogolyubov transformation $\left[  \hat{U}(q)  \right]_{ij} = \tilde{u}_{ij}$, $\left[  \hat{V}(q)  \right]_{ij} = \tilde{v}_{ij}$.  Their rather tedious calculation is placed in the Supplemental Material~\cite{SupplementalMaterialArXiv}.

The final result for the quantized transverse Hamiltonian reads
\begin{equation} \label{HTrans-app}
\mathcal{H}_{T} = \Delta E^{(T)}_0 + \sum_{\tor{q},s}  \left[
\hbar \tred{\omega^{(\gl)}_{sq}} \hat{\alpha}^{\dagger}_{s,q} \hat{\alpha}_{s,q} 
+  \hbar \tred{\omega^{(\gf)}_{sq}} \hat{\beta}^{\dagger}_{s,q} \hat{\beta}_{s,q} 
\right],
\end{equation}
where $\Delta E^{(T)}_0$ is the vacuum energy. The index $s=+,-$ marks the left or right circularly polarized transverse modes, respectively.  The creation(annihilation) Bose-operators $\hat{\alpha}^{\dagger}_{s,q}$ ($\hat{\alpha}_{s,q}$)  and $\hat{\beta}^{\dagger}_{s,q}$ ($\hat{\beta}_{s,q}$) are related, respectively, to the \tred{gapless (gl) and gapful (gf) branches} of the elementary excitation spectrum
\begin{widetext}
\begin{equation}   \label{wtran1} 
 \left[ \tred{\omega^{(\gf/\gl)}_{+q}} \right]^2 = \frac12 \left[   \Omega^{(\varphi)2}_{1q} +   \Omega^{(u)2}_{1q}  - 2K_{1q} \Omega^{(\varphi)}_{1q}   \right] \pm \frac12 \left\{  \left[  \Omega^{(\varphi)2}_{1q}  - \Omega^{(u)2}_{1q} - 2K_{1q} \Omega^{(\varphi)}_{1q} \right]^2  + 16 \Omega^{(u)}_{1q}   \Omega^{(\varphi)}_{1q}  \left( K_{0q} - K_{2q}\right)^2  \right\}^{\frac12},
\end{equation}
\begin{equation}  \label{wtran2}
 \left[ \tred{\omega^{(\gf/\gl)}_{-q}} \right]^2 = \frac12 \left[  \Omega^{(\varphi)2}_{1q}  +   \Omega^{(u)2}_{1q}  + 2K_{1q} \Omega^{(\varphi)}_{1q}   \right] \pm \frac12 \left\{  \left[  \Omega^{(\varphi)2}_{1q}  -  \Omega^{(u)2}_{1q} + 2K_{1q} \Omega^{(\varphi)}_{1q} \right]^2  + 16 \Omega^{(u)}_{1q}   \Omega^{(\varphi)}_{1q}  \left( K_{0q} +  K_{2q}\right)^2  \right\}^{\frac12}. 
\end{equation} 
 
The spectra (\ref{wtran1},\ref{wtran2}) coincide exactly with that of the classical theory\cite{Kishine2020}
\begin{equation}
\left[ \tred{\omega^{(\gf/\gl)}_{\pm q}} \right]^2 = \frac{1}{2\rho j_1}  \left[  b_{\pm}(q) + j_1 a_{\pm} (q)  
\pm   \sqrt{\left( b_{\pm} (q) - j_1 a_{\pm}(q) \right)^2 + 4 j_1 \Delta^2_{\pm} (q) } \right],
\end{equation}
with $a_{\pm}(q)=A_{55}q^2$,  $b_{\pm}(q) = B_{44}q^2 \mp 2 (C_{44} -C_{74})q + (A_{44}+ A_{55}-2A_{47})$  and $\Delta_{\pm}(q) = C_{74}q^2 \mp (A_{47}-A_{55})q$. 

The quantization procedure yields the following expressions for the fields of microdisplacements ($i=1,2$)
$$
\hat{u}_{iq} = \sqrt{\frac{\hbar}{2\rho V_0 \Omega^{(u)}_{1q}}} \left\{
\left[ \tilde{u}_{i1}(q) + \tilde{v}_{i1}(q) \right] \hat{\alpha}_{+,q}  + \left[ \tilde{u}_{i2}(q) + \tilde{v}_{i2}(q) \right] \hat{\alpha}_{-,q}  
+ \left[ \tilde{u}_{i3}(q) + \tilde{v}_{i3}(q) \right] \hat{\beta}_{+,q} 
\right.
$$
$$
+ \left[ \tilde{u}_{i4}(q) + \tilde{v}_{i4}(q) \right] \hat{\beta}_{-,q}  + \left[  \tilde{u}^{\ast}_{i1}(-q) + \tilde{v}^{\ast}_{i1}(-q)  \right]   \hat{\alpha}^{\dagger}_{+,-q} + \left[  \tilde{u}^{\ast}_{i2}(-q) + \tilde{v}^{\ast}_{i2}(-q)  \right]   \hat{\alpha}^{\dagger}_{-,-q}
$$
\begin{equation} \label{QFu}
\left.
 + \left[  \tilde{u}^{\ast}_{i3}(-q) + \tilde{v}^{\ast}_{i3}(-q)  \right]   \hat{\beta}^{\dagger}_{+,-q} + \left[  \tilde{u}^{\ast}_{i4}(-q) + \tilde{v}^{\ast}_{i4}(-q)  \right]   \hat{\beta}^{\dagger}_{-,-q}
\right\}
\end{equation}
and microrotations 
$$
\hat{\varphi}_{iq} =  \sqrt{\frac{\hbar}{2 \rho V_0 j_1 \Omega^{(\varphi)}_{1q}}} \left\{
\left[ \tilde{u}_{i+2,1}(q) + \tilde{v}_{i+2,1}(q) \right] \hat{\alpha}_{+,q}  + \left[ \tilde{u}_{i+2,2}(q) + \tilde{v}_{i+2,2}(q) \right] \hat{\alpha}_{-,q}  
+ \left[ \tilde{u}_{i+2,3}(q) + \tilde{v}_{i+2,3}(q) \right] \hat{\beta}_{+,q} 
\right.
$$
$$
+ \left[ \tilde{u}_{i+2,4}(q) + \tilde{v}_{i+2,4}(q) \right] \hat{\beta}_{-,q} 
+ \left[  \tilde{u}^{\ast}_{i+2,1}(-q) + \tilde{v}^{\ast}_{i+2,1}(-q)  \right]   \hat{\alpha}^{\dagger}_{+,-q} + \left[  \tilde{u}^{\ast}_{i+2,2}(-q) + \tilde{v}^{\ast}_{i+2,2}(-q)  \right]   \hat{\alpha}^{\dagger}_{-,-q}
$$
\begin{equation}  \label{QFphi}
\left.
 + \left[  \tilde{u}^{\ast}_{i+2,3}(-q) + \tilde{v}^{\ast}_{i+2,3}(-q)  \right]   \hat{\beta}^{\dagger}_{+,-q} + \left[  \tilde{u}^{\ast}_{i+2,4}(-q) + \tilde{v}^{\ast}_{i+2,4}(-q)  \right]   \hat{\beta}^{\dagger}_{-,-q}
\right\}.
\end{equation}

Then, the time-dependent  circular components, $\hat{u}_{\pm,q} = \hat{u}_{1q} \pm i \hat{u}_{2q}$ and $\hat{\varphi}_{\pm,q} = \hat{\varphi}_{1q} \pm i \hat{\varphi}_{2q}$, are given by
\begin{equation}  \label{u+}
\hat{u}_{+,q}(t) = \sqrt{\frac{\hbar}{2\rho V_0 \Omega^{(u)}_{1q}}} \left\{ \tred{U^{(\gl)}_{+}}(q) \hat{\alpha}_{+,q}(t)  + \tred{U^{(\gf)}_{+}}(q) \hat{\beta}_{+,q} (t) + \tred{\overstar{U}^{(\gl)}_{-}}(-q) \hat{\alpha}^{\dagger}_{-,-q}(t) + \tred{\overstar{U}^{(\gf)}_{-}}(-q) \hat{\beta}^{\dagger}_{-,-q}(t)  \right\}, 
\end{equation}  
\begin{equation}  \label{u-}
\hat{u}_{-,q}(t) = \sqrt{\frac{\hbar}{2\rho V_0 \Omega^{(u)}_{1q}}} \left\{ \tred{U^{(\gl)}_{-}}(q) \hat{\alpha}_{-,q}(t)  + \tred{U^{(\gf)}_{-}}(q) \hat{\beta}_{-,q} (t) + \tred{\overstar{U}^{(\gl)}_{+}}(-q) \hat{\alpha}^{\dagger}_{+,-q}(t) + \tred{\overstar{U}^{(\gf)}_{+}}(-q) \hat{\beta}^{\dagger}_{+,-q}(t)  \right\}, 
\end{equation}  
\begin{equation} \label{phi+}
\hat{\varphi}_{+,q} (t) =  \sqrt{\frac{\hbar}{2 \rho V_0 j_1 \Omega^{(\varphi)}_{1q}}} \left\{ \tred{\Phi^{(\gl)}_{+}} (q)  \hat{\alpha}_{+,q}(t) +  \tred{\Phi^{(\gf)}_{+}} (q)  \hat{\beta}_{+,q}(t) +  \tred{\overstar{\Phi}^{(\gl)}_{-}}(-q) \hat{\alpha}^{\dagger}_{-,-q}(t) + \tred{\overstar{\Phi}^{(\gf)}_{-}}(-q) \hat{\beta}^{\dagger}_{-,-q}(t)  \right\},   
\end{equation}  
\begin{equation}  \label{phi-}
\hat{\varphi}_{-,q} (t) =  \sqrt{\frac{\hbar}{2 \rho V_0 j_1 \Omega^{(\varphi)}_{1q}}} \left\{ \tred{\Phi^{(\gl)}_{-}} (q)  \hat{\alpha}_{-,q}(t) +  \tred{\Phi^{(\gf)}_{-}} (q)  \hat{\beta}_{-,q}(t) +  \tred{\overstar{\Phi}^{(\gl)}_{+}}(-q) \hat{\alpha}^{\dagger}_{+,-q}(t) + \tred{\overstar{\Phi}^{(\gf)}_{+}}(-q) \hat{\beta}^{\dagger}_{+,-q}(t)  \right\}.  
\end{equation}  

Given
\begin{equation} \label{HTrans-app-2}
\mathcal{H}_{T} = \Delta E^{(T)}_0 + \sum_q \left[
\hbar \tred{\omega^{(\gl)}_{+}}(q) \alpha^{\dagger}_{+,q} \alpha_{+,q} 
+ \hbar \tred{\omega^{(\gl)}_{-}}(q) \alpha^{\dagger}_{-,q} \alpha_{-,q} 
+  \hbar \tred{\omega^{(\gf)}_{+}}(q) \beta^{\dagger}_{+,q} \beta_{+,q} 
+  \hbar \tred{\omega^{(\gf)}_{-}}(q) \beta^{\dagger}_{-,q} \beta_{-,q} 
\right],
\end{equation}
the time dependence of the coordinates is determined by the time dependencies of the creation/annihilation operators
\begin{equation}
\hat{\alpha}_{+,q}(t)  = e^{-i \tred{\omega^{(\gl)}_{+}}(q) t} \hat{\alpha}_{+,q}, \quad   
\hat{\beta}_{+,q}(t)  = e^{-i \tred{\omega^{(\gf)}_{+}}(q) t} \hat{\beta}_{+,q}, \quad
\hat{\alpha}^{\dagger}_{-,-q}(t) =  e^{i \tred{\omega^{(\gl)}_{+}}(q) t} \hat{\alpha}^{\dagger}_{-,-q}, \quad
\hat{\beta}^{\dagger}_{-,-q}(t) = e^{i \tred{\omega^{(\gf)}_{+}}(q) t} \hat{\beta}^{\dagger}_{-,-q},  
\end{equation}
\begin{equation}
\hat{\alpha}_{-,q}(t)  = e^{-i \tred{\omega^{(\gl)}_{-}}(q) t} \hat{\alpha}_{-,q}, \quad
\hat{\beta}_{-,q}(t)  = e^{-i \tred{\omega^{(\gf)}_{-}}(q) t} \hat{\beta}_{-,q}, \quad
\hat{\alpha}^{\dagger}_{+,-q}(t) =  e^{i \tred{\omega^{(\gl)}_{-}}(q) t} \hat{\alpha}^{\dagger}_{+,-q}, \quad
\hat{\beta}^{\dagger}_{+,-q}(t) = e^{i \tred{\omega^{(\gf)}_{-}}(q) t} \hat{\beta}^{\dagger}_{+,-q}.   
\end{equation}

Here,
\begin{equation}  
\tred{U^{(\gl)}_{+}}(q) = \tilde{u}_{11}(q) +  \tilde{v}_{11}(q)  + i \tilde{u}_{21}(q) + i \tilde{v}_{21}(q), \quad 
\tred{U^{(\gl)}_{-}}(q) = \tilde{u}_{12}(q) +  \tilde{v}_{12}(q)  -  i \tilde{u}_{22}(q) -  i \tilde{v}_{22}(q), 
\end{equation}
\begin{equation}  \label{LB3}
\tred{U^{(\gf)}_{+}}(q) = \tilde{u}_{13}(q) +  \tilde{v}_{13}(q)  + i \tilde{u}_{23}(q) + i \tilde{v}_{23}(q),  \quad 
\tred{U^{(\gf)}_{-}}(q) = \tilde{u}_{14}(q) +  \tilde{v}_{14}(q)  -  i \tilde{u}_{24}(q) -  i \tilde{v}_{24}(q),   
\end{equation}
\begin{equation}  \label{LB5}
\tred{\Phi^{(\gl)}_{+}} (q)  = \tilde{u}_{31}(q) + \tilde{v}_{31}(q) + i \tilde{u}_{41}(q) +  i \tilde{v}_{41}(q),  \quad 
\tred{\Phi^{(\gl)}_{-}} (q)  = \tilde{u}_{32}(q) + \tilde{v}_{32}(q) - i \tilde{u}_{42}(q) -  i \tilde{v}_{42}(q), 
\end{equation}
\begin{equation}  \label{LB7}
\tred{\Phi^{(\gf)}_{+}} (q)  = \tilde{u}_{33}(q) + \tilde{v}_{33}(q) + i \tilde{u}_{43}(q) + i \tilde{v}_{43}(q),  \quad 
\tred{\Phi^{(\gf)}_{-}} (q)  = \tilde{u}_{34}(q) + \tilde{v}_{34}(q) - i \tilde{u}_{44}(q) - i \tilde{v}_{44}(q),   
\end{equation}
and the relationships 
\begin{equation} 
\tilde{u}_{12}(q) + \tilde{v}_{12}(q) + i \tilde{u}_{22}(q) + i \tilde{v}_{22}(q) =0,  \qquad 
\tilde{u}_{14}(q) + \tilde{v}_{14}(q) + i \tilde{u}_{24}(q) + i \tilde{v}_{24}(q) =0, 
\end{equation}
\begin{equation} 
\tilde{u}_{11}(q) + \tilde{v}_{11}(q) - i \tilde{u}_{21}(q) - i \tilde{v}_{21}(q) =0,   \qquad 
\tilde{u}_{13}(q) + \tilde{v}_{13}(q) - i \tilde{u}_{23}(q) - i \tilde{v}_{23}(q) =0,    
\end{equation}
\begin{equation}
\tilde{u}_{32}(q) + \tilde{v}_{32}(q) + i \tilde{u}_{42}(q) + i \tilde{v}_{42}(q) =0, \qquad 
\tilde{u}_{34}(q) + \tilde{v}_{34}(q) + i \tilde{u}_{44}(q) + i \tilde{v}_{44}(q) =0, 
\end{equation}
\begin{equation}
\tilde{u}_{31}(q) + \tilde{v}_{31}(q) - i \tilde{u}_{41}(q) - i \tilde{v}_{41}(q) =0,  \qquad 
\tilde{u}_{33}(q) + \tilde{v}_{33}(q) - i \tilde{u}_{43}(q) - i \tilde{v}_{43}(q) =0   
\end{equation}
are accounted for.  

\end{widetext}

\section{Electromagnetic fields in piezoelectric micropolar medium}
\label{app:em}

It is well known that acoustic phonons, which are periodic density modulation, make periodic electric fields in piezoelectric crystals. This effect underlies the piezoelectric mechanism of electron-phonon interaction, which can occur in crystals with a lack of an inversion center.\cite{Mahan1990}  Below, we demonstrate that elastic waves in a micropolar medium generate a periodic electric polarization field, which leads to the emergence of circularly polarized magnetic fields, whose sum has elliptical polarization. 

In order to establish a coupling between  the \tred{micropolar deformation fields} and the electric polarization  $\boldsymbol{P}$, we use the equations of motion of the micropolar piezoelectric medium\cite{Eringen1999,Ivanova2013}
\begin{equation}
\rho \ddot{u}_l = \rho f_l + \frac{\partial \sigma_{kl}}{\partial x_k} - \frac12 \left(\boldsymbol{\nabla}\times\boldsymbol{q} \right)_l,  
\end{equation}
\begin{equation}
\rho j_{lk} \ddot{\varphi}_{k} = \rho l_l + q_l -  \left(\boldsymbol{\nabla}\times\boldsymbol{m} \right)_l.
\end{equation}
Here  $\hat{\boldsymbol\sigma}$ is the symmetric part of stress tensor, $\boldsymbol{q}$ is the vector characterizing the antisymmetric part of stress tensor, $\boldsymbol{m}$ is the vector characterizing the antisymmetric part of momentum stress tensor, $\rho$ is the mass density and $\hat{\boldsymbol{j}}$ is the inertia tensor per unit mass. 

The energy balance  equation reads 
\begin{equation}
\rho \dot{\varepsilon} =  \sigma_{kl} \dot{\varepsilon}_{kl}  - \boldsymbol{q} \cdot\dot{\boldsymbol\Theta} + \boldsymbol{m} \cdot\dot{\boldsymbol\gamma} + \frac{\partial h_k}{\partial x_k} + \mathcal{Q},  \end{equation}
where $\boldsymbol{h}$ is the heat flow vector, $\mathcal{Q}$ is  the rate of energy supply from an external source. The strain vectors $\boldsymbol\Theta$ and $\boldsymbol\gamma$ are connected with translational displacements and microrotations by the relations
\begin{equation}
\boldsymbol\Theta = \boldsymbol\varphi - \frac12\boldsymbol{\nabla}\times\boldsymbol{u}, \qquad 
\boldsymbol\gamma =\boldsymbol{\nabla}\times\boldsymbol\varphi. 
\end{equation}

The rate of work done by an external electric field $\boldsymbol{E}$ is $\mathcal{Q} = \boldsymbol{E} \cdot  \dot{\boldsymbol{P}}$ with $\boldsymbol{P}$ being the electric polarization. 
  
Therefore, the inner energy $U$ becomes
\begin{equation}
dU =  \sigma_{kl} d \varepsilon_{kl} - q_k d \Theta_k + m_k d \gamma_k + E_k d P_k. 
\end{equation}

When the electric field is maintained constant by experimental conditions, it is convenient to introduce the electric enthalpy $H_{E} = U - \boldsymbol{E} \cdot  \boldsymbol{P}$, for which
\begin{equation}
dH_{E} =  \sigma_{kl} d \varepsilon_{kl} - q_k d \Theta_k + m_k d \gamma_k -   P_k   dE_k. 
\end{equation}
 
The $\hat{\boldsymbol\sigma}$, $\boldsymbol{q}$, $\boldsymbol{m}$ and $\boldsymbol{P}$  are obtained as partial derivatives of  the relevant potential 
$$
\sigma_{kl} = \frac{\partial H_E}{\partial  \varepsilon_{kl}}, \qquad  q_k = - \frac{\partial H_E}{\partial \Theta_k}, 
$$
\begin{equation} \label{CE_DHE}
m_k = \frac{\partial H_E}{\partial \gamma_k},  \qquad P_k = - \frac{\partial H_E}{\partial E_k}.
\end{equation}

The general form of the functional that allows the derivation of the constitutive equations for a piezoelectric material is written as\cite{Ivanova2012} 
$$
H_E = \frac12 \varepsilon_{\lambda} \hat{C}_{\lambda \mu} \varepsilon_{\mu} - \frac12 E_k \hat{\kappa}_{kl} E_l + \frac12 \Theta_k \hat{Y}_{kl} \Theta_l  
$$
$$
+  \frac12 \gamma_k \hat{\chi}_{kl} \gamma_l + \varepsilon_{\lambda} \hat{N}_{\lambda k} \Theta_k  - E_k \hat{d}_{k\lambda} \varepsilon_{\lambda}  
$$
\begin{equation} \label{Hentalp}
- E_k \hat{X}_{kl} \Theta_l - E_k \hat{\Gamma}^{(1)}_{kl} \gamma_l + \varepsilon_{\lambda} \hat{\Gamma}^{(2)}_{\lambda l} \gamma_l  + \Theta_k \hat{\Gamma}^{(3)}_{kl} \gamma_l,   
\end{equation}
where $k,l=1,2,3$,  and the Voigt notations are used,  $\lambda=1,2,\ldots,6$,  so that  $\varepsilon_1=\varepsilon_{xx}$, $\varepsilon_2=\varepsilon_{yy}$, $\varepsilon_3=\varepsilon_{zz}$, $\varepsilon_4=2\varepsilon_{yz}$, $\varepsilon_5=2\varepsilon_{xz}$, $\varepsilon_6=2\varepsilon_{xy}$.  

The tensors $\hat{C}$, $\hat{\kappa}$, $\hat{Y}$, $\hat{\chi}$, $\hat{N}$, $\hat{d}$, $\hat{X}$, $\hat{\Gamma}^{(1,2,3)}$ specify physical properties of the material.    Besides the permittivity tensor $\hat{\kappa}$ and the conventional tensor of piezoelectric moduli $\hat{d}$, two additional piezoelectric  tensors are introduced, $\hat{X}$ and $\hat{\Gamma}^{(1)}$,  which are related with the rotational degrees of freedom, $\Theta_l$ and $\gamma_l$, respectively.   Certain constraints are imposed on these tensors, which arise from the condition of positive definiteness of the internal energy $U$.

  \tor{Inserting} (\ref{Hentalp}) into the last relationship in (\ref{CE_DHE}), we are immediately led to the  constitutive relations 
\begin{equation}  \label{CED}
P_k = \hat{\kappa}_{kl} E_l + \hat{d}_{k \lambda} \varepsilon_{\lambda} + \hat{X}_{kl} \Theta_{l} + \hat{\Gamma}^{(1)}_{kl} \gamma_l. 
\end{equation}

\tb{Throughout this work we have in mind, as one representative example, a hexagonal crystal of the CrNb${}_3$S${}_6$ type with space group $P6_322$~\cite{Kishine2020}, although the formalism applies to any crystal of the point group $622$.} The piezoelectric tensors of the point group $622$ have been tabulated in Ref.\cite{Sirotin1982} and appear as 
 $$
\hat{d} = \left(
\begin{array}{cccccc}
0 &  0  &  0  &  d_{14} & 0   &  0 \\
0 &  0 &   0 &  0 &  - d_{14} &  0 \\
0 & 0 & 0 & 0 & 0 & 0  
\end{array}
\right), 
$$
\begin{equation}
\hat{X} = \left(
\begin{array}{ccc}
X_{11} & 0 & 0 \\
0 & X_{11} & 0 \\
0 & 0 & X_{33} 
\end{array}
\right),  \quad
\hat{\Gamma}^{(1)} = \left(
\begin{array}{ccc}
\Gamma^{(1)}_{11} & 0 & 0 \\
0 & \Gamma^{(1)}_{11} & 0 \\
0 & 0 & \Gamma^{(1)}_{33}
\end{array}
\right),  
\end{equation}
the permittivity tensor has the diagonal form
\begin{equation}
\hat{\kappa} = \left(
\begin{array}{ccc}
\kappa_{11} & 0 & 0 \\
0 & \kappa_{11} & 0 \\
0 & 0 & \kappa_{33}
\end{array}
\right). 
\end{equation}

In the absence of the electric field, micropolar plane waves propagating along the crystalline $[001]$ axis  give rise to the electrical polarization $\boldsymbol{P}$ with two completely decoupled circularly polarized transverse components, $P_{\pm} = P_1 \pm i P_2$, 
$$
 P_{\pm} (z,t) = \mp i d_{14} \partial_z u_{\pm}(z,t) + X_{11} \varphi_{\pm}(z,t) 
$$
\begin{equation}
\mp \frac{i}{2} X_{11} \partial_z u_{\pm} (z,t)  \pm i \Gamma^{(1)}_{11} \partial_z \varphi_{\pm}(z,t), 
\end{equation}
and with the $z$-component
\begin{equation}
P_3(z,t) = X_{33} \varphi_3(z,t). 
\end{equation}
Here, $u_{\pm} = u_1 \pm i u_2$ and $\varphi_{\pm}= \varphi_1 \pm i \varphi_2$. 

In the momentum representation  we have

\begin{widetext}
\begin{equation}  \label{P+P-}
\hat{P}_{\pm,q}(t) = \pm q \left( d_{14} + \frac12 X_{11} \right) \hat{u}_{\pm,q}(t)  + \left( X_{11} \mp q \Gamma^{(1)}_{11} \right) \hat{\varphi}_{\pm,q}(t).  
\end{equation}

Using (\ref{u+}-\ref{phi-}), this result is expressed in the second-quantization form
$$
\hat{P}_{\pm,q}(t) = \pm q \left( d_{14} + \frac12 X_{11} \right)
$$
$$
\times 
 \sqrt{\frac{\hbar}{2\rho V_0 \Omega^{(u)}_{1q}}} \left\{ \tred{U^{(\gl)}_{\pm}}(q) e^{-i \tred{\omega^{(\gl)}_{\pm}}(q) t}  \hat{\alpha}_{\pm,q}  + \tred{U^{(\gf)}_{\pm}}(q) e^{-i \tred{\omega^{(\gf)}_{\pm}}(q) t}  \hat{\beta}_{\pm,q}  + \tred{\overstar{U}^{(\gl)}_{\mp}}(-q)  e^{i \tred{\omega^{(\gl)}_{\pm}}(q) t}  \hat{\alpha}^{\dagger}_{\mp,-q} + \tred{\overstar{U}^{(\gf)}_{\mp}}(-q)  e^{i \tred{\omega^{(\gf)}_{\pm}}(q) t} \hat{\beta}^{\dagger}_{\mp,-q}  \right\}
$$
$$
+ \left( X_{11} \mp q \Gamma^{(1)}_{11} \right)
$$
\begin{equation} \label{Ppk}
\times  \sqrt{\frac{\hbar}{2 \rho V_0 j_1 \Omega^{(\varphi)}_{1q}}} \left\{ \tred{\Phi^{(\gl)}_{\pm}} (q)  e^{-i \tred{\omega^{(\gl)}_{\pm}}(q) t}  \hat{\alpha}_{\pm,q} +  \tred{\Phi^{(\gf)}_{\pm}} (q)  e^{-i \tred{\omega^{(\gf)}_{\pm}}(q) t} \hat{\beta}_{\pm,q} +  \tred{\overstar{\Phi}^{(\gl)}_{\mp}}(-q) e^{i \tred{\omega^{(\gl)}_{\pm}}(q) t} \hat{\alpha}^{\dagger}_{\mp,-q} + \tred{\overstar{\Phi}^{(\gf)}_{\mp}}(-q) e^{i \tred{\omega^{(\gf)}_{\pm}}(q) t}  \hat{\beta}^{\dagger}_{\mp,-q}  \right\}.    
\end{equation}

\end{widetext}

From Maxwell's equations
\begin{equation}
\boldsymbol{\nabla}\cdot \boldsymbol{B}=0, \quad
\boldsymbol{\nabla}\times \boldsymbol{B} = \mu_0 \dot{\boldsymbol{P}}, 
\end{equation}
we have in the momentum representation
\begin{equation}
\left( \boldsymbol{q} \cdot \boldsymbol{B} (\boldsymbol{q},t)\right) = 0, 
\end{equation}
\begin{equation}
\boldsymbol{B} (\boldsymbol{q},t) = i \frac{\mu_0}{q^2} \left[ \boldsymbol{q} \times \dot{\boldsymbol{P}} (\boldsymbol{q},t)  \right].
\end{equation}

Then, we obtain only the non-zero transverse components of the magnetic field 
\begin{equation}
 B_1 (q,t) = - i \frac{\mu_0}{q} \dot{P}_2 (q,t),  \qquad B_2 (q,t) =  i \frac{\mu_0}{q} \dot{P}_1 (q,t),  
\end{equation}
or, in the circular basis, $B_{\pm}=\tor{B_1} \pm i B_2$,  
\begin{equation}
B_{+}(q,t) = - \frac{\mu_0}{q} \dot{P}_{+} (q,t), \qquad B_{-}(q,t) =  \frac{\mu_0}{q} \dot{P}_{-} (q,t). 
\end{equation}

Using Eq. (\ref{P+P-}), we obtain the relationship between the magnetic fields and  the \tred{micropolar deformation fields}  
\begin{widetext}
\begin{equation}
\hat{B}_{\pm,q}(t) = \mp \frac{\mu_0}{q} \left[ \pm  q \left( d_{14} + \frac12 X_{11} \right) \dot{\hat{u}}_{\pm,q}(t)  + \left( X_{11} \mp  q \Gamma^{(1)}_{11} \right) \dot{\hat{\varphi}}_{\pm,q}(t) \right].  
\end{equation}

The expressions  (\ref{u+}-\ref{phi-}) provide the second-quantization form of the magnetic fields
$$
\hat{B}_{\pm,q}(t) = - \mu_0 \left( d_{14} + \frac12 X_{11} \right)  
$$
$$
\times  \sqrt{\frac{\hbar}{2\rho V_0 \Omega^{(u)}_{1q}}} \left\{ -i \tred{\omega^{(\gl)}_{\pm}}(q)  \tred{U^{(\gl)}_{\pm}}(q) e^{-i \tred{\omega^{(\gl)}_{\pm}}(q) t}  \hat{\alpha}_{\pm,q}  -i \tred{\omega^{(\gf)}_{\pm}}(q) \tred{U^{(\gf)}_{\pm}}(q) e^{-i \tred{\omega^{(\gf)}_{\pm}}(q) t}  \hat{\beta}_{\pm,q}  
\right. 
$$
$$
\left. 
+ i \tred{\omega^{(\gl)}_{\pm}}(q) \tred{\overstar{U}^{(\gl)}_{\mp}}(-q)  e^{i \tred{\omega^{(\gl)}_{\pm}}(q) t}  \hat{\alpha}^{\dagger}_{\mp,-q} + i \tred{\omega^{(\gf)}_{\pm}}(q) \tred{\overstar{U}^{(\gf)}_{\mp}}(-q)  e^{i \tred{\omega^{(\gf)}_{\pm}}(q) t} \hat{\beta}^{\dagger}_{\mp,-q}  \right\}
$$
$$
 \mp \frac{\mu_0}{q}  \left( X_{11} \mp q \Gamma^{(1)}_{11} \right)  
$$
$$
\times  \sqrt{\frac{\hbar}{2 \rho V_0 j_1 \Omega^{(\varphi)}_{1q}}} \left\{ -i \tred{\omega^{(\gl)}_{\pm}}(q)  \tred{\Phi^{(\gl)}_{\pm}} (q)  e^{-i \tred{\omega^{(\gl)}_{\pm}}(q) t}  \hat{\alpha}_{\pm,q} -i \tred{\omega^{(\gf)}_{\pm}}(q)  \tred{\Phi^{(\gf)}_{\pm}} (q)  e^{-i \tred{\omega^{(\gf)}_{\pm}}(q) t} \hat{\beta}_{\pm,q}   \right.
$$
\begin{equation} \label{Bpk}
+ \left.   i \tred{\omega^{(\gl)}_{\pm}}(q) \tred{\overstar{\Phi}^{(\gl)}_{\mp}}(-q) e^{i \tred{\omega^{(\gl)}_{\pm}}(q) t} \hat{\alpha}^{\dagger}_{\mp,-q} + i \tred{\omega^{(\gf)}_{\pm}}(q)  \tred{\overstar{\Phi}^{(\gf)}_{\mp}}(-q) e^{i \tred{\omega^{(\gf)}_{\pm}}(q) t}  \hat{\beta}^{\dagger}_{\mp,-q}  \right\},   
\end{equation}
\end{widetext}

\tor{It is easy to verify} 
\begin{equation}
\hat{B}^{\dagger}_{+,q}(t) = \hat{B}_{-,-q}(t),  \qquad 
\hat{B}^{\dagger}_{-,q}(t) = \hat{B}_{+,-q}(t).  
\end{equation}

If we define the state with the magnetic field $B_{\pm}$,  $\left| B_{\pm,q}(t) \right\rangle = \hat{B}^{\dagger}_{\pm,q}(t)  \left| 0 \right\rangle$, where $\left| 0 \right\rangle$ is the phonon vacuum,  we obtain the norm squared  at $T=0$
\begin{widetext}
\begin{equation}
\left\langle B_{\pm,q}(t) \right.  \left| B_{\pm,q}(t) \right\rangle = \left\langle 0 \right|  \hat{B}_{\pm,q}(t)  \hat{B}_{\mp,-q}(t)  \left| 0 \right\rangle
 = \frac{\hbar \mu^2_0}{2 \rho V_0 q^2} \sum_{\tred{s=\gl,\gf}} \left[ \omega^{(s)}_{\pm} (q) \right]^2  \mathcal{F}^{(s)}_{\pm}(q)
\end{equation}
with
\begin{equation}
\mathcal{F}^{(s)}_{\pm}(q)  
 =     
 \frac{q^2}{\Omega^{(u)}_{1q}} \left| U^{(s)}_{\pm} (q) \right|^2   \left( d_{14} + \frac12 X_{11} \right)^2   
\end{equation}
$$
\pm \frac{q}{\sqrt{j_1 \Omega^{(u)}_{1q} \Omega^{(\varphi)}_{1q}}}  \left( U^{(s)}_{\pm} (q) { \overstar{\Phi}^{(s)}_{\pm} (q)} +   \Phi^{(s)}_{\pm} (q) { \overstar{U}^{(s)}_{\pm} (q)} \right)   \left( d_{14} + \frac12 X_{11} \right)  \left( X_{11} \mp  q \Gamma^{(1)}_{11} \right)  
+  \frac{1}{j_1 \Omega^{(\varphi)}_{1q}} \left| \Phi^{(s)}_{\pm} (q) \right|^2  \left( X_{11} \mp q \Gamma^{(1)}_{11} \right)^2.
$$

The difference between the squares of the amplitudes of the left- and right-circular components of the magnetic field is
\begin{equation}  \label{DiffB}
\left\langle B_{+,q}(t) \right.  \left| B_{+,q}(t) \right\rangle  -  \left\langle B_{-,q}(t) \right.  \left| B_{-,q}(t) \right\rangle 
 = \frac{\hbar \mu^2_0}{2 \rho V_0 q^2} \sum_{\tred{s=\gl,\gf}} \left[ \omega^{(s)}_{+} (q) \right]^2  
\left\{ \frac{q^2}{\Omega^{(u)}_{1q}} \left| U^{(s)}_{+} (q) \right|^2   \left( d_{14} + \frac12 X_{11} \right)^2     
 \right. 
\end{equation}
$$
\left. 
+ \frac{q}{\sqrt{j_1 \Omega^{(u)}_{1q} \Omega^{(\varphi)}_{1q}}}  \left( U^{(s)}_{+} (q) { \overstar{\Phi}^{(s)}_{+} (q)} +   \Phi^{(s)}_{+} (q) { \overstar{U}^{(s)}_{+} (q)} \right)   \left( d_{14} + \frac12 X_{11} \right)  \left( X_{11} -  q \Gamma^{(1)}_{11} \right)  
+  \frac{1}{j_1 \Omega^{(\varphi)}_{1q}} \left| \Phi^{(s)}_{+} (q) \right|^2  \left( X_{11} - q \Gamma^{(1)}_{11} \right)^2   
\right\}
$$
$$
-  \frac{\hbar \mu^2_0}{2 \rho V_0 q^2} \sum_{\tred{s=\gl,\gf}} \left[ \omega^{(s)}_{-} (q) \right]^2  
\left\{ \frac{q^2}{\Omega^{(u)}_{1q}} \left| U^{(s)}_{-} (q) \right|^2   \left( d_{14} + \frac12 X_{11} \right)^2     
 \right. 
$$
$$
\left. 
- \frac{q}{\sqrt{j_1 \Omega^{(u)}_{1q} \Omega^{(\varphi)}_{1q}}}  \left( U^{(s)}_{-} (q) { \overstar{\Phi}^{(s)}_{-} (q)} +   \Phi^{(s)}_{-} (q) { \overstar{U}^{(s)}_{-} (q)} \right)   \left( d_{14} + \frac12 X_{11} \right)  \left( X_{11} +  q \Gamma^{(1)}_{11} \right)  
+  \frac{1}{j_1 \Omega^{(\varphi)}_{1q}} \left| \Phi^{(s)}_{-} (q) \right|^2  \left( X_{11} + q \Gamma^{(1)}_{11} \right)^2   
\right\}.
$$

\tb{For the numerical evaluation of the long-wavelength scaling behavior, we use the same representative parameter set as in Fig.~\ref{fig:Bpm}, i.e., the values listed in Sec.~III.}

\tmag{To provide further insight into the wave-number dependence shown in Fig.~\ref{fig:Bpm}, we analyze the long-wavelength limit ($q \to 0$) of the fields. Numerical calculations reveal that for the \tred{gapless} branches, only the contribution proportional to $|\tred{U_{\pm}^{(\gl)}}(0)|^2$ remains. \tb{Consequently, the squared single-phonon amplitude, $\langle B_{\pm,q}|B_{\pm,q}\rangle\propto[\omega_{\pm}^{(\gl)}(q)]^2|U_{\pm}^{(\gl)}(q)|^2/\Omega_{1q}^{(u)}$, vanishes linearly in this regime, $\langle B_{\pm,q}|B_{\pm,q}\rangle\sim|q|$, because $\omega_{\pm}^{(\gl)}(q) \sim |q|$ and $\Omega_{1q}^{(u)} \sim |q|$; the field amplitude itself therefore behaves regularly as $B^{(1)}_{\pm}(q)\sim|q|^{1/2}$.}}

\tmag{Conversely, for the \tred{gapful} branches, the long-wavelength contribution to the magnetic field is strictly controlled by $|\tred{\Phi_{\pm}^{(\gf)}}(0)|^2$. Because both $\Omega_{1q}^{(\varphi)}$ and $\tred{\omega_{\pm}^{(\gf)}}(q)$ remain finite as $q \to 0$, this specific scaling behavior directly ensures the formal $1/q$ divergence of the single-phonon magnetic-field amplitude, $|B_{\pm, q}| \sim 1/q$, near the zone center.
}

\section{Zilch}
\label{app:zilch}

Lipkin's zilch, known as optical chirality, is determined by  
\begin{equation}
Z(z,t)=\boldsymbol{B}(z,t) \boldsymbol{\nabla}\times\boldsymbol{B}(z,t) = \mu_0 \boldsymbol{B}(z,t) \dot{\boldsymbol{P}}(z,t) 
= \frac{\mu_0}{2} \left( B_{+}(z,t) \dot{P}_{-}(z,t) +  B_{-}(z,t) \dot{P}_{+}(z,t) \right).
\end{equation}

Using the Fourier transform
$$
Z(z,t) = \int \frac{dq}{2\pi}  Z_q(t) e^{iqz},
$$
we obtain
\begin{equation}
\hat{Z}_q(t) = \frac{\mu_0}{4\pi} \int dk \, \left[   \hat{B}_{+,k}(t) \dot{\hat{P}}_{-,q-k}(t) +  \hat{B}_{-,k}(t) \dot{\hat{P}}_{+,q-k}(t)  \right],
\end{equation}
in which the use of the second-quantized operator forms  (\ref{Ppk},\ref{Bpk})  is assumed.

The direct calculation yields the time-independent observable for the ground state at $T=0$
$$
 \left\langle 0 \right|  \hat{Z}_q (t)  \left| 0 \right\rangle =   \delta_{q,0} \,  \frac{\mu^2_0}{4\pi}  \int dk \frac{1}{2\rho V_0 k} \left\{
\sum_{\tred{s=\gl,\gf}} \left[ \omega^{(s)}_{-} (k) \right]^2  
\left[ \frac{k^2}{\Omega^{(u)}_{1k}} \left| U^{(s)}_{-} (k) \right|^2   \left( d_{14} + \frac12 X_{11} \right)^2     
 \right. 
\right. 
$$
$$
\left. 
- \frac{k}{\sqrt{j_1 \Omega^{(u)}_{1k} \Omega^{(\varphi)}_{1k}}}  \left( U^{(s)}_{-} (k) { \overstar{\Phi}^{(s)}_{-} (k)} +   \Phi^{(s)}_{-} (k) { \overstar{U}^{(s)}_{-} (k)} \right)   \left( d_{14} + \frac12 X_{11} \right)  \left( X_{11} +  k \Gamma^{(1)}_{11} \right)  
+  \frac{1}{j_1 \Omega^{(\varphi)}_{1k}} \left| \Phi^{(s)}_{-} (k) \right|^2  \left( X_{11} + k \Gamma^{(1)}_{11} \right)^2   
\right]
$$
$$
- \sum_{\tred{s=\gl,\gf}} \left[ \omega^{(s)}_{+} (k) \right]^2  
\left[ \frac{k^2}{\Omega^{(u)}_{1k}} \left| U^{(s)}_{+} (k) \right|^2   \left( d_{14} + \frac12 X_{11} \right)^2     
 \right. 
$$
$$
\left.
\left. 
+ \frac{k}{\sqrt{j_1 \Omega^{(u)}_{1k} \Omega^{(\varphi)}_{1k}}}  \left( U^{(s)}_{+} (k) { \overstar{\Phi}^{(s)}_{+} (k)} +   \Phi^{(s)}_{+} (k) { \overstar{U}^{(s)}_{+} (k)} \right)   \left( d_{14} + \frac12 X_{11} \right)  \left( X_{11} -  k \Gamma^{(1)}_{11} \right)  
+  \frac{1}{j_1 \Omega^{(\varphi)}_{1k}} \left| \Phi^{(s)}_{+} (k) \right|^2  \left( X_{11} - k \Gamma^{(1)}_{11} \right)^2   
\right]
\right\}.
$$

By comparing this expression with the magnetic field difference (\ref{DiffB}), we obtain finally
\begin{equation}
 \left\langle 0 \right|  \hat{Z}_q   \left| 0 \right\rangle =  -   \frac{\delta_{q,0}}{4\pi \hbar}  \,  \int dk \,  k \left(  \left\langle B_{+,k} \right.  \left| B_{+,k} \right\rangle  -  \left\langle B_{-,k} \right.  \left| B_{-,k} \right\rangle   \right).  
\end{equation}

\end{widetext}

\section{Micropolar phonon assisted electron spin polarization: Floquet high-frequency expansion}
\label{app:floquet}

Consider the emergence of a nonzero electronic spin polarization in the atomic limit, when each individual electron spin is coupled to the phonon field of a micropolar medium through the effective ac magnetic fields $\boldsymbol{B}_{\pm}(t)$  rotating oppositely with the frequencies $\omega_{\pm}$. \tmag{Here, the magnetic field amplitudes $\boldsymbol{B}_{\pm}$ correspond to 
the classical field amplitudes induced by the coherent chiral phonon 
states, whose microscopic, single-phonon operator structure was rigorously 
derived in Appendix~\ref{app:em}.} \tor{For a coherent state with occupation number $N_{\pm}(\boldsymbol{q})$, these amplitudes scale as $B_{\pm}\propto\sqrt{N_{\pm}(\boldsymbol{q})}\,B^{(1)}_{\pm}(q)$, where $B^{(1)}_{\pm}(q)$ is the single-phonon rms amplitude defined in Appendix~\ref{app:em}; this is the relation underlying Eq.~(\ref{Beff}) of the main text.}

The model Hamiltonian of the problem reads  
\begin{align}  \label{HamOneSpin}
\hat{H}(t)=\mu_B \boldsymbol{\sigma}\cdot \boldsymbol{B}_{+}(t)
+\mu_B \boldsymbol{\sigma}\cdot \boldsymbol{B}_{-}(t),
\end{align}
where the Pauli matrix $\boldsymbol{\sigma}$ represents the electron  spin, $\mu_B$ is the Bohr magneton. 

The clockwise and anticlockwise constituents of the elliptically polarized magnetic field are given by 
\begin{align}
\boldsymbol{B}_+(t)&=B_+ \left\{\cos(\omega_+ t),\sin(\omega_+ t),0 \right\},\\
\boldsymbol{B}_-(t)&=B_- \left\{\cos(\omega_- t+\phi_0),-\sin(\omega_- t+\phi_0),0\right\},
\end{align}
where $B_\pm$ are magnitudes of the ac magnetic fields, and an initial phase $\phi_0$ is introduced for  generality.

Given the frequencies $\omega_{\pm}$ are almost degenerate, one may define $\omega_{\pm} = \omega \pm \delta \omega$  with $\omega = (\omega_++\omega_-)/2$ and $\delta\omega=(\omega_+-\omega_-)/2$.

Substituting $\omega$ and $\delta\omega$ into (\ref{HamOneSpin}) the Hamiltonian can be rewritten as 
$$
\hat{H}(t)=\mu_B
\left[
\sigma_+(B_+e^{-i\omega_+t}+B_- e^{i(\omega_-t+\phi_0)})
\right.
$$
\begin{align}
\left.
+\sigma_-(B_+e^{i\omega_+t}+B_-e^{-i(\omega_-t+\phi_0)})
\right]
\end{align}
with  $\sigma_\pm= \left(  \sigma_x\pm i\sigma_y \right)/2$.

Our goal is to apply the Floquet high-frequency expansion to this time-dependent system to gain insight into its  slow spin dynamics.   The true period \(T\) defined by \(\hat{H}(t)=\hat{H}(t+T)\) is very long
because of quasi-degeneracy of  $\omega_\pm$, $T\sim 1/\delta\omega$. We adopt the adiabatic approximation ignoring the slow time dependence of the factors \(e^{\pm i\delta\omega t}\),
and perform a high-frequency expansion with respect to the fast terms \(e^{\pm i\omega t}\).

This leads to the form 
$$
\hat{H}(t)=\mu_B \left[
\sigma_+\!\left(B_+^{(-\delta)}e^{-i\omega t}+B_-^{(-\delta)}e^{i\omega t}e^{i\phi_0}\right)
\right.
$$
\begin{align} \label{Hperiod}
\left. 
+\sigma_-\!\left(B_+^{(+\delta)}e^{i\omega t}+B_-^{(+\delta)}e^{-i\omega t}e^{-i\phi_0}\right)
\right],
\end{align}
where the \tor{notations} $B_\pm^{(\pm\delta)}=B_\pm e^{\pm i\delta\omega t}$ are introduced. 

Next, it is necessary to solve the Schr\"odinger equation for the time-periodic system
\begin{align} \label{TDSE}
i\hbar\frac{\partial}{\partial t}\Psi(t)=\hat{H}(t)\Psi(t). 
\end{align}

 According to Floquet theorem, the solution has the form $\Psi(t)=e^{-i\varepsilon t/\hbar}\Phi(t)$ with $\Phi(t)$ being a periodic function, $\Phi(t)=\Phi(t+T)$.  The real $\varepsilon$ is called the Floquet quasienergy.

From the periodicity of $\hat{H}(t)$ and $\Phi(t)$, we can define their Fourier transforms, respectively,
\begin{align} \label{FTHPSI}
\hat{H}(t)=\sum_{m\in\mathbb{Z}}e^{-im\omega t}\hat{H}_m,\quad
\Phi(t)=\sum_{m\in\mathbb{Z}}e^{-im\omega t}\Phi_m,
\end{align} 
where $m$ takes integer values.

The inverse transformations are given by
\begin{align}
\hat{H}_m=\frac{1}{T}\int_0^T\!dt\,e^{im\omega t}\hat{H}(t),\quad
\Phi_m=\frac{1}{T}\int_0^T\!dt\,e^{im\omega t}\Phi(t).
\end{align}

Plugging (\ref{FTHPSI}) into the time-dependent Schr\"odinger equation (\ref{TDSE}) yields the eigenvalue problem for the quasienergy $\varepsilon$
\begin{align}
\sum_{n\in\mathbb{Z}}(\hat{H}_{m-n}-m \hbar \omega\delta_{mn})\Phi_n=\varepsilon\Phi_m.
\label{eq:FloquetMap}
\end{align}

\begin{widetext}
We  can write it explicitly as

\begin{equation}
\begin{pmatrix}
\ddots & \ddots & \ddots  & \ddots &  & &  \\
\cdots  & \hat{H}_0-2\hbar\omega & \hat{H}_1 & \hat{H}_2  & \hat{H}_3  & \cdots & \\
\cdots  & \hat{H}_{-1} & \hat{H}_0-\hbar\omega & \hat{H}_1 & \hat{H}_2 & \hat{H}_3 &  \cdots \\
\cdots  & \hat{H}_{-2} & \hat{H}_{-1} & \hat{H}_0 & \hat{H}_1 & \hat{H}_2 &  \cdots \\
\cdots  & \hat{H}_{-3} & \hat{H}_{-2} & \hat{H}_{-1} & \hat{H}_0 + \hbar \omega & \hat{H}_1 &  \cdots \\
 & \cdots & \hat{H}_{-3} & \hat{H}_{-2} & \hat{H}_{-1} & \hat{H}_0 + 2 \hbar \omega &  \cdots \\
 & & & \ddots & \ddots & \ddots & \ddots 
\end{pmatrix}
\begin{pmatrix}
\vdots \\ \Phi_2 \\ \Phi_1 \\ \Phi_0 \\ \Phi_{-1} \\ \Phi_{-2} \\ \vdots
\end{pmatrix}
= \varepsilon
\begin{pmatrix}
\vdots \\ \Phi_2 \\ \Phi_1 \\ \Phi_0 \\ \Phi_{-1} \\ \Phi_{-2} \\ \vdots
\end{pmatrix}. 
\label{eq:FloquetMap2}
\end{equation}
\begin{figure}[htbp]
  \begin{center}
  \includegraphics[width=\linewidth]{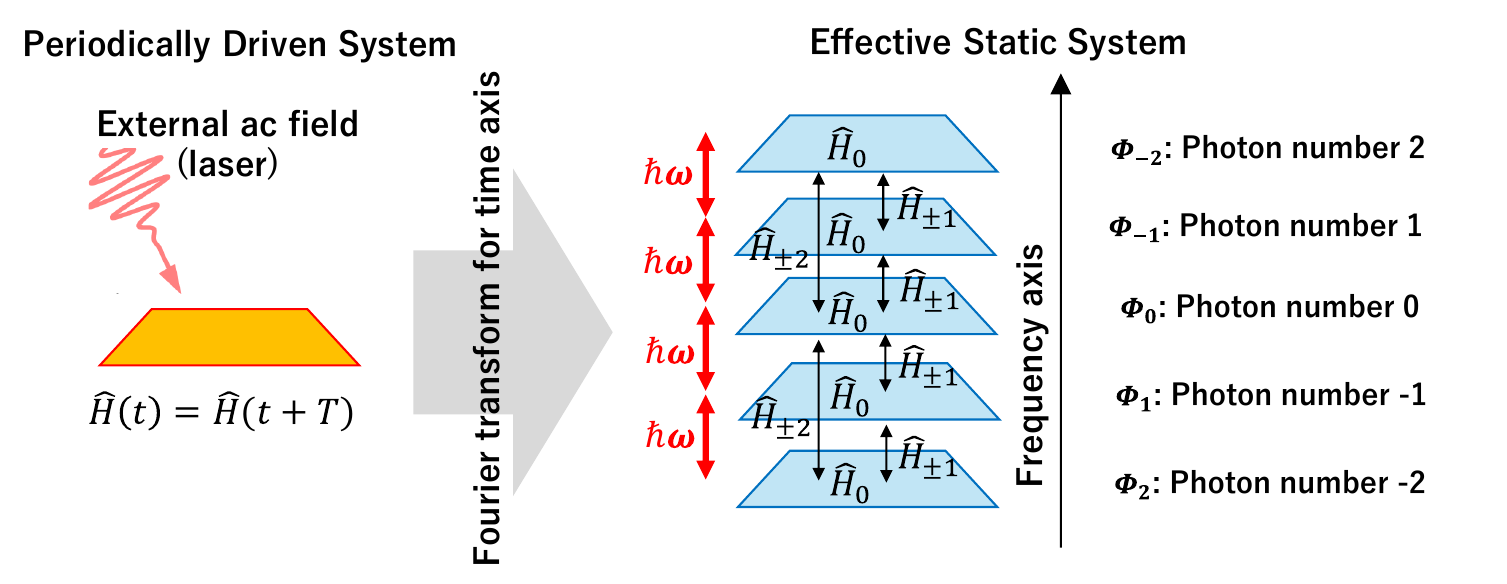}
  \caption{Schematic Image of an original time-periodic system and the corresponding effective static system of Eqs.~\eqref{eq:FloquetMap} and ~\eqref{eq:FloquetMap2}.} 
  \label{fig:Floquet}
  \end{center}
\end{figure}
\end{widetext}
The diagonal part of the matrix on the left-hand side is given by $\hat{H}_0 + n \hbar \omega$. Therefore, the corresponding wave function $\Phi_n$ may be viewed as the $n$-photon dressed state if to suppose that the time periodicity of the Hamiltonian comes from interaction with an external electromagnetic wave of the frequency $\omega$ [see Fig.~\ref{fig:Floquet}].

To derive an effective Hamiltonian describing the slow dynamics of the system within the framework of the high-frequency expansion, it is assumed that the single-photon energy is much larger than the system's characteristic energy scale, i.e. $\hbar \omega \gg \hat{H}_{\pm n} $. 

Then, the effective Hamiltonian in the subspace of a fixed photon number can be derived via degenerate perturbation theory, where the unperturbed part is the diagonal photon energy $m\hbar \omega$ and $\hat{H}_{\pm n}$ serve as perturbative terms \cite{Bukov2014,Eckardt2015,Eckardt2017,Oka2019,Sato2025}. 

The resulting high-frequency expansion takes the form 
\begin{align}  \label{FloquetExpan}
\hat{H}_{\mathrm{eff}}=\sum_{k=0}^{\infty}\hat{H}^{(k)},
\end{align}
where each term is of order $(1/\omega)^k$.  Explicit expressions for the lower-order $\hat{H}^{(k)}$ are given in \tor{the Supplemental Material~\cite{SupplementalMaterialArXiv}}.

To obtain the effective Floquet Hamiltonian from the time-periodic Hamiltonian (\ref{Hperiod}), we ignore the slow time dependence of  $B_\pm^{(\pm\delta)}$. Then, $\hat{H}(t)$ includes only terms proportional $e^{\pm i \omega t}$. In this case, only the Fourier components  $\hat{H}_{\pm 1}$ survive, and we have
\begin{align}
\hat{H}_1&=\mu_B(B_+^{(-\delta)}\sigma_++B_-^{(+\delta)}\sigma_-e^{-i\phi_0}),\\
\hat{H}_{-1}&=\mu_B(B_+^{(+\delta)}\sigma_-+B_-^{(-\delta)}\sigma_+e^{i\phi_0}).
\end{align}
Apparently, $\hat{H}_{-1} = (\hat{H}_1)^\dagger$.

The leading term of the effective Floquet Hamiltonian~\cite{Takayoshi2014,Sato2016} is
 \begin{align}
\hat{H}^{(1)}&=\tb{-\frac{[\hat{H}_1,\hat{H}_{-1}]}{\hbar\omega}}
=\tb{-\frac{\mu_B^2}{\hbar\omega}}(B_+^2-B_-^2)\sigma_z.
\end{align}
Thus an effective Zeeman field along $z$ appears. 

 The second-order term vanishes when $\hat{H}_0=0$. Using $[\sigma_z,\sigma_\pm]=\pm 2 \sigma_\pm$, we obtain the third-order term 
\begin{align}
\hat{H}^{(3)}
= \tb{\frac{\mu_B^4}{(\hbar\omega)^3}} (B_+^4 - B_-^4) \sigma_z.
\end{align}

Therefore, the effective Floquet Hamiltonian up to  $O(1/\omega^3)$ reads 
\begin{align} \label{HFloquetTot}
\hat{H}_{\mathrm{eff}}
=\tb{-\frac{\mu_B^2}{\hbar\omega}}(B_+^2-B_-^2)\sigma_z
+\tb{\frac{\mu_B^4(B_+^4-B_-^4)}{(\hbar\omega)^3}}\sigma_z.
\end{align}
\tb{As a check, for $B_-=0$ the Hamiltonian (\ref{HamOneSpin}) is exactly solvable in the frame rotating with the field, and the quasienergies are $\varepsilon_{\uparrow,\downarrow}=\pm[\hbar\omega/2-\sqrt{(\hbar\omega/2)^2+\mu_B^2B_+^2}\,]=\mp[\mu_B^2B_+^2/\hbar\omega-\mu_B^4B_+^4/(\hbar\omega)^3+\cdots]$, in agreement with Eq.~(\ref{HFloquetTot}).}
  
This result clearly indicates that an effective Zeeman interaction and magnetization along the $z$-axis appear
when the amplitudes of clockwise and anticlockwise magnetic field \tor{constituents} differ, i.e. $B_+ \ne B_-$.

\section{Inverse Zeeman phonon splitting of electron spin bands}
\label{app:split}

Below we show that another manifestation of the elliptic polarization of the magnetic field of piezoelectric origin is the effect that is the inverse of the phonon Zeeman splitting one. This phenomenon results from Rashba-like nonreciprocal splitting in the transverse branches of the phonon spectrum, which gives rise to Zeeman-like splitting of the electronic subbands due to electron-phonon coupling.  

The dynamical piezoelectric mechanism  assumed here operates only when the induced piezoelectric polarization is not efficiently screened by conduction electrons. This requirement imposes a quantitative condition
that the frequency of the \tred{gapful branch}, 
$$
\omega (q) = \left( \tred{\omega^{(\gf)}_{+}} (q) +  \tred{\omega^{(\gf)}_{-}}(q) \right)/2,
$$
must be comparable to or greater than the plasma frequency of the electron gas, $\omega_p$. Using the \tred{gapful-branch frequencies} discussed in this work, typically in the range of 6-12 THz, the condition $\omega(q) \geq \omega_p$ can be satisfied for carrier densities below $\sim 10^{23}$ m${}^{-3}$  corresponding to weakly doped or insulating materials. Therefore, our theoretical framework naturally  applies not only to piezo-active insulators but also to  lightly doped semiconductors where screening is insufficient.

To start, we specify the magnetic field $\boldsymbol{B}=\boldsymbol{\nabla}\times\boldsymbol{A}$ in terms of the vector potential $\boldsymbol{A}$. Their Fourier \tor{transforms} are related by   
\begin{equation} \label{A_PZ}
  A_{\pm} \left( q \right) = \mp \frac{1}{q} B_{\pm} \left( q \right),  \quad A_3 \left(q \right) =0,
\end{equation}
where the \tor{Coulomb} gauge $\boldsymbol{\nabla}\cdot \boldsymbol{A}=0$ is assumed. 

Similarly, the electric charge $\rho =  \boldsymbol{\nabla}\cdot \boldsymbol{P}$,  induced by the polarization $\boldsymbol{P}$,  determines the electric scalar potential $\phi$ via the Poisson's equation, $\Delta \phi = - \rho/\epsilon_0$, where $\epsilon_0$ is the vacuum permittivity. This yields
\begin{equation} \label{Phi_PZ}
 \phi  \left( q \right) = \frac{i}{\epsilon_0 q} X_{33} \varphi_3 \left( q \right). 
\end{equation} 

The electron-phonon interaction can be derived  from the Hamiltonian
\begin{equation} \label{Hel}
\mathcal{H}_{el} = \sum_i \left[\frac{1}{2m} \left( \boldsymbol{p}_i +  e \boldsymbol{A}(\boldsymbol{r}_i) \right)^2 -  e \phi (\boldsymbol{r}_i) - \boldsymbol{m}(\boldsymbol{r}_i)  \cdot \boldsymbol{B} (\boldsymbol{r}_i) \right].   
\end{equation} 
Here, the last term  accounts for the coupling between the magnetic field  and the electron magnetic moment $\boldsymbol{m}(\boldsymbol{r}_i) = g_e\mu_{\textrm{B}} \boldsymbol{s}_i=\frac12 g_e\mu_{\textrm{B}} \boldsymbol\sigma_i$. The sum is taken over all doped electrons with the \tor{instantaneous} positions $\boldsymbol{r}_i$ and the momenta $\boldsymbol{p}_i$. 

The second-quantized form of the Hamiltonian (\ref{Hel}) in the momentum representation  reads
\begin{widetext}
\begin{equation} \label{Hel1}
\mathcal{H}_{el} = \sum_{\boldsymbol{k}\sigma} \varepsilon_{\boldsymbol{k}} c^{\dagger}_{\boldsymbol{k}\sigma} c_{\boldsymbol{k}\sigma} +  \frac{\hbar e}{m} \sum_{\boldsymbol{k}\boldsymbol{q}\sigma} \left( \boldsymbol{k} \cdot \boldsymbol{A} (\boldsymbol{q}) \right)  c^{\dagger}_{\boldsymbol{k}+\boldsymbol{q} \sigma} c_{\boldsymbol{k}\sigma} - e   \sum_{\boldsymbol{k}\boldsymbol{q}\sigma}  \phi (\boldsymbol{q}) c^{\dagger}_{\boldsymbol{k}+\boldsymbol{q} \sigma} c_{\boldsymbol{k}\sigma} 
 -\frac12 g_e \mu_B   \sum_{\boldsymbol{k}\boldsymbol{q}} \left(  B_{-}(\boldsymbol{q}) c^{\dagger}_{\boldsymbol{k}+\boldsymbol{q} \uparrow} c_{\boldsymbol{k}\downarrow}  +  B_{+}(\boldsymbol{q}) c^{\dagger}_{\boldsymbol{k}+\boldsymbol{q} \downarrow} c_{\boldsymbol{k}\uparrow}  \right), 
\end{equation}
\end{widetext}
where $\varepsilon_{\boldsymbol{k}}$ is the energy of the itinerant electron with the wave vector $\boldsymbol{k}$. Here, the Fourier transforms of the electromagnetic fields are used and the term proportional to $\boldsymbol{A}^2$ is neglected. Provided that electrons travel along the $z$-axis, when  $\boldsymbol{k} || \boldsymbol{q}$, the second term  vanishes for the \tor{Coulomb} gauge $\boldsymbol{\nabla}\cdot\boldsymbol{A}=0$.

 For simplicity, we set $X_{33}=0$, thereby eliminating the scalar potential, so the Hamiltonian (\ref{Hel1}) reduces to 
\begin{widetext}
\begin{equation} \label{Hel2}
\mathcal{H}_{el} = \sum_{\boldsymbol{k}\sigma} \varepsilon_{\boldsymbol{k}} c^{\dagger}_{\boldsymbol{k}\sigma} c_{\boldsymbol{k}\sigma} 
 -\frac12 g_e \mu_B   \sum_{\boldsymbol{k}\boldsymbol{q}} \left(  B_{-}(\boldsymbol{q}) c^{\dagger}_{\boldsymbol{k}+\boldsymbol{q} \uparrow} c_{\boldsymbol{k}\downarrow}  +  B_{+}(\boldsymbol{q}) c^{\dagger}_{\boldsymbol{k}+\boldsymbol{q} \downarrow} c_{\boldsymbol{k}\uparrow}  \right).
\end{equation}
\end{widetext}
It describes spin-flip electron scattering processes occurring under the potential generated by the circular components of the phonon-induced magnetic field. 

In order to find the one-particle electron spectrum, the retarded  Green function $G^R_{\boldsymbol{k}\sigma,\boldsymbol{k}^{'}\sigma^{'}}(t) = -i \theta(t) \left\langle \left\{ c_{\boldsymbol{k}\sigma}(t), c^{\dagger}_{\boldsymbol{k}^{'}\sigma^{'}}  \right\} \right\rangle$ is introduced, whose equation of motion in the frequency representation takes the form   
\begin{widetext}
\begin{equation}
 \hbar \delta_{\boldsymbol{k}\boldsymbol{k}^{'}} \delta_{\sigma \sigma^{'}} =  \left( \hbar \omega  -  \varepsilon_{\boldsymbol{k}} \right)  G^R_{\boldsymbol{k}\sigma,\boldsymbol{k}^{'}\sigma^{'}}(\omega)
+ \frac12 g_e \mu_B  \sum_{\boldsymbol{q}}
\left(
\delta_{\sigma \uparrow}  B_{-}(\boldsymbol{q})  G^R_{\boldsymbol{k}-\boldsymbol{q}\downarrow,\boldsymbol{k}^{'}\sigma^{'}}(\omega) +  \delta_{\sigma \downarrow}  B_{+}(\boldsymbol{q})  G^R_{\boldsymbol{k}-\boldsymbol{q}\uparrow,\boldsymbol{k}^{'}\sigma^{'}}(\omega)
\right). 
\end{equation}

In the mean-field approximation\cite{Lee1974}, components of the magnetic field with a specific momentum $\boldsymbol{Q}$ are singled out, which leads to the coupled equations describing the spin-flip processes for spin-up and spin-down electrons, respectively
\begin{equation}
\left(
\begin{array}{cc}
\hbar \omega - \varepsilon_{\boldsymbol{k}} &  \frac12 g_e \mu_B  B_{-}(-\boldsymbol{Q})  \\
 \frac12 g_e \mu_B  B_{+}(\boldsymbol{Q})  &   \hbar \omega - \varepsilon_{\boldsymbol{k}+\boldsymbol{Q}}
\end{array}
\right) 
\left( 
\begin{array}{c}
G^R_{\boldsymbol{k}\uparrow,\boldsymbol{k}\uparrow} (\omega)  \\
G^R_{\boldsymbol{k}+\boldsymbol{Q}\downarrow,\boldsymbol{k}\uparrow}   (\omega) 
\end{array}
 \right) = 
\left(
\begin{array}{c}
\hbar  \\
0
\end{array}
\right),
\end{equation}
\begin{equation}
\left(
\begin{array}{cc}
\hbar \omega - \varepsilon_{\boldsymbol{k}} &  \frac12 g_e \mu_B  B_{+}(-\boldsymbol{Q})  \\
\frac12 g_e \mu_B  B_{-}(\boldsymbol{Q})    &  \hbar \omega - \varepsilon_{\boldsymbol{k}+\boldsymbol{Q}} 
\end{array}
\right) 
\left( 
\begin{array}{c}
G^R_{\boldsymbol{k}\downarrow,\boldsymbol{k}\downarrow} (\omega)  \\
G^R_{\boldsymbol{k}+\boldsymbol{Q}\uparrow,\boldsymbol{k}\downarrow}   (\omega) 
\end{array}
 \right) =
\left(
\begin{array}{c}
\hbar  \\
0
\end{array}
\right). 
\end{equation}

Thus, the energies of the electronic subbands are 
\begin{equation} \label{w12p0}
\hbar \omega_{1,2} (\boldsymbol{k}) = \frac12 \left(  \varepsilon_{\boldsymbol{k}} + \varepsilon_{\boldsymbol{k}+\boldsymbol{Q}} \right) \pm \left[  \frac14 \left( \varepsilon_{\boldsymbol{k}} - \varepsilon_{\boldsymbol{k}+\boldsymbol{Q}} \right)^2 + \tor{\left( \tfrac12 g_e \mu_B \right)^2} \left| B_{+} (\boldsymbol{Q}) \right|^2  \right]^{\frac12}, 
\end{equation} 
\begin{equation} \label{w34p0}
\hbar \omega_{3,4} (\boldsymbol{k}) = \frac12 \left(  \varepsilon_{\boldsymbol{k}} + \varepsilon_{\boldsymbol{k}+\boldsymbol{Q}} \right) \pm \left[  \frac14 \left( \varepsilon_{\boldsymbol{k}} - \varepsilon_{\boldsymbol{k}+\boldsymbol{Q}} \right)^2 + \tor{\left( \tfrac12 g_e \mu_B \right)^2} \left| B_{-} (\boldsymbol{Q}) \right|^2  \right]^{\frac12}, 
\end{equation}  
where the property $B_{\pm}(\boldsymbol{Q})^{\ast} = B_{\mp}(-\boldsymbol{Q})$ was used.

When $X_{33}$ is nonzero, the one-electron spectrum is slightly modified by the electric scalar potential
\begin{equation} \label{w12}
\hbar \omega_{1,2} (\boldsymbol{k}) \approx \frac12 \left(  \varepsilon_{\boldsymbol{k}} + \varepsilon_{\boldsymbol{k}+\boldsymbol{Q}} \right) \pm \left[  \frac14 \left( \varepsilon_{\boldsymbol{k}} - \varepsilon_{\boldsymbol{k}+\boldsymbol{Q}} \right)^2 + e^2 \left|  \phi (\boldsymbol{Q}) \right|^2 + \tor{\left( \tfrac12 g_e \mu_B \right)^2} \left| B_{+} (\boldsymbol{Q}) \right|^2  \right]^{\frac12}, 
\end{equation}
\begin{equation}  \label{w34}
\hbar \omega_{3,4} (\boldsymbol{k}) \approx \frac12 \left(  \varepsilon_{\boldsymbol{k}} + \varepsilon_{\boldsymbol{k}+\boldsymbol{Q}} \right) \pm \left[  \frac14 \left( \varepsilon_{\boldsymbol{k}} - \varepsilon_{\boldsymbol{k}+\boldsymbol{Q}} \right)^2 + e^2 \left|  \phi (\boldsymbol{Q}) \right|^2 + \tor{\left( \tfrac12 g_e \mu_B \right)^2} \left| B_{-} (\boldsymbol{Q}) \right|^2  \right]^{\frac12}, 
\end{equation}
\end{widetext}
where terms containing powers of the piezoelectric coefficients higher than the second order are neglected.

As is clearly seen (see Fig.~\ref{fig:lattice}), the difference $\left| B_{+} (\boldsymbol{Q}) \right|^2 \not= \left| B_{-} (\boldsymbol{Q}) \right|^2$ gives rise to an additional Zeeman-like splitting of the doubly degenerate electronic spectrum, which may be regarded as the inverse effect to that of phonon Zeeman splitting.

Indeed, the  retarded Green functions
$$
G^R_{\boldsymbol{k}\uparrow,\boldsymbol{k}\uparrow} (\omega + i \delta) = \frac{\hbar \omega - \varepsilon_{\boldsymbol{k}+\boldsymbol{Q}}}{\left[ \hbar \omega_{1} (\boldsymbol{k}) - \hbar \omega_{2} (\boldsymbol{k}) \right]} 
$$
\begin{equation}
\times \left(
\frac{1}{\omega - \omega_{1} (\boldsymbol{k}) + i \delta} - \frac{1}{\omega - \omega_{2} (\boldsymbol{k}) + i \delta} 
\right), 
\end{equation}  
$$
G^R_{\boldsymbol{k}\downarrow,\boldsymbol{k}\downarrow} (\omega + i \delta) = \frac{\hbar \omega - \varepsilon_{\boldsymbol{k}+\boldsymbol{Q}}}{\left[ \hbar \omega_{3} (\boldsymbol{k}) - \hbar \omega_{4} (\boldsymbol{k}) \right]} 
$$
\begin{equation}
\times \left(
\frac{1}{\omega - \omega_{3} (\boldsymbol{k}) + i \delta} - \frac{1}{\omega - \omega_{4} (\boldsymbol{k}) + i \delta} 
\right).  
\end{equation}  
determine the spectral densities via the relation
\begin{widetext}
$$
 \mathcal{J} _{ c_{\boldsymbol{k}\uparrow} c^{\dagger}_{\boldsymbol{k}\uparrow}} (\omega)  = -i n_{F}(\omega) 
 \left[  G_{\boldsymbol{k} \uparrow,\boldsymbol{k}\uparrow}  (\omega - i \delta) -  G_{\boldsymbol{k}\uparrow,\boldsymbol{k}\uparrow}  (\omega + i \delta) \right] 
$$ 
\begin{equation}
= \frac{2\pi}{\hbar \omega_{1} (\boldsymbol{k}) - \hbar \omega_{2} (\boldsymbol{k})} \left[
n_{F}(\omega_{1} (\boldsymbol{k})) \left( \hbar \omega - \varepsilon_{\boldsymbol{k}+\boldsymbol{Q}} \right) \delta \left( \omega -  \omega_{1} (\boldsymbol{k}) \right) 
- n_{F}(\omega_{2} (\boldsymbol{k})) \left( \hbar \omega - \varepsilon_{\boldsymbol{k}+\boldsymbol{Q}} \right) \delta \left( \omega -  \omega_{2} (\boldsymbol{k}) \right)
\right]
\end{equation}
and the similar expression for $ \mathcal{J} _{ c_{\boldsymbol{k}\downarrow} c^{\dagger}_{\boldsymbol{k}\downarrow}} (\omega)$.   Here, $n_{F}(\omega) = \left[ \exp(\beta \hbar \omega) +1 \right]^{-1}$  is the Fermi-Dirac distribution function. 

\begin{figure*}[h]
\includegraphics[scale=0.8]{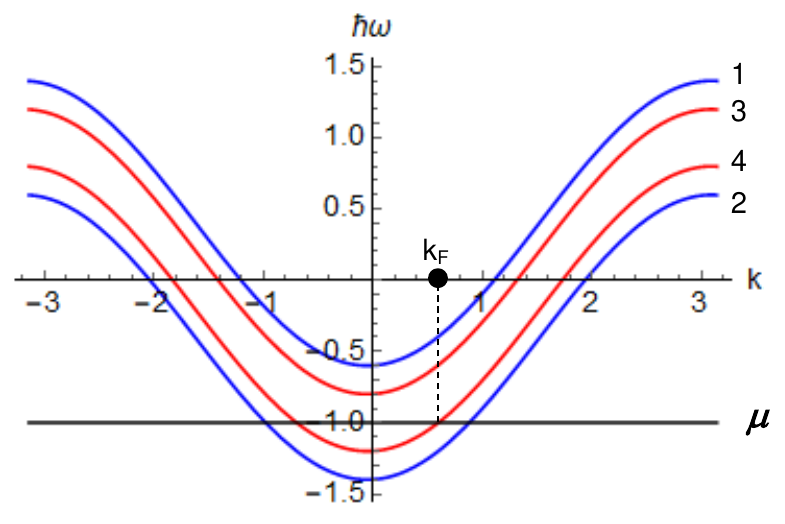}
\caption{Splitting of the electron subbands when the lowest ones, 2 and 4,  are only filled.}
\label{fig:lattice}
\end{figure*} 

The spectral densities specify the occupation numbers for spin-up and spin-down electrons, respectively, 
\begin{equation}
\left\langle c^{\dagger}_{\boldsymbol{k}\uparrow} c_{\boldsymbol{k}\uparrow}  \right\rangle  =  \int^{+\infty}_{-\infty} \frac{d \omega}{2\pi} \mathcal{J} _{ c_{\boldsymbol{k}\uparrow} c^{\dagger}_{\boldsymbol{k}\uparrow}} (\omega)  
= \frac{ 
n_{F}(\omega_{1} (\boldsymbol{k})) \left[ \hbar \omega_{1} (\boldsymbol{k}) - \varepsilon_{\boldsymbol{k}+\boldsymbol{Q}} \right]   
- n_{F}(\omega_{2} (\boldsymbol{k})) \left[ \hbar \omega_{2} (\boldsymbol{k}) - \varepsilon_{\boldsymbol{k}+\boldsymbol{Q}} \right] 
}{\hbar \omega_{1} (\boldsymbol{k}) - \hbar \omega_{2} (\boldsymbol{k})},  
\end{equation} 
\begin{equation}
\left\langle c^{\dagger}_{\boldsymbol{k}\downarrow} c_{\boldsymbol{k}\downarrow}  \right\rangle  =  \int^{+\infty}_{-\infty} \frac{d \omega}{2\pi} \mathcal{J} _{ c_{\boldsymbol{k}\downarrow} c^{\dagger}_{\boldsymbol{k}\downarrow}} (\omega)  
= \frac{ 
n_{F}(\omega_{3} (\boldsymbol{k})) \left[ \hbar \omega_{3} (\boldsymbol{k}) - \varepsilon_{\boldsymbol{k}+\boldsymbol{Q}} \right]   
- n_{F}(\omega_{4} (\boldsymbol{k})) \left[ \hbar \omega_{4} (\boldsymbol{k}) - \varepsilon_{\boldsymbol{k}+\boldsymbol{Q}} \right] 
}{\hbar \omega_{3} (\boldsymbol{k}) - \hbar \omega_{4} (\boldsymbol{k})}.  
\end{equation} 

Typically, all four subbands are filled up to the common chemical potential $\mu$. \textcolor{black}{For simplicity, we consider the case of weakly doped conduction bands when the lowest electronic levels are filled only,}  $n_{F}(\omega_{1} (\boldsymbol{k})) = n_{F}(\omega_{3} (\boldsymbol{k})) = 0$ and $n_{F}(\omega_{2} (\boldsymbol{k})) = n_{F}(\omega_{4} (\boldsymbol{k})) = 1$  (see Fig.~\ref{fig:lattice}).  Then, the direct calculation leads to the nonzero net magnetization
$$
\left\langle  m_z \right\rangle = \frac12  g_e \mu_B \sum_{\boldsymbol{k}} \left( \left\langle c^{\dagger}_{\boldsymbol{k}\uparrow} c_{\boldsymbol{k}\uparrow}  \right\rangle -  \left\langle c^{\dagger}_{\boldsymbol{k}\downarrow} c_{\boldsymbol{k}\downarrow}  \right\rangle \right) 
$$
\begin{equation}
= \frac14 g_e \mu_B \sum_{\boldsymbol{k}} \left\{
\frac{\varepsilon_{\boldsymbol{k}+\boldsymbol{Q}}  - \varepsilon_{\boldsymbol{k}}}{\left[   \left( \varepsilon_{\boldsymbol{k}+\boldsymbol{Q}} - \varepsilon_{\boldsymbol{k}}  \right)^2 +   \left(g_e \mu_B \right)^2  \left| B_{+} (\boldsymbol{Q}) \right|^2  \right]^{\frac12}}
-
\frac{\varepsilon_{\boldsymbol{k}+\boldsymbol{Q}}  - \varepsilon_{\boldsymbol{k}}}{\left[   \left(  \varepsilon_{\boldsymbol{k}+\boldsymbol{Q}} - \varepsilon_{\boldsymbol{k}} \right)^2 +   \left(g_e \mu_B \right)^2  \left| B_{-} (\boldsymbol{Q}) \right|^2  \right]^{\frac12}}
\right\}. 
\end{equation}

\end{widetext}

Assuming that the difference $\left| B_{+} (\boldsymbol{Q}) \right|^2 - \left| B_{-} (\boldsymbol{Q}) \right|^2$ is small, one gets
\begin{equation} \label{MzDGF}
\left\langle  m_z \right\rangle \approx \frac18 \left( g_e \mu_B \right)^3  S(\boldsymbol{Q}) \left(  \left| B_{+} (\boldsymbol{Q}) \right|^2 - \left| B_{-} (\boldsymbol{Q}) \right|^2 \right), 
\end{equation}
where the factor
\begin{equation}
 S(\boldsymbol{Q}) = \sum_{\boldsymbol{k} (k<k_{\textrm{F}})} \frac{\varepsilon_{\boldsymbol{k}} - \varepsilon_{\boldsymbol{k}+\boldsymbol{Q}} }{\left[   \left(    \varepsilon_{\boldsymbol{k}} - \varepsilon_{\boldsymbol{k}+\boldsymbol{Q}} \right)^2 +   \left(g_e \mu_B \right)^2  \left| B_{-} (\boldsymbol{Q}) \right|^2  \right]^{\frac32}}.
\end{equation}
Here, the sum over $\boldsymbol{k}$ is limited to the value $k_{\textrm{F}}$ corresponding to the chemical potential $\mu$.

\tmag{The results of this section formally agree with the high-frequency Floquet 
description under the condition that the energy scale of single-electron excitations 
satisfies $\lvert\varepsilon_{\boldsymbol{k}} - \varepsilon_{\boldsymbol{k}+\boldsymbol{Q}}\rvert \ll \hbar\omega$, 
i.e., when the splitting of the electronic subbands is much smaller than the 
phonon energy. In this regime, the linear approximation 
$\varepsilon_{\boldsymbol{k}} - \varepsilon_{\boldsymbol{k}+\boldsymbol{Q}} \approx -\hbar(\boldsymbol{v}_{\text{F}}\cdot\boldsymbol{Q})$ 
holds, restricting the transitions caused by the magnetic fields $\boldsymbol{B}_{\pm}(\boldsymbol{Q})$ 
to a small wavenumber, $Q \ll \omega/v_{\text{F}}$, where $\boldsymbol{v}_{\text{F}}$ 
is the Fermi velocity. This physically implies that the temporal dynamics of the 
moving electron are predominantly controlled by a continuous adiabatic adjustment to 
the rotating field direction, rather than by scattering processes on the spatially 
inhomogeneous background.}

\textcolor{black}{Assuming a symmetric band with the bandwidth $2 \hbar Q v_{\boldsymbol{F}}$ around the Fermi energy $\varepsilon_{\boldsymbol{F}}$, one may simplify $S(\boldsymbol{Q})$
\begin{equation}
 S(\boldsymbol{Q}) \approx - 2 \hbar v_{\boldsymbol{F}} Q  \rho(\varepsilon_{\boldsymbol{F}}) 
 \frac{\hbar \left( \boldsymbol{v}_{\textrm{F}} \boldsymbol{Q} \right)}{\left[ \hbar^2 \left( \boldsymbol{v}_{\textrm{F}} \boldsymbol{Q} \right)^2   +   \left(g_e \mu_B \right)^2  \left| B_{-} (\boldsymbol{Q}) \right|^2  \right]^{\frac32}},
\end{equation}
where $\rho(\varepsilon_{\boldsymbol{F}})$ is the electronic density of states at the Fermi energy.
}

\section{The interplay between the bulk Rashba effect and the inverse Zeeman phonon splitting phenomenon}
\label{app:rashba}

In non-centrosymmetric crystals with broken inversion symmetry, the bulk Rashba effect is allowed. To gain insight into how the momentum-dependent splitting of spin bands affects the inverse chiral phonon Zeeman effect, we include the Rashba term in the Hamiltonian (\ref{Hel2})
\begin{widetext}
\begin{equation} \label{RashbaHam}
\mathcal{H}_{el} = \sum_{\boldsymbol{k}\sigma} \varepsilon_{\boldsymbol{k}} c^{\dagger}_{\boldsymbol{k}\sigma} c_{\boldsymbol{k}\sigma} 
 -\frac12 g_e \mu_B   \sum_{\boldsymbol{k}\boldsymbol{q}} \left(  B_{-}(\boldsymbol{q}) c^{\dagger}_{\boldsymbol{k}+\boldsymbol{q} \uparrow} c_{\boldsymbol{k}\downarrow}  +  B_{+}(\boldsymbol{q}) c^{\dagger}_{\boldsymbol{k}+\boldsymbol{q} \downarrow} c_{\boldsymbol{k}\uparrow}  \right)
+ \alpha_{\textrm{R}} \sum_{\boldsymbol{k} \sigma \sigma^{'}} \left( -k_x \hat{\sigma}^y_{\sigma \sigma^{'}} + k_y \hat{\sigma}^x_{\sigma \sigma^{'}}\right) c^{\dagger}_{\boldsymbol{k}\sigma} c_{\boldsymbol{k}\sigma^{'}},
\end{equation}
where $\alpha_{\textrm{R}}$ is the Rashba coupling, and $\hat{\sigma}^{x,y}$ are the Pauli matrices. 

The equation of motion for the retarded  Green function  reads
$$
 \hbar \delta_{\boldsymbol{k}\boldsymbol{k}^{'}} \delta_{\sigma \sigma^{'}} =  \left( \hbar \omega  -  \varepsilon_{\boldsymbol{k}} \right)  G^R_{\boldsymbol{k}\sigma,\boldsymbol{k}^{'}\sigma^{'}}(\omega)
- \alpha_{\textrm{R}}  \left( -k_x \hat{\sigma}^y_{\sigma \bar{\sigma}} + k_y \hat{\sigma}^x_{\sigma \bar{\sigma}}\right) G^R_{\boldsymbol{k}\bar{\sigma},\boldsymbol{k}^{'}\sigma^{'}}(\omega)
$$
\begin{equation}
+ \frac12 g_e \mu_B  \sum_{\boldsymbol{q}}
\left(
\delta_{\sigma \uparrow}  B_{-}(\boldsymbol{q})  G^R_{\boldsymbol{k}-\boldsymbol{q}\downarrow,\boldsymbol{k}^{'}\sigma^{'}}(\omega) +  \delta_{\sigma \downarrow}  B_{+}(\boldsymbol{q})  G^R_{\boldsymbol{k}-\boldsymbol{q}\uparrow,\boldsymbol{k}^{'}\sigma^{'}}(\omega)
\right). 
\end{equation} 

Within the mean-field approximation, the components of the magnetic field with  a specific  momentum Q are singled out, leading to a set of coupled equations that describe spin-flip processes for spin-up and spin-down electrons, respectively
\begin{equation}  \label{Gup}
\left(
\begin{array}{cccc}
\hbar \omega - \varepsilon_{\boldsymbol{k}} & -i \alpha_{\textrm{R}} k^{(-)} & 0 &  \frac12 g_e \mu_B  B_{-}(-\boldsymbol{Q})  \\
i \alpha_{\textrm{R}} k^{(+)}  & \hbar \omega - \varepsilon_{\boldsymbol{k}} &  \frac12 g_e \mu_B  B_{+}(-\boldsymbol{Q}) & 0 \\
0 & \frac12 g_e \mu_B  B_{-}(\boldsymbol{Q})  &  \hbar \omega - \varepsilon_{\boldsymbol{k}+\boldsymbol{Q}} & -i \alpha_{\textrm{R}} (k+Q)^{(-)} \\
 \frac12 g_e \mu_B  B_{+}(\boldsymbol{Q})  & 0 & i \alpha_{\textrm{R}} (k+Q)^{(+)} &   \hbar \omega - \varepsilon_{\boldsymbol{k}+\boldsymbol{Q}}
\end{array}
\right) 
\left( 
\begin{array}{c}
G^R_{\boldsymbol{k}\uparrow,\boldsymbol{k}\uparrow} (\omega)  \\
G^R_{\boldsymbol{k}\downarrow,\boldsymbol{k}\uparrow} (\omega)  \\
G^R_{\boldsymbol{k}+\boldsymbol{Q}\uparrow,\boldsymbol{k}\uparrow}   (\omega) \\
G^R_{\boldsymbol{k}+\boldsymbol{Q}\downarrow,\boldsymbol{k}\uparrow}   (\omega) 
\end{array}
 \right) = 
\left(
\begin{array}{c}
\hbar  \\
0 \\
0 \\
0
\end{array}
\right),
\end{equation}
\begin{equation}  \label{Gdown}
\left(
\begin{array}{cccc}
\hbar \omega - \varepsilon_{\boldsymbol{k}} &  i \alpha_{\textrm{R}} k^{(+)} & 0 &  \frac12 g_e \mu_B  B_{+}(-\boldsymbol{Q})  \\
- i \alpha_{\textrm{R}} k^{(-)}  & \hbar \omega - \varepsilon_{\boldsymbol{k}} &  \frac12 g_e \mu_B  B_{-}(-\boldsymbol{Q}) & 0 \\
0 & \frac12 g_e \mu_B  B_{+}(\boldsymbol{Q})  &  \hbar \omega - \varepsilon_{\boldsymbol{k}+\boldsymbol{Q}} & i \alpha_{\textrm{R}} (k+Q)^{(+)} \\
 \frac12 g_e \mu_B  B_{-}(\boldsymbol{Q})  & 0 & - i \alpha_{\textrm{R}} (k+Q)^{(-)} &   \hbar \omega - \varepsilon_{\boldsymbol{k}+\boldsymbol{Q}}
\end{array}
\right) 
\left( 
\begin{array}{c}
G^R_{\boldsymbol{k}\downarrow,\boldsymbol{k}\downarrow} (\omega)  \\
G^R_{\boldsymbol{k}\uparrow,\boldsymbol{k}\downarrow} (\omega)  \\
G^R_{\boldsymbol{k}+\boldsymbol{Q}\downarrow,\boldsymbol{k}\downarrow}   (\omega) \\
G^R_{\boldsymbol{k}+\boldsymbol{Q}\uparrow,\boldsymbol{k}\downarrow}   (\omega) 
\end{array}
 \right) = 
\left(
\begin{array}{c}
\hbar  \\
0 \\
0 \\
0
\end{array}
\right).
\end{equation}
Here $k^{(\pm)} = k_x \pm i k_y$ and $(k+Q)^{(\pm)} = (k+Q)_x \pm i (k+Q)_y$. 

In the small $\alpha_{\textrm{R}}$ limit, the matrices in Eqs. (\ref{Gup}) and (\ref{Gdown}) can be presented as
\begin{equation}
\left(
\begin{array}{cccc}
\hbar \omega - \varepsilon_{\boldsymbol{k}} & -i \alpha_{\textrm{R}} k^{(-)} & 0 &  \frac12 g_e \mu_B  B_{-}(-\boldsymbol{Q})  \\
i \alpha_{\textrm{R}} k^{(+)}  & \hbar \omega - \varepsilon_{\boldsymbol{k}} &  \frac12 g_e \mu_B  B_{+}(-\boldsymbol{Q}) & 0 \\
0 & \frac12 g_e \mu_B  B_{-}(\boldsymbol{Q})  &  \hbar \omega - \varepsilon_{\boldsymbol{k}+\boldsymbol{Q}} & -i \alpha_{\textrm{R}} (k+Q)^{(-)} \\
 \frac12 g_e \mu_B  B_{+}(\boldsymbol{Q})  & 0 & i \alpha_{\textrm{R}} (k+Q)^{(+)} &   \hbar \omega - \varepsilon_{\boldsymbol{k}+\boldsymbol{Q}}
\end{array}
\right)  = \hat{A}_{\uparrow} - \alpha_{\textrm{R}} \hat{B}_{\uparrow}, 
\end{equation}
\begin{equation}
\left(
\begin{array}{cccc}
\hbar \omega - \varepsilon_{\boldsymbol{k}} &  i \alpha_{\textrm{R}} k^{(+)} & 0 &  \frac12 g_e \mu_B  B_{+}(-\boldsymbol{Q})  \\
- i \alpha_{\textrm{R}} k^{(-)}  & \hbar \omega - \varepsilon_{\boldsymbol{k}} &  \frac12 g_e \mu_B  B_{-}(-\boldsymbol{Q}) & 0 \\
0 & \frac12 g_e \mu_B  B_{+}(\boldsymbol{Q})  &  \hbar \omega - \varepsilon_{\boldsymbol{k}+\boldsymbol{Q}} & i \alpha_{\textrm{R}} (k+Q)^{(+)} \\
 \frac12 g_e \mu_B  B_{-}(\boldsymbol{Q})  & 0 & - i \alpha_{\textrm{R}} (k+Q)^{(-)} &   \hbar \omega - \varepsilon_{\boldsymbol{k}+\boldsymbol{Q}}
\end{array}
\right)  =  \hat{A}_{\downarrow} - \alpha_{\textrm{R}} \hat{B}_{\downarrow}. 
\end{equation}

Here,
\begin{equation}
\hat{A}_{\uparrow} =
\left(
\begin{array}{cccc}
\hbar \omega - \varepsilon_{\boldsymbol{k}} & 0 & 0 &  \frac12 g_e \mu_B  B^{\ast}_{+}(\boldsymbol{Q})  \\
0  & \hbar \omega - \varepsilon_{\boldsymbol{k}} &  \frac12 g_e \mu_B  B^{\ast}_{-}(\boldsymbol{Q}) & 0 \\
0 & \frac12 g_e \mu_B  B_{-}(\boldsymbol{Q})  &  \hbar \omega - \varepsilon_{\boldsymbol{k}+\boldsymbol{Q}} & 0 \\
 \frac12 g_e \mu_B  B_{+}(\boldsymbol{Q})  & 0 & 0 &   \hbar \omega - \varepsilon_{\boldsymbol{k}+\boldsymbol{Q}}
\end{array}
\right),  
\end{equation}
\begin{equation}
\hat{B}_{\uparrow} = 
\left(
\begin{array}{cccc}
0  &  i  k^{(-)} & 0 &  0  \\
- i  k^{(+)}  & 0 &  0 & 0 \\
0 & 0 & 0 & i  (k+Q)^{(-)} \\
 0  & 0 & - i  (k+Q)^{(+)} &   0
\end{array}
\right), 
\end{equation}
\begin{equation}
\hat{A}_{\downarrow} =
\left(
\begin{array}{cccc}
\hbar \omega - \varepsilon_{\boldsymbol{k}} & 0 & 0 &  \frac12 g_e \mu_B  B^{\ast}_{-}(\boldsymbol{Q})  \\
0  & \hbar \omega - \varepsilon_{\boldsymbol{k}} &  \frac12 g_e \mu_B  B^{\ast}_{+}(\boldsymbol{Q}) & 0 \\
0 & \frac12 g_e \mu_B  B_{+}(\boldsymbol{Q})  &  \hbar \omega - \varepsilon_{\boldsymbol{k}+\boldsymbol{Q}} & 0 \\
 \frac12 g_e \mu_B  B_{-}(\boldsymbol{Q})  & 0 & 0 &   \hbar \omega - \varepsilon_{\boldsymbol{k}+\boldsymbol{Q}}
\end{array}
\right),  
\end{equation}
\begin{equation}
\hat{B}_{\downarrow} = 
\left(
\begin{array}{cccc}
0  &  - i  k^{(+)} & 0 &  0  \\
 i  k^{(-)}  & 0 &  0 & 0 \\
0 & 0 & 0 & - i  (k+Q)^{(+)} \\
 0  & 0 &  i  (k+Q)^{(-)} &   0
\end{array}
\right). 
\end{equation}

To find the Green functions, we use the series 
\begin{equation}
\left(  \hat{A}_s -  \alpha_{\textrm{R}} \hat{B}_s \right)^{-1}  = \hat{A}^{-1}_s  +  \alpha_{\textrm{R}} \hat{A}^{-1}_s \hat{B}_s \hat{A}^{-1}_s +  \alpha^2_{\textrm{R}} \hat{A}^{-1}_s \hat{B}_s \hat{A}^{-1}_s \hat{B}_s \hat{A}^{-1}_s + \ldots
\end{equation}
with $s =\uparrow, \downarrow$.  

A direct calculation yields
\begin{equation}
\hat{A}^{-1}_{\uparrow} =  
\left(
\begin{array}{cccc}
 \frac{\hbar \omega - \varepsilon_{\boldsymbol{k}+\boldsymbol{Q}}}{\Delta_{12}} & 0 & 0 &  -\frac12 g_e \mu_B  \frac{B^{\ast}_{+}(\boldsymbol{Q})}{\Delta_{12}}  \\
0  &  \frac{\hbar \omega - \varepsilon_{\boldsymbol{k}+\boldsymbol{Q}}}{\Delta_{34}} &  - \frac12 g_e \mu_B  \frac{B^{\ast}_{-}(\boldsymbol{Q})}{\Delta_{34}} & 0 \\
0 &  - \frac12 g_e \mu_B  \frac{B_{-}(\boldsymbol{Q})}{\Delta_{34}}   &  \frac{\hbar \omega - \varepsilon_{\boldsymbol{k}}}{\Delta_{34}} & 0 \\
-\frac12 g_e \mu_B  \frac{B_{+}(\boldsymbol{Q})}{\Delta_{12}}  & 0 & 0 &   \frac{\hbar \omega - \varepsilon_{\boldsymbol{k}}}{\Delta_{12}}
\end{array}
\right),  
\end{equation}
\begin{equation}
\hat{A}^{-1}_{\downarrow} =  
\left(
\begin{array}{cccc}
 \frac{\hbar \omega - \varepsilon_{\boldsymbol{k}+\boldsymbol{Q}}}{\Delta_{34}} & 0 & 0 &  -\frac12 g_e \mu_B  \frac{B^{\ast}_{-}(\boldsymbol{Q})}{\Delta_{34}}  \\
0  &  \frac{\hbar \omega - \varepsilon_{\boldsymbol{k}+\boldsymbol{Q}}}{\Delta_{12}} &  - \frac12 g_e \mu_B  \frac{B^{\ast}_{+}(\boldsymbol{Q})}{\Delta_{12}} & 0 \\
0 &  - \frac12 g_e \mu_B  \frac{B_{+}(\boldsymbol{Q})}{\Delta_{12}}   &  \frac{\hbar \omega - \varepsilon_{\boldsymbol{k}}}{\Delta_{12}} & 0 \\
-\frac12 g_e \mu_B  \frac{B_{-}(\boldsymbol{Q})}{\Delta_{34}}  & 0 & 0 &   \frac{\hbar \omega - \varepsilon_{\boldsymbol{k}}}{\Delta_{34}}
\end{array}
\right),  
\end{equation}
where 
\begin{equation}
\Delta_{12} = \left[ \hbar \omega - \hbar \omega_1 (\boldsymbol{k}) \right] \left[ \hbar \omega - \hbar \omega_2 (\boldsymbol{k}) \right],
\end{equation}
\begin{equation}
\Delta_{34} = \left[ \hbar \omega - \hbar \omega_3 (\boldsymbol{k}) \right] \left[ \hbar \omega - \hbar \omega_4 (\boldsymbol{k}) \right].
\end{equation}
The energies $\omega_i (\boldsymbol{k})$ ($i=1,...4$) of the electronic subbands are given by Eqs. (\ref{w12p0},\ref{w34p0}). 

From Eqs. (\ref{Gup},\ref{Gdown}), it follows that
\begin{equation}
G^R_{\boldsymbol{k}\uparrow,\boldsymbol{k}\uparrow} (\omega) \approx G^{R(0)}_{\boldsymbol{k}\uparrow,\boldsymbol{k}\uparrow} (\omega)  + G^{R(2)}_{\boldsymbol{k}\uparrow,\boldsymbol{k}\uparrow} (\omega), 
\end{equation}
\begin{equation}
G^R_{\boldsymbol{k}\downarrow,\boldsymbol{k}\downarrow} (\omega) \approx G^{R(0)}_{\boldsymbol{k}\downarrow,\boldsymbol{k}\downarrow} (\omega)  + G^{R(2)}_{\boldsymbol{k}\downarrow,\boldsymbol{k}\downarrow} (\omega),
\end{equation}
wherein contributions are limited to terms up to second order in the small parameter $\alpha_{\textrm{R}}$
\begin{equation}
G^{R(0)}_{\boldsymbol{k}\uparrow,\boldsymbol{k}\uparrow} (\omega) = \frac{\hbar \left( \hbar \omega - \varepsilon_{\boldsymbol{k}+\boldsymbol{Q}} \right)}{\Delta_{12}}, 
\qquad 
G^{R(0)}_{\boldsymbol{k}\downarrow,\boldsymbol{k}\downarrow} (\omega) = \frac{\hbar \left( \hbar \omega - \varepsilon_{\boldsymbol{k}+\boldsymbol{Q}} \right)}{\Delta_{34}},
\end{equation}
$$
G^{R(2)}_{\boldsymbol{k}\uparrow,\boldsymbol{k}\uparrow} (\omega) = \frac{\hbar \alpha^2_{\textrm{R}}}{\Delta^2_{12}\Delta_{34}} \left[
 k^2_{\perp} \left( \hbar \omega - \varepsilon_{\boldsymbol{k}+\boldsymbol{Q}} \right)^3  
-  \left( \frac12 \tor{g_e} \mu_B \right)^2 k^{(+)} (k+Q)^{(+)} B_{-}(\boldsymbol{Q})B^{\ast}_{+}(\boldsymbol{Q}) \left( \hbar \omega - \varepsilon_{\boldsymbol{k}+\boldsymbol{Q}} \right)
\right.
$$
\begin{equation} \label{GF2up}
\left.
-  \left( \frac12 \tor{g_e} \mu_B \right)^2  k^{(-)} (k+Q)^{(-)} B_{+}(\boldsymbol{Q})B^{\ast}_{-}(\boldsymbol{Q}) \left( \hbar \omega - \varepsilon_{\boldsymbol{k}+\boldsymbol{Q}} \right)
+ (k+Q)^2_{\perp} \left( \frac12 \tor{g_e} \mu_B \right)^2  \left| B_{+}(\boldsymbol{Q}) \right|^2  \left( \hbar \omega - \varepsilon_{\boldsymbol{k}} \right)
\right],
\end{equation}
$$
G^{R(2)}_{\boldsymbol{k}\downarrow,\boldsymbol{k}\downarrow} (\omega) = \frac{\hbar \alpha^2_{\textrm{R}}}{\Delta_{12}\Delta^2_{34}} \left[
 k^2_{\perp} \left( \hbar \omega - \varepsilon_{\boldsymbol{k}+\boldsymbol{Q}} \right)^3  
-  \left( \frac12 \tor{g_e} \mu_B \right)^2 k^{(+)} (k+Q)^{(+)} B_{-}(\boldsymbol{Q})B^{\ast}_{+}(\boldsymbol{Q}) \left( \hbar \omega - \varepsilon_{\boldsymbol{k}+\boldsymbol{Q}} \right)
\right.
$$
\begin{equation}   \label{GF2down}
\left.
-  \left( \frac12 \tor{g_e} \mu_B \right)^2  k^{(-)} (k+Q)^{(-)} B_{+}(\boldsymbol{Q})B^{\ast}_{-}(\boldsymbol{Q}) \left( \hbar \omega - \varepsilon_{\boldsymbol{k}+\boldsymbol{Q}} \right)
+ (k+Q)^2_{\perp} \left( \frac12 \tor{g_e} \mu_B \right)^2  \left| B_{-}(\boldsymbol{Q}) \right|^2  \left( \hbar \omega - \varepsilon_{\boldsymbol{k}} \right)
\right]. 
\end{equation}
Here, $k^2_{\perp} = k^2_x +k^2_y$.

Introducing the notations 
$$
\mathcal{Z}_{\uparrow} (\boldsymbol{k},\omega) = k^2_{\perp} \left( \hbar \omega - \varepsilon_{\boldsymbol{k}+\boldsymbol{Q}} \right)^3  
-  \left( \frac12 \tor{g_e} \mu_B \right)^2 k^{(+)} (k+Q)^{(+)} B_{-}(\boldsymbol{Q})B^{\ast}_{+}(\boldsymbol{Q}) \left( \hbar \omega - \varepsilon_{\boldsymbol{k}+\boldsymbol{Q}} \right)
$$
\begin{equation}
-  \left( \frac12 \tor{g_e} \mu_B \right)^2  k^{(-)} (k+Q)^{(-)} B_{+}(\boldsymbol{Q})B^{\ast}_{-}(\boldsymbol{Q}) \left( \hbar \omega - \varepsilon_{\boldsymbol{k}+\boldsymbol{Q}} \right)
+ (k+Q)^2_{\perp} \left( \frac12 \tor{g_e} \mu_B \right)^2  \left| B_{+}(\boldsymbol{Q}) \right|^2  \left( \hbar \omega - \varepsilon_{\boldsymbol{k}} \right), 
\end{equation}
$$
\mathcal{Z}_{\downarrow} (\boldsymbol{k},\omega) =  k^2_{\perp} \left( \hbar \omega - \varepsilon_{\boldsymbol{k}+\boldsymbol{Q}} \right)^3  
-  \left( \frac12 \tor{g_e} \mu_B \right)^2 k^{(+)} (k+Q)^{(+)} B_{-}(\boldsymbol{Q})B^{\ast}_{+}(\boldsymbol{Q}) \left( \hbar \omega - \varepsilon_{\boldsymbol{k}+\boldsymbol{Q}} \right)
$$
\begin{equation}
-  \left( \frac12 \tor{g_e} \mu_B \right)^2  k^{(-)} (k+Q)^{(-)} B_{+}(\boldsymbol{Q})B^{\ast}_{-}(\boldsymbol{Q}) \left( \hbar \omega - \varepsilon_{\boldsymbol{k}+\boldsymbol{Q}} \right)
+ (k+Q)^2_{\perp} \left( \frac12 \tor{g_e} \mu_B \right)^2  \left| B_{-}(\boldsymbol{Q}) \right|^2  \left( \hbar \omega - \varepsilon_{\boldsymbol{k}} \right), 
\end{equation}
and using the partial fraction decompositions
\begin{equation}
\frac{1}{\left( \omega - \omega_1 (\boldsymbol{k}) \right)^2 \left( \omega - \omega_2 (\boldsymbol{k}) \right)^2 \left( \omega - \omega_3 (\boldsymbol{k}) \right) \left( \omega - \omega_4 (\boldsymbol{k}) \right)}
=  \sum_{i=1}^4 \frac{P_i(\boldsymbol{k})}{\omega - \omega_i(\boldsymbol{k})} + \frac{S_1(\boldsymbol{k})}{\left( \omega - \omega_1 (\boldsymbol{k}) \right)^2} + \frac{S_2 (\boldsymbol{k})}{\left( \omega - \omega_2 (\boldsymbol{k}) \right)^2}, 
\end{equation}
\begin{equation}
\frac{1}{\left( \omega - \omega_1 (\boldsymbol{k}) \right) \left( \omega - \omega_2 (\boldsymbol{k}) \right) \left( \omega - \omega_3 (\boldsymbol{k}) \right)^2 \left( \omega - \omega_4 (\boldsymbol{k}) \right)^2}
=  \sum_{i=1}^4 \frac{R_i (\boldsymbol{k})}{\omega - \omega_i (\boldsymbol{k})} + \frac{S_3 (\boldsymbol{k})}{\left( \omega - \omega_3 (\boldsymbol{k}) \right)^2} + \frac{S_4(\boldsymbol{k})}{\left( \omega - \omega_4 (\boldsymbol{k})\right)^2},
\end{equation}
the second-order Green functions (\ref{GF2up},\ref{GF2down}) are found to be
\begin{equation}
G^{R(2)}_{\boldsymbol{k}\uparrow,\boldsymbol{k}\uparrow} (\omega) =  \frac{\alpha^2_{\textrm{R}}}{\hbar^5} \mathcal{Z}_{\uparrow} (\boldsymbol{k},\omega)  \left[   \sum_{i=1}^4 \frac{P_i(\boldsymbol{k})}{\omega - \omega_i(\boldsymbol{k})} + \frac{S_1(\boldsymbol{k})}{\left( \omega - \omega_1 (\boldsymbol{k})\right)^2} + \frac{S_2(\boldsymbol{k})}{\left( \omega - \omega_2 (\boldsymbol{k})\right)^2} \right], 
\end{equation}
\begin{equation}
G^{R(2)}_{\boldsymbol{k}\downarrow,\boldsymbol{k}\downarrow} (\omega) =  \frac{\alpha^2_{\textrm{R}}}{\hbar^5} \mathcal{Z}_{\downarrow} (\boldsymbol{k},\omega) \left[  \sum_{i=1}^4 \frac{R_i(\boldsymbol{k})}{\omega - \omega_i(\boldsymbol{k})} + \frac{S_3(\boldsymbol{k})}{\left( \omega - \omega_3 (\boldsymbol{k}) \right)^2} + \frac{S_4(\boldsymbol{k})}{\left( \omega - \omega_4(\boldsymbol{k}) \right)^2} \right]. 
\end{equation}
The coefficients $P_i(\boldsymbol{k})$, $R_i(\boldsymbol{k})$, and $S_i(\boldsymbol{k})$ are expressed in terms of the differences between the eigenfrequencies,  $\omega_{ij}(\boldsymbol{k}) = \omega_i (\boldsymbol{k}) - \omega_j (\boldsymbol{k})$, 
\begin{equation}
P_1(\boldsymbol{k}) = - \frac{1}{\omega^3_{12}(\boldsymbol{k})\omega^2_{13}(\boldsymbol{k}) \omega^2_{14}(\boldsymbol{k})} \left[ \omega_{12}(\boldsymbol{k}) \left( \omega_{13}(\boldsymbol{k}) + \omega_{14}(\boldsymbol{k}) \right)+ 2 \omega_{13}(\boldsymbol{k}) \omega_{14}(\boldsymbol{k})  \right], \quad 
\end{equation}
\begin{equation}
P_2(\boldsymbol{k}) = - \frac{1}{\omega^3_{12}(\boldsymbol{k})\omega^2_{23}(\boldsymbol{k}) \omega^2_{24}(\boldsymbol{k})} \left[ \omega_{12}(\boldsymbol{k}) \left( \omega_{23}(\boldsymbol{k}) + \omega_{24}(\boldsymbol{k}) \right) - 2 \omega_{23}(\boldsymbol{k}) \omega_{24}(\boldsymbol{k})  \right]
\end{equation}
\begin{equation}
P_3(\boldsymbol{k}) = \frac{1}{\omega_{34}(\boldsymbol{k})\omega^2_{13}(\boldsymbol{k})\omega^2_{23}(\boldsymbol{k})}, 
\qquad
P_4(\boldsymbol{k}) =  - \frac{1}{\omega_{34}(\boldsymbol{k})\omega^2_{14}(\boldsymbol{k})\omega^2_{24}(\boldsymbol{k})},
\end{equation} 
\begin{equation}
R_1(\boldsymbol{k}) = \frac{1}{\omega_{12}(\boldsymbol{k})\omega^2_{13}(\boldsymbol{k})\omega^2_{14}(\boldsymbol{k})}, \qquad
R_2(\boldsymbol{k}) = - \frac{1}{\omega_{12}(\boldsymbol{k})\omega^2_{23}(\boldsymbol{k})\omega^2_{24}(\boldsymbol{k})}, 
\end{equation} 
\begin{equation}
R_3(\boldsymbol{k}) = \frac{1}{\omega^3_{34}(\boldsymbol{k})\omega^2_{13}(\boldsymbol{k}) \omega^2_{23}(\boldsymbol{k})} \left[ \omega_{34}(\boldsymbol{k}) \left( \omega_{13}(\boldsymbol{k}) + \omega_{23}(\boldsymbol{k}) \right) - 2 \omega_{23}(\boldsymbol{k}) \omega_{13}(\boldsymbol{k})  \right], 
\end{equation} 
\begin{equation}
R_4(\boldsymbol{k}) = \frac{1}{\omega^3_{34}(\boldsymbol{k})\omega^2_{14}(\boldsymbol{k}) \omega^2_{24}(\boldsymbol{k})} \left[ \omega_{34}(\boldsymbol{k}) \left( \omega_{14}(\boldsymbol{k}) + \omega_{24}(\boldsymbol{k}) \right) + 2 \omega_{14}(\boldsymbol{k}) \omega_{24}(\boldsymbol{k})  \right],  
\end{equation} 
\begin{equation}
S_1(\boldsymbol{k}) = \frac{1}{\omega^2_{12}(\boldsymbol{k}) \omega_{13}(\boldsymbol{k}) \omega_{14}(\boldsymbol{k})}, \quad S_2(\boldsymbol{k}) = \frac{1}{\omega^2_{12}(\boldsymbol{k}) \omega_{23}(\boldsymbol{k}) \omega_{24}(\boldsymbol{k})}, 
\end{equation} 
\begin{equation}
S_3(\boldsymbol{k}) = \frac{1}{\omega^2_{34}(\boldsymbol{k}) \omega_{13}(\boldsymbol{k}) \omega_{23}(\boldsymbol{k})}, \quad S_4(\boldsymbol{k}) = \frac{1}{\omega^2_{34}(\boldsymbol{k}) \omega_{14}(\boldsymbol{k}) \omega_{24}(\boldsymbol{k})}.
\end{equation}

Making use of the identities
\begin{equation}
\frac{1}{x \pm i \epsilon} = \textrm{P} \frac{1}{x} \mp i \pi \delta (x),  \qquad 
\frac{1}{\left(x \pm i \epsilon\right)^2} = \textrm{P} \frac{1}{x^2} \pm i \pi \delta^{'} (x),
\end{equation}
the jumps of the second-order Green functions across the real $\omega$-axis are
\begin{equation}
G^{R(2)}_{\boldsymbol{k}\uparrow,\boldsymbol{k}\uparrow} (\omega - i \epsilon) - G^{R(2)}_{\boldsymbol{k}\uparrow,\boldsymbol{k}\uparrow} (\omega + i \epsilon) = 2\pi i \frac{\alpha^2_{\textrm{R}}}{\hbar^5} \mathcal{Z}_{\uparrow} (\boldsymbol{k},\omega)  \left[ \sum_{i=1}^4 P_i(\boldsymbol{k}) \delta (\omega - \omega_i(\boldsymbol{k}))  -  \sum_{i=1,2} S_i(\boldsymbol{k}) \delta^{'}  (\omega - \omega_i(\boldsymbol{k})) \right], 
\end{equation}
\begin{equation}
G^{R(2)}_{\boldsymbol{k}\downarrow,\boldsymbol{k}\downarrow} (\omega - i \epsilon) - G^{R(2)}_{\boldsymbol{k}\downarrow,\boldsymbol{k}\downarrow} (\omega + i \epsilon) = 2\pi i \frac{\alpha^2_{\textrm{R}}}{\hbar^5} \mathcal{Z}_{\downarrow} (\boldsymbol{k},\omega)  \left[ \sum_{i=1}^4 R_i(\boldsymbol{k}) \delta (\omega - \omega_i(\boldsymbol{k}))  -  \sum_{i=3,4} S_i(\boldsymbol{k}) \delta^{'}  (\omega - \omega_i(\boldsymbol{k})) \right]. 
\end{equation}

The corresponding spectral densities
\begin{equation}
\mathcal{J}^{(2)} _{ c_{\boldsymbol{k}\uparrow} c^{\dagger}_{\boldsymbol{k}\uparrow}} (\omega) = 2\pi \frac{\alpha^2_{\textrm{R}}}{\hbar^5} \mathcal{Z}_{\uparrow} (\boldsymbol{k},\omega)  n_{F}(\omega) \left[ \sum_{i=1}^4 P_i(\boldsymbol{k}) \delta (\omega - \omega_i(\boldsymbol{k}))  -  \sum_{i=1,2} S_i(\boldsymbol{k}) \delta^{'}  (\omega - \omega_i(\boldsymbol{k})) \right],
\end{equation}
\begin{equation}
 \mathcal{J}^{(2)} _{ c_{\boldsymbol{k}\downarrow} c^{\dagger}_{\boldsymbol{k}\downarrow}} (\omega) = 2\pi \frac{\alpha^2_{\textrm{R}}}{\hbar^5} \mathcal{Z}_{\downarrow} (\boldsymbol{k},\omega)  n_{F}(\omega) \left[ \sum_{i=1}^4 R_i(\boldsymbol{k}) \delta (\omega - \omega_i(\boldsymbol{k}))  -  \sum_{i=3,4} S_i(\boldsymbol{k}) \delta^{'}  (\omega - \omega_i(\boldsymbol{k})) \right] 
\end{equation}
determine the thermal averages of the fermion occupation numbers
\begin{equation}
\left\langle c^{\dagger}_{\boldsymbol{k}\uparrow} c_{\boldsymbol{k}\uparrow}  \right\rangle^{(2)}  =  \int^{+\infty}_{-\infty} \frac{d \omega}{2\pi} \mathcal{J}^{(2)} _{ c_{\boldsymbol{k}\uparrow} c^{\dagger}_{\boldsymbol{k}\uparrow}} (\omega)  
= \frac{\alpha^2_{\textrm{R}}}{\hbar^5} 
\left\{
\sum_{i=1}^4 P_i (\boldsymbol{k}) \mathcal{Z}_{\uparrow} (\boldsymbol{k},\omega_i)   n_{F}(\omega_i(\boldsymbol{k}))  + \sum_{i=1,2} S_i(\boldsymbol{k}) \frac{\partial}{\partial \omega}  \left[ \mathcal{Z}_{\uparrow} (\boldsymbol{k},\omega)   n_{F}(\omega) \right]_{\omega = \omega_i(\boldsymbol{k})}
\right\},
\end{equation}
\begin{equation}
\left\langle c^{\dagger}_{\boldsymbol{k}\downarrow} c_{\boldsymbol{k}\downarrow}  \right\rangle^{(2)}  =  \int^{+\infty}_{-\infty} \frac{d \omega}{2\pi} \mathcal{J}^{(2)}_{ c_{\boldsymbol{k}\downarrow} c^{\dagger}_{\boldsymbol{k}\downarrow}} (\omega)  
= \frac{\alpha^2_{\textrm{R}}}{\hbar^5} 
\left\{
\sum_{i=1}^4 R_i(\boldsymbol{k}) \mathcal{Z}_{\downarrow} (\boldsymbol{k},\omega_i)   n_{F}(\omega_i(\boldsymbol{k}))  + \sum_{i=3,4} S_i(\boldsymbol{k}) \frac{\partial}{\partial \omega}  \left[ \mathcal{Z}_{\downarrow} (\boldsymbol{k},\omega)   n_{F}(\omega) \right]_{\omega = \omega_i(\boldsymbol{k})}
\right\}, 
\end{equation}
where
\begin{equation}
\left. \frac{\partial}{\partial \omega} \left[ \mathcal{Z}_s (\boldsymbol{k},\omega)   n_{F}(\omega) \right] \right|_{\omega = \omega_i(\boldsymbol{k})} = \left. \frac{\partial}{\partial \omega} \left[ \mathcal{Z}_s (\boldsymbol{k},\omega)    \right] \right|_{\omega = \omega_i(\boldsymbol{k})} n_{F}(\omega_i(\boldsymbol{k})) - \beta \hbar  \mathcal{Z}_s (\boldsymbol{k},\omega_i(\boldsymbol{k}))  n_{F}(\omega_i(\boldsymbol{k})) \left[ 1 - n_{F}(\omega_i(\boldsymbol{k}))  \right].
\end{equation}
The second-order correction to net magnetization is obtained from the difference of number of particles with up- and down-spin orientations
\begin{equation}
\left\langle  m_z \right\rangle^{(2)} = \frac12  g_e \mu_B \sum_{\boldsymbol{k}} \left( \left\langle c^{\dagger}_{\boldsymbol{k}\uparrow} c_{\boldsymbol{k}\uparrow}  \right\rangle^{(2)} -  \left\langle c^{\dagger}_{\boldsymbol{k}\downarrow} c_{\boldsymbol{k}\downarrow}  \right\rangle^{(2)} \right). 
\end{equation}

At $T=0$, when $n_{F}(\omega_1(\boldsymbol{k}))=n_{F}(\omega_3(\boldsymbol{k}))=0$ and $n_{F}(\omega_2(\boldsymbol{k}))=n_{F}(\omega_4(\boldsymbol{k}))=1$, one readily obtains
$$
\left\langle  m_z \right\rangle^{(2)} 
 = \frac{g_e \mu_B \alpha^2_{\textrm{R}}}{2 \hbar^5} \sum_{\boldsymbol{k}} 
\left[   
P_2 (\boldsymbol{k})  \mathcal{Z}_{\uparrow} (\boldsymbol{k},\omega_2(\boldsymbol{k}))  -  R_2 (\boldsymbol{k})  \mathcal{Z}_{\downarrow} (\boldsymbol{k},\omega_2(\boldsymbol{k}))   
\right. 
$$  
\begin{equation}
\left.
+  P_4 (\boldsymbol{k})  \mathcal{Z}_{\uparrow} (\boldsymbol{k},\omega_4(\boldsymbol{k}))  -  R_4 (\boldsymbol{k})  \mathcal{Z}_{\downarrow} (\boldsymbol{k},\omega_4(\boldsymbol{k})) 
+ \tor{S_2} (\boldsymbol{k}) \mathcal{Z}^{'}_{\uparrow} (\boldsymbol{k},\omega_2(\boldsymbol{k}))  - \tor{S_4} (\boldsymbol{k}) \mathcal{Z}^{'}_{\downarrow} (\boldsymbol{k},\omega_4(\boldsymbol{k})) 
\right],
\end{equation}
where the prime means the derivative with respect to $\omega$.

\end{widetext}

\bibliographystyle{apsrev4-2}
\bibliography{references,SMextra}

\clearpage
\onecolumngrid
\begin{center}
{\large\bfseries Supplemental Material for\\
``Inverse Chiral Phonon Zeeman Effect in Noncentrosymmetric Crystals''}\\[6pt]
Jun-ichiro Kishine, A.~S.~Ovchinnikov, Masahiro Sato, G.~N.~Makarov, and A.~D.~Lyakhov
\end{center}
\vspace{8pt}
\twocolumngrid
\setcounter{section}{0}
\setcounter{subsection}{0}
\setcounter{equation}{0}
\setcounter{figure}{0}
\setcounter{table}{0}
\renewcommand{\thesection}{S\arabic{section}}
\renewcommand{\thesubsection}{\thesection.\arabic{subsection}}
\renewcommand{\theequation}{S\arabic{equation}}
\renewcommand{\thefigure}{S\arabic{figure}}
\renewcommand{\thetable}{S\arabic{table}}
\makeatletter
\def\@hangfrom@section#1#2#3{\@hangfrom{#1#2}\MakeTextUppercase{#3}}%
\let\@sectioncntformat\relax
\@removefromreset{equation}{section}
\makeatother
This Supplemental Material contains the explicit Bogolyubov transformations diagonalizing the longitudinal and transverse micropolar Hamiltonians, and the general form of the Floquet high-frequency expansion. The physical discussion is contained in the main article and its Appendices.

\section{Bogolyubov transformation for the longitudinal modes}
\label{app:bogoL}

To find the $uv$-coefficients of the Bogolyubov transformations the following  retarded Green functions are introduced
\begin{equation}  \label{Gaa1}
G^{-+}_{aa} (k,t) = - i \theta(t) \left\langle \left[ a_{3k}(t), a^{\dagger}_{3k}(0) \right] \right\rangle, 
\end{equation}
\begin{equation}
G^{++}_{aa} (k,t) = - i \theta(t) \left\langle \left[ a^{\dagger}_{3-k}(t), a^{\dagger}_{3k}(0) \right] \right\rangle, 
\end{equation}
\begin{equation}
G^{-+}_{ba} (k,t) = - i \theta(t) \left\langle \left[ b_{3k}(t), a^{\dagger}_{3k}(0) \right] \right\rangle, 
\end{equation}
\begin{equation}  \label{Gba1}
G^{++}_{ba} (k,t) = - i \theta(t) \left\langle \left[ b^{\dagger}_{3-k}(t), a^{\dagger}_{3k}(0) \right] \right\rangle.  
\end{equation}

\begin{widetext}

Equations of motion for Fourier transforms of these Green functions take the form
\begin{equation}
\left(
\begin{array}{cccc}
\Omega^{(u)}_{3k} - \omega & 0 & K_{3k} & K_{3k} \\
0 & \Omega^{(u)}_{3k} +  \omega &  K_{3k} & K_{3k} \\
K_{3k} & K_{3k} & \Omega^{(\varphi)}_{3k} - \omega & 0  \\
K_{3k} & K_{3k} & 0 &  \Omega^{(\varphi)}_{3k} + \omega   \\
\end{array}
\right)
\left(
\begin{array}{c}
G^{-+}_{aa} (k,\omega)  \\
G^{++}_{aa} (k,\omega)  \\
G^{-+}_{ba} (k,\omega)   \\
G^{++}_{ba} (k,\omega) 
\end{array}
\right)
=
\left(
\begin{array}{c}
-1  \\ 
0 \\
0 \\
0
\end{array}
\right),
\end{equation}
from which we find
\begin{equation} \label{Gmpaa1}
G^{-+}_{aa} (k,\omega) = \frac{\left( \omega+  \Omega^{(u)}_{3k} \right) \left(  \tred{\omega^2} - \left[ \Omega^{(\varphi)}_{3k} \right]^2 \right) + 2 K^2_{3k} \Omega^{(\varphi)}_{3k}}{\left(  \tred{\omega^2} - \left[ \tred{\omega^{(\gf)}_{3k}}\right]^2\right) \left(  \tred{\omega^2} - \left[ \tred{\omega^{(\gl)}_{3k}}\right]^2\right)},
\end{equation}
\begin{equation} \label{Gppaa1}
G^{++}_{aa} (k,\omega) = -  \frac{2K^2_{3k}  \Omega^{(\varphi)}_{3k} }{\left(  \tred{\omega^2} - \left[ \tred{\omega^{(\gf)}_{3k}}\right]^2\right) \left(  \tred{\omega^2} - \left[ \tred{\omega^{(\gl)}_{3k}}\right]^2\right)},
\end{equation}
\begin{equation} \label{Gmpba1}
G^{-+}_{ba} (k,\omega) =   \frac{K_{3k}  \left(\Omega^{(u)}_{3k} + \omega \right)  \left( \Omega^{(\varphi)}_{3k} + \omega \right) }{\left(  \tred{\omega^2} - \left[ \tred{\omega^{(\gf)}_{3k}}\right]^2\right) \left(  \tred{\omega^2} - \left[ \tred{\omega^{(\gl)}_{3k}}\right]^2\right)},
\end{equation}
\begin{equation} \label{Gppba1}
G^{++}_{ba} (k,\omega) =   \frac{K_{3k}  \left(\Omega^{(u)}_{3k} + \omega \right)  \left( \Omega^{(\varphi)}_{3k} - \omega \right) }{\left(  \tred{\omega^2} - \left[ \tred{\omega^{(\gf)}_{3k}}\right]^2\right) \left(  \tred{\omega^2} - \left[ \tred{\omega^{(\gl)}_{3k}}\right]^2\right)}, 
\end{equation}

The poles at real frequencies yield the spectrum $\tred{\omega^{(\gf/\gl)}_{3k}}$  of the longitudinal micropolar elastic waves. 

Similarly, one can define the retarded Green functions 
$$
G^{-+}_{\alpha \alpha} (k,t) = - i \theta(t) \left\langle  \left[ \alpha_{3k}(t), \alpha^{\dagger}_{3k}(0)  \right] \right\rangle,
$$
\begin{equation}
G^{+-}_{\alpha \alpha} (k,t) = - i \theta(t) \left\langle  \left[ \alpha^{\dagger}_{3-k}(t), \alpha_{3-k}(0)  \right] \right\rangle,
\end{equation}
\begin{equation}
G^{-+}_{\beta \beta} (k,t) = - i \theta(t) \left\langle  \left[ \beta_{3k}(t), \beta^{\dagger}_{3k}(0)  \right] \right\rangle,
\end{equation}
\begin{equation}
G^{+-}_{\beta \beta} (k,t) = - i \theta(t) \left\langle  \left[ \beta^{\dagger}_{3-k}(t), \beta_{3-k}(0)  \right] \right\rangle   
\end{equation}
related to the diagonalized Hamiltonian
\begin{equation}  \label{Hdiag-app-2}
\tilde{\mathcal{H}}_L =  \Delta E_0 + \sum_k \left( \hbar \tred{\omega^{(\gl)}_{3k}}  \alpha^{\dagger}_{3k} \alpha_{3k}  + \hbar \tred{\omega^{(\gf)}_{3k}}  \beta^{\dagger}_{3k} \beta_{3k}  \right).    
\end{equation}

The explicit form of their Fourier transforms can be easily found to be
\begin{equation}  \label{DGaa}
G^{-+}_{\alpha \alpha} (k,\omega) = \frac{1}{\omega - \tred{\omega^{(\gl)}_{3k}}},  \qquad 
G^{+-}_{\alpha \alpha} (k,\omega) = -  \frac{1}{\omega + \tred{\omega^{(\gl)}_{3-k}}},
\end{equation}
\begin{equation}  \label{DGbb}
G^{-+}_{\beta \beta} (k,\omega) = \frac{1}{\omega - \tred{\omega^{(\gf)}_{3k}}},  \qquad 
G^{+-}_{\beta \beta} (k,\omega) = -  \frac{1}{\omega + \tred{\omega^{(\gf)}_{3-k}}}.  
\end{equation}

Using the Bogolyubov $uv$-transformation
\begin{equation} \label{aDecomp-app-2}
a_{3k} = u_{11}(k) \alpha_{3k} + u_{12}(k) \beta_{3k} + v^{\ast}_{11}(-k) \alpha^{\dagger}_{3-k} + v^{\ast}_{12}(-k)  \beta^{\dagger}_{3-k}, 
\end{equation}
\begin{equation}  \label{bDecomp-app-2}
b_{3k} = u_{21}(k) \alpha_{3k} + u_{22}(k) \beta_{3k} + v^{\ast}_{21}(-k) \alpha^{\dagger}_{3-k} + v^{\ast}_{22}(-k)  \beta^{\dagger}_{3-k},   
\end{equation}
one can establish the relationships between the Green functions (\ref{Gaa1}-\ref{Gba1}) and (\ref{DGaa}-\ref{DGbb})
\begin{equation}  \label{Gmpaa2}
G^{-+}_{aa} (k,\omega) = \frac{\left| u_{11}(k)\right|^2}{\omega - \tred{\omega^{(\gl)}_{3k}}} + \frac{\left| u_{12}(k)\right|^2}{\omega - \tred{\omega^{(\gf)}_{3k}}}
-  \frac{\left| v_{11}(-k)\right|^2}{\omega + \tred{\omega^{(\gl)}_{3-k}}} -  \frac{\left| v_{12}(-k)\right|^2}{\omega + \tred{\omega^{(\gf)}_{3-k}}}, 
\end{equation}
\begin{equation}  \label{Gppaa2}
G^{++}_{aa} (k,\omega) =  - \frac{ u^{\ast}_{11}(-k) v_{11} (-k)}{\omega + \tred{\omega^{(\gl)}_{3-k}}} -
\frac{ u^{\ast}_{12}(-k) v_{12} (-k)}{\omega + \tred{\omega^{(\gf)}_{3-k}}}
+  \frac{ u^{\ast}_{11}(k) v_{11} (k)}{\omega - \tred{\omega^{(\gl)}_{3k}}}
+  \frac{ u^{\ast}_{12}(k) v_{12} (k)}{\omega - \tred{\omega^{(\gf)}_{3k}}}, 
\end{equation}
\begin{equation}  \label{Gmpba2}
G^{-+}_{ba} (k,\omega) =   \frac{ u^{\ast}_{11}(k) u_{21} (k)}{\omega - \tred{\omega^{(\gl)}_{3k}}} -
\frac{ v^{\ast}_{21}(-k) v_{11} (-k)}{\omega + \tred{\omega^{(\gl)}_{3-k}}}
+  \frac{ u^{\ast}_{12}(k) u_{22} (k)}{\omega - \tred{\omega^{(\gf)}_{3k}}}
-  \frac{ v^{\ast}_{22}(-k) v_{12} (-k)}{\omega + \tred{\omega^{(\gf)}_{3-k}}}, 
\end{equation}
\begin{equation}  \label{Gppba2}
G^{++}_{ba} (k,\omega) =   - \frac{ u^{\ast}_{21}(-k) v_{11} (-k)}{\omega + \tred{\omega^{(\gl)}_{3-k}}} 
- \frac{ u^{\ast}_{22}(-k) v_{12} (-k)}{\omega + \tred{\omega^{(\gf)}_{3-k}}}
+  \frac{ u^{\ast}_{11}(k) v_{21} (k)}{\omega - \tred{\omega^{(\gl)}_{3k}}}
+  \frac{ u^{\ast}_{12}(k) v_{22} (k)}{\omega - \tred{\omega^{(\gf)}_{3k}}}.
\end{equation}

A direct  comparison between these expressions and those given by Eqs. (\ref{Gmpaa1}-\ref{Gppba1}) leads to 
\begin{equation} \label{u11}
 u_{11}(k)  =  \left[
\frac{\left( \tred{\omega^{(\gl)}_{3k}} +  \Omega^{(u)}_{3k} \right) \left(  \left[  \tred{\omega^{(\gl)}_{3k}} \right]^2 - \left[ \Omega^{(\varphi)}_{3k} \right]^2 \right) + 2 K^2_{3k} \Omega^{(\varphi)}_{3k}}{2\tred{\omega^{(\gl)}_{3k}} \left(  \left[  \tred{\omega^{(\gl)}_{3k}} \right]^2  -  \left[  \tred{\omega^{(\gf)}_{3k}} \right]^2   \right)}
\right]^{\frac12} e^{i\phi_1(k)},   
\end{equation}
\begin{equation}  \label{u12}
u_{12}(k) =  \left[
\frac{\left( \tred{\omega^{(\gf)}_{3k}} +  \Omega^{(u)}_{3k} \right) \left(  \left[  \tred{\omega^{(\gf)}_{3k}} \right]^2 - \left[ \Omega^{(\varphi)}_{3k} \right]^2 \right) + 2 K^2_{3k} \Omega^{(\varphi)}_{3k}}{2\tred{\omega^{(\gf)}_{3k}} \left(  \left[  \tred{\omega^{(\gf)}_{3k}} \right]^2  -  \left[  \tred{\omega^{(\gl)}_{3k}} \right]^2   \right)}
\right]^{\frac12} e^{i\phi_2(k)},   
\end{equation}
\begin{equation}  \label{u21}
u_{21}(k) =  \frac{K_{3k}  \left(\Omega^{(u)}_{3k} + \tred{\omega^{(\gl)}_{3k}} \right)  \left( \Omega^{(\varphi)}_{3k} + \tred{\omega^{(\gl)}_{3k}} \right) }{\left[  2\tred{\omega^{(\gl)}_{3k}} \left(  \left[  \tred{\omega^{(\gl)}_{3k}} \right]^2  -  \left[  \tred{\omega^{(\gf)}_{3k}} \right]^2   \right)  \left\{ \left( \tred{\omega^{(\gl)}_{3k}} +  \Omega^{(u)}_{3k} \right) \left(  \left[  \tred{\omega^{(\gl)}_{3k}} \right]^2 - \left[ \Omega^{(\varphi)}_{3k} \right]^2 \right) + 2 K^2_{3k} \Omega^{(\varphi)}_{3k} \right\} \right]^{\frac12}} \, \,  e^{i\phi_1(k)},  
\end{equation} 
\begin{equation}  \label{u22}
u_{22}(k) =  \frac{K_{3k}  \left(\Omega^{(u)}_{3k} + \tred{\omega^{(\gf)}_{3k}} \right)  \left( \Omega^{(\varphi)}_{3k} + \tred{\omega^{(\gf)}_{3k}} \right) }{\left[  2\tred{\omega^{(\gf)}_{3k}} \left(  \left[  \tred{\omega^{(\gf)}_{3k}} \right]^2  -  \left[  \tred{\omega^{(\gl)}_{3k}} \right]^2   \right)  \left\{ \left( \tred{\omega^{(\gf)}_{3k}} +  \Omega^{(u)}_{3k} \right) \left(  \left[  \tred{\omega^{(\gf)}_{3k}} \right]^2 - \left[ \Omega^{(\varphi)}_{3k} \right]^2 \right) + 2 K^2_{3k} \Omega^{(\varphi)}_{3k} \right\} \right]^{\frac12}} \, \,  e^{i\phi_2(k)}.   
\end{equation} 
\begin{equation}  \label{v11}
v_{11}(k) = -   \frac{2K^2_{3k}  \Omega^{(\varphi)}_{3k} }{\left[  2\tred{\omega^{(\gl)}_{3k}} \left(  \left[  \tred{\omega^{(\gl)}_{3k}} \right]^2  -  \left[  \tred{\omega^{(\gf)}_{3k}} \right]^2   \right)  \left\{ \left( \tred{\omega^{(\gl)}_{3k}} +  \Omega^{(u)}_{3k} \right) \left(  \left[  \tred{\omega^{(\gl)}_{3k}} \right]^2 - \left[ \Omega^{(\varphi)}_{3k} \right]^2 \right) + 2 K^2_{3k} \Omega^{(\varphi)}_{3k} \right\} \right]^{\frac12}}    \, \,  e^{i\phi_1(k)},  
\end{equation} 
\begin{equation}  \label{v12}
v_{12}(k)  = -   \frac{2K^2_{3k}  \Omega^{(\varphi)}_{3k} }{\left[  2\tred{\omega^{(\gf)}_{3k}} \left(  \left[  \tred{\omega^{(\gf)}_{3k}} \right]^2  -  \left[  \tred{\omega^{(\gl)}_{3k}} \right]^2   \right)  \left\{ \left( \tred{\omega^{(\gf)}_{3k}} +  \Omega^{(u)}_{3k} \right) \left(  \left[  \tred{\omega^{(\gf)}_{3k}} \right]^2 - \left[ \Omega^{(\varphi)}_{3k} \right]^2 \right) + 2 K^2_{3k} \Omega^{(\varphi)}_{3k} \right\} \right]^{\frac12}}    \, \,  e^{i\phi_2(k)},  
\end{equation}
\begin{equation}  \label{v21}
v_{21}(k) =  \frac{K_{3k}  \left(\Omega^{(u)}_{3k} + \tred{\omega^{(\gl)}_{3k}} \right)  \left( \Omega^{(\varphi)}_{3k} - \tred{\omega^{(\gl)}_{3k}} \right) }{\left[  2\tred{\omega^{(\gl)}_{3k}} \left(  \left[  \tred{\omega^{(\gl)}_{3k}} \right]^2  -  \left[  \tred{\omega^{(\gf)}_{3k}} \right]^2   \right)  \left\{ \left( \tred{\omega^{(\gl)}_{3k}} +  \Omega^{(u)}_{3k} \right) \left(  \left[  \tred{\omega^{(\gl)}_{3k}} \right]^2 - \left[ \Omega^{(\varphi)}_{3k} \right]^2 \right) + 2 K^2_{3k} \Omega^{(\varphi)}_{3k} \right\} \right]^{\frac12}} \, \,  e^{i\phi_1(k)},  
\end{equation} 
\begin{equation}  \label{v22}
v_{22}(k) =  \frac{K_{3k}  \left(\Omega^{(u)}_{3k} + \tred{\omega^{(\gf)}_{3k}} \right)  \left( \Omega^{(\varphi)}_{3k} - \tred{\omega^{(\gf)}_{3k}} \right) }{\left[  2\tred{\omega^{(\gf)}_{3k}} \left(  \left[  \tred{\omega^{(\gf)}_{3k}} \right]^2  -  \left[  \tred{\omega^{(\gl)}_{3k}} \right]^2   \right)  \left\{ \left( \tred{\omega^{(\gf)}_{3k}} +  \Omega^{(u)}_{3k} \right) \left(  \left[  \tred{\omega^{(\gf)}_{3k}} \right]^2 - \left[ \Omega^{(\varphi)}_{3k} \right]^2 \right) + 2 K^2_{3k} \Omega^{(\varphi)}_{3k} \right\} \right]^{\frac12}} \, \,  e^{i\phi_2(k)},   
\end{equation} 
Here $\phi_{1,2}(k)$ are arbitrary real parameters, which were taken zero in numerical estimations.   

\section{Bogolyubov transformation for the transverse modes}
\label{app:bogoT}

The $uv$-transformation is introduced in the following form
\begin{equation}  \label{BogTran}
\left(
\begin{array}{c}
a_{1k} \\
a_{2k} \\
b_{1k}  \\
b_{2k} \\
a^{\dagger}_{1-k} \\
a^{\dagger}_{2-k} \\
b^{\dagger}_{1-k} \\
b^{\dagger}_{2-k} \\
\end{array}
\right) 
=
\left(
\begin{array}{cccccccc}
\tilde{u}_{11}(k) & \tilde{u}_{12}(k) & \tilde{u}_{13}(k) &  \tilde{u}_{14}(k) &  \tilde{v}^{\ast}_{11}(-k) & \tilde{v}^{\ast}_{12}(-k) &  \tilde{v}^{\ast}_{13}(-k) & \tilde{v}^{\ast}_{14}(-k) \\ 
\tilde{u}_{21}(k) & \tilde{u}_{22}(k) & \tilde{u}_{23}(k) &  \tilde{u}_{24}(k) &  \tilde{v}^{\ast}_{21}(-k) & \tilde{v}^{\ast}_{22}(-k) &  \tilde{v}^{\ast}_{23}(-k) & \tilde{v}^{\ast}_{24}(-k) \\ 
\tilde{u}_{31}(k) & \tilde{u}_{32}(k) & \tilde{u}_{33}(k) &  \tilde{u}_{34}(k) &  \tilde{v}^{\ast}_{31}(-k) & \tilde{v}^{\ast}_{32}(-k) &  \tilde{v}^{\ast}_{33}(-k) & \tilde{v}^{\ast}_{34}(-k) \\ 
\tilde{u}_{41}(k) & \tilde{u}_{42}(k) & \tilde{u}_{43}(k) &  \tilde{u}_{44}(k) &  \tilde{v}^{\ast}_{41}(-k) & \tilde{v}^{\ast}_{42}(-k) &  \tilde{v}^{\ast}_{43}(-k) & \tilde{v}^{\ast}_{44}(-k) \\  
\tilde{v}_{11}(k) & \tilde{v}_{12}(k) & \tilde{v}_{13}(k) &  \tilde{v}_{14}(k) & \tilde{u}^{\ast}_{11}(-k) & \tilde{u}^{\ast}_{12}(-k) & \tilde{u}^{\ast}_{13}(-k) & \tilde{u}^{\ast}_{14}(-k) \\
\tilde{v}_{21}(k) & \tilde{v}_{22}(k) & \tilde{v}_{23}(k) &  \tilde{v}_{24}(k) & \tilde{u}^{\ast}_{21}(-k) & \tilde{u}^{\ast}_{22}(-k) & \tilde{u}^{\ast}_{23}(-k) & \tilde{u}^{\ast}_{24}(-k) \\
\tilde{v}_{31}(k) & \tilde{v}_{32}(k) & \tilde{v}_{33}(k) &  \tilde{v}_{34}(k) & \tilde{u}^{\ast}_{31}(-k) & \tilde{u}^{\ast}_{32}(-k) & \tilde{u}^{\ast}_{33}(-k) & \tilde{u}^{\ast}_{34}(-k) \\
\tilde{v}_{41}(k) & \tilde{v}_{42}(k) & \tilde{v}_{43}(k) &  \tilde{v}_{44}(k) & \tilde{u}^{\ast}_{41}(-k) & \tilde{u}^{\ast}_{42}(-k) & \tilde{u}^{\ast}_{43}(-k) & \tilde{u}^{\ast}_{44}(-k) 
\end{array}
\right)
\left(
\begin{array}{c}
\alpha_{+,k} \\
\alpha_{-,k} \\
\beta_{+,k}  \\
\beta_{-,k} \\
\alpha^{\dagger}_{+,-k} \\
\alpha^{\dagger}_{-,-k} \\
\beta^{\dagger}_{+,-k} \\
\beta^{\dagger}_{-,-k} \\
\end{array}
\right).   
\end{equation}

In terms of the new $\alpha$, $\beta$ quasiparticles  the initial Hamiltonian becomes diagonal
\begin{equation} \label{HTrans-app-3}
\mathcal{H}_{T} = \Delta E^{(T)}_0 + \sum_k \left[
\hbar \tred{\omega^{(\gl)}_{+}}(k) \alpha^{\dagger}_{+,k} \alpha_{+,k} 
+ \hbar \tred{\omega^{(\gl)}_{-}}(k) \alpha^{\dagger}_{-,k} \alpha_{-,k} 
+  \hbar \tred{\omega^{(\gf)}_{+}}(k) \beta^{\dagger}_{+,k} \beta_{+,k} 
+  \hbar \tred{\omega^{(\gf)}_{-}}(k) \beta^{\dagger}_{-,k} \beta_{-,k} 
\right],
\end{equation}
where the frequencies $\tred{\omega^{(\gl)}_{\pm}}$,  $\tred{\omega^{(\gf)}_{\pm}}$   have to be determined from the poles of the retarded  Green functions
\begin{equation}  \label{Ga1a1}
G^{-+}_{a_1 a_1}(k,t) = - i \theta(t) \langle \left[  a_{1k}(t), a^{\dagger}_{1k}(0) \right] \rangle,  \qquad 
G^{++}_{a_1 a_1}(k,t) = - i \theta(t) \langle \left[  a^{\dagger}_{1-k}(t), a^{\dagger}_{1k}(0) \right] \rangle, 
\end{equation}
\begin{equation}
G^{-+}_{a_2 a_1}(k,t) = - i \theta(t) \langle \left[  a_{2k}(t), a^{\dagger}_{1k}(0) \right] \rangle,  \qquad 
G^{++}_{a_2 a_1}(k,t) = - i \theta(t) \langle \left[  a^{\dagger}_{2-k}(t), a^{\dagger}_{1k}(0) \right] \rangle, 
\end{equation}
\begin{equation}
G^{-+}_{b_1 a_1}(k,t) = - i \theta(t) \langle \left[  b_{1k}(t), a^{\dagger}_{1k}(0) \right] \rangle,   \qquad 
G^{++}_{b_1 a_1}(k,t) = - i \theta(t) \langle \left[  b^{\dagger}_{1-k}(t), a^{\dagger}_{1k}(0) \right] \rangle, 
\end{equation}
\begin{equation}  \label{Gb2a1}
G^{-+}_{b_2 a_1}(k,t) = - i \theta(t) \langle \left[  b_{2k}(t), a^{\dagger}_{1k}(0) \right] \rangle,   \qquad 
G^{++}_{b_2 a_1}(k,t) = - i \theta(t) \langle \left[  b^{\dagger}_{2-k}(t), a^{\dagger}_{1k}(0) \right] \rangle.  
\end{equation}  

 The coupled set of their equations of motion are written most conveniently in the Fourier representation
\begin{equation}  \label{EOMT}
\left(
\begin{array}{cccccccc}
\Omega^{(u)}_{1k} - \omega &  0 & 0 & 0 & K_{0k} & K_{0k} & - i K_{2k} & - i K_{2k} \\
0 & \Omega^{(u)}_{1k} + \omega & 0 & 0 &  K_{0k} & K_{0k} &  - i K_{2k} & - i K_{2k} \\
0 & 0 & \Omega^{(u)}_{2k} - \omega & 0 &   i K_{2k} &   i K_{2k} &  K_{0k} & K_{0k} \\
0 & 0 & 0 & \Omega^{(u)}_{2k} + \omega &  i K_{2k} &   i K_{2k} &  K_{0k} & K_{0k} \\
K_{0k} & K_{0k} & - i K_{2k} & - i K_{2k} & \Omega^{(\varphi)}_{1k} - \omega & 0 & - iK_{1k} & - iK_{1k}  \\
K_{0k} & K_{0k} &  - i K_{2k} & - i K_{2k} & 0 & \Omega^{(\varphi)}_{1k} + \omega & - iK_{1k} & - iK_{1k} \\
 i K_{2k} &   i K_{2k} &  K_{0k} & K_{0k} &   iK_{1k} &  iK_{1k}  & \Omega^{(\varphi)}_{1k} - \omega &   0 \\
 i K_{2k} &   i K_{2k} & K_{0k} & K_{0k} &  iK_{1k} &  iK_{1k}  & 0   & \Omega^{(\varphi)}_{1k} + \omega
\end{array}
\right)
\left(
\begin{array}{c}
G^{-+}_{a_1 a_1}(k,\omega) \\
G^{++}_{a_1 a_1}(k,\omega) \\
G^{-+}_{a_2 a_1}(k,\omega) \\
G^{++}_{a_2 a_1}(k,\omega) \\
G^{-+}_{b_1 a_1}(k,\omega) \\
G^{++}_{b_1 a_1}(k,\omega) \\
G^{-+}_{b_2 a_1}(k,\omega) \\
G^{++}_{b_2 a_1}(k,\omega) \\
\end{array}
\right)
=
\left(
\begin{array}{c}
-1 \\
0 \\
0 \\
0 \\
0\\
0\\
0\\
0
\end{array}
\right). 
\end{equation}

The zeros of the determinant  of the left-hand side matrix, which can be transformed to the form
\begin{equation}
\Delta (k,\omega) = \left( \omega^2 -  \left[\tred{\omega^{(\gl)}_{+,k}}\right]^2  \right)  \left( \omega^2 -  \left[\tred{\omega^{(\gf)}_{+,k}}\right]^2  \right) \left( \omega^2 -  \left[\tred{\omega^{(\gl)}_{-,k}}\right]^2  \right)  \left( \omega^2 -  \left[\tred{\omega^{(\gf)}_{-,k}}\right]^2  \right), 
\end{equation}
give the frequencies of the micropolar elastic waves. 

The Green functions themselves can be constructed via the minors  of the $\mu\nu$-elements, $\Delta_{\mu\nu}$,     
\begin{equation}  \label{Ga1a1old}
G^{-+}_{a_1 a_1}(k,\omega) = - \frac{\Delta_{11}(k,\omega)}{\Delta(k,\omega)},  \qquad 
G^{++}_{a_1 a_1}(k,\omega)  = \frac{\Delta_{12}(k,\omega)}{\Delta(k,\omega)}, 
\end{equation}
\begin{equation}
G^{-+}_{a_2 a_1}(k,\omega)  =  - \frac{\Delta_{13}(k,\omega)}{\Delta(k,\omega)},  \qquad 
G^{++}_{a_2 a_1}(k,\omega)  = \frac{\Delta_{14}(k,\omega)}{\Delta(k,\omega)}, 
\end{equation}
\begin{equation}
G^{-+}_{b_1 a_1}(k,\omega) = - \frac{\Delta_{15}(k,\omega)}{\Delta(k,\omega)},  \qquad 
G^{++}_{b_1 a_1}(k,\omega) = \frac{\Delta_{16}(k,\omega)}{\Delta(k,\omega)}, 
\end{equation}
\begin{equation}  \label{Gb2a1old}
G^{-+}_{b_2 a_1}(k,\omega) = - \frac{\Delta_{17}(k,\omega)}{\Delta(k,\omega)},  \qquad 
G^{++}_{b_2 a_1}(k,\omega)  = \frac{\Delta_{18}(k,\omega)}{\Delta(k,\omega)}. 
\end{equation}

Intrducing the retarded Green functions built up from the quasiparticles of the diagonalized Hamiltonian~(\ref{HTrans-app-3})
\begin{equation}
G^{-+}_{\alpha \alpha} (+,k,t) = - i \theta(t) \left\langle  \left[ \alpha_{+,k}(t), \alpha^{\dagger}_{+,k}(0)  \right] \right\rangle, \quad 
G^{+-}_{\alpha \alpha} (+,k,t) = - i \theta(t) \left\langle  \left[ \alpha^{\dagger}_{+,-k}(t), \alpha_{+,-k}(0)  \right] \right\rangle,
\end{equation}
\begin{equation}
G^{-+}_{\alpha \alpha} (-,k,t) = - i \theta(t) \left\langle  \left[ \alpha_{-,k}(t), \alpha^{\dagger}_{-,k}(0)  \right] \right\rangle, \quad 
G^{+-}_{\alpha \alpha} (-,k,t) = - i \theta(t) \left\langle  \left[ \alpha^{\dagger}_{-,-k}(t), \alpha_{-,-k}(0)  \right] \right\rangle,
\end{equation}
\begin{equation}
G^{-+}_{\beta \beta} (+,k,t) = - i \theta(t) \left\langle  \left[ \beta_{+,k}(t), \beta^{\dagger}_{+,k}(0)  \right] \right\rangle,  \quad 
G^{+-}_{\beta \beta} (+,k,t) = - i \theta(t) \left\langle  \left[ \beta^{\dagger}_{+,-k}(t), \beta_{+,-k}(0)  \right] \right\rangle,
\end{equation}
\begin{equation}
G^{-+}_{\beta \beta} (-,k,t) = - i \theta(t) \left\langle  \left[ \beta_{-,k}(t), \beta^{\dagger}_{-,k}(0)  \right] \right\rangle,  \quad 
G^{+-}_{\beta \beta} (-,k,t) = - i \theta(t) \left\langle  \left[ \beta^{\dagger}_{-,-k}(t), \beta_{-,-k}(0)  \right] \right\rangle. 
\end{equation}

Their Fourier transforms can be easily found to be
\begin{equation}
G^{-+}_{\alpha \alpha} (+,k,\omega) = \frac{1}{\omega - \tred{\omega^{(\gl)}_{+}}(k)}, \qquad 
G^{+-}_{\alpha \alpha} (+,k,\omega) = - \frac{1}{\omega +  \tred{\omega^{(\gl)}_{-}}(k)},
\end{equation}
\begin{equation}
G^{-+}_{\alpha \alpha} (-,k,\omega) =  \frac{1}{\omega -  \tred{\omega^{(\gl)}_{-}}(k)},   \qquad 
G^{+-}_{\alpha \alpha} (-,k,\omega) =  - \frac{1}{\omega +  \tred{\omega^{(\gl)}_{+}}(k)},
\end{equation}
\begin{equation}
G^{-+}_{\beta \beta} (+,k,\omega) = \frac{1}{\omega - \tred{\omega^{(\gf)}_{+}}(k)},   \qquad 
G^{+-}_{\beta \beta} (+,k,\omega) = - \frac{1}{\omega +  \tred{\omega^{(\gf)}_{-}}(k)},
\end{equation}
\begin{equation}
G^{-+}_{\beta \beta} (-,k,\omega) =  \frac{1}{\omega -  \tred{\omega^{(\gf)}_{-}}(k)},   \qquad 
G^{+-}_{\beta \beta} (-,k,\omega) =  - \frac{1}{\omega +  \tred{\omega^{(\gf)}_{+}}(k)}. 
\end{equation}

The Bogolyubov transformation (\ref{BogTran}) enables us to relate these Green functions to that of interacting quasiparticles (\ref{Ga1a1}-\ref{Gb2a1})  
$$
G^{-+}_{a_1 a_1} (k,\omega) =  \frac{\left| \tilde{u}_{11}(k) \right|^2}{\omega - \tred{\omega^{(\gl)}_{+}}(k)}  +   \frac{\left| \tilde{u}_{12}(k) \right|^2}{\omega - \tred{\omega^{(\gl)}_{-}}(k)} 
+  \frac{\left| \tilde{u}_{13}(k) \right|^2}{\omega - \tred{\omega^{(\gf)}_{+}}(k)} +  \frac{\left| \tilde{u}_{14}(k) \right|^2}{\omega - \tred{\omega^{(\gf)}_{-}}(k)} 
$$
\begin{equation}  \label{Ga1Ga1new}
-   \frac{\left| \tilde{v}_{11}(-k) \right|^2}{\omega + \tred{\omega^{(\gl)}_{-}}(k)} -   \frac{\left| \tilde{v}_{12}(-k) \right|^2}{\omega + \tred{\omega^{(\gl)}_{+}}(k)} 
-  \frac{\left| \tilde{v}_{13}(-k) \right|^2}{\omega + \tred{\omega^{(\gf)}_{-}}(k)} -   \frac{\left| \tilde{v}_{14}(-k) \right|^2}{\omega + \tred{\omega^{(\gf)}_{+}}(k)},  
\end{equation}
$$
G^{++}_{a_1 a_1} (k,\omega) =  \frac{\tilde{v}_{11}(k) \tilde{u}_{11}^{\ast}(k)}{\omega - \tred{\omega^{(\gl)}_{+}}(k)}  +   \frac{\tilde{v}_{12}(k) \tilde{u}_{12}^{\ast}(k)}{\omega - \tred{\omega^{(\gl)}_{-}}(k)} 
+  \frac{\tilde{v}_{13}(k) \tilde{u}_{13}^{\ast}(k)}{\omega - \tred{\omega^{(\gf)}_{+}}(k)} +  \frac{\tilde{v}_{14}(k) \tilde{u}_{14}^{\ast}(k)}{\omega - \tred{\omega^{(\gf)}_{-}}(k)} 
$$
\begin{equation}
-   \frac{\tilde{v}_{11}(-k) \tilde{u}_{11}^{\ast}(-k)}{\omega + \tred{\omega^{(\gl)}_{-}}(k)} -   \frac{\tilde{v}_{12}(-k) \tilde{u}_{12}^{\ast}(-k)}{\omega + \tred{\omega^{(\gl)}_{+}}(k)} 
-  \frac{\tilde{v}_{13}(-k) \tilde{u}_{13}^{\ast}(-k)}{\omega + \tred{\omega^{(\gf)}_{-}}(k)} -   \frac{\tilde{v}_{14}(-k) \tilde{u}_{14}^{\ast}(-k)}{\omega + \tred{\omega^{(\gf)}_{+}}(k)},  
\end{equation}
$$
G^{-+}_{a_2 a_1} (k,\omega) =  \frac{\tilde{u}_{21}(k) \tilde{u}_{11}^{\ast}(k)}{\omega - \tred{\omega^{(\gl)}_{+}}(k)}  +   \frac{\tilde{u}_{22}(k) \tilde{u}_{12}^{\ast}(k)}{\omega - \tred{\omega^{(\gl)}_{-}}(k)} 
+  \frac{\tilde{u}_{23}(k) \tilde{u}_{13}^{\ast}(k)}{\omega - \tred{\omega^{(\gf)}_{+}}(k)} +  \frac{\tilde{u}_{24}(k) \tilde{u}_{14}^{\ast}(k)}{\omega - \tred{\omega^{(\gf)}_{-}}(k)} 
$$
\begin{equation}
-   \frac{\tilde{v}^{\ast}_{21}(-k) \tilde{v}_{11}(-k)}{\omega + \tred{\omega^{(\gl)}_{-}}(k)} -   \frac{\tilde{v}^{\ast}_{22}(-k) \tilde{v}_{12}(-k)}{\omega + \tred{\omega^{(\gl)}_{+}}(k)} 
-  \frac{\tilde{v}^{\ast}_{23}(-k) \tilde{v}_{13}(-k)}{\omega + \tred{\omega^{(\gf)}_{-}}(k)} -   \frac{\tilde{v}^{\ast}_{24}(-k) \tilde{v}_{14}(-k)}{\omega + \tred{\omega^{(\gf)}_{+}}(k)},  
\end{equation}
$$
G^{++}_{a_2 a_1} (k,\omega) =  \frac{\tilde{v}_{21}(k) \tilde{u}_{11}^{\ast}(k)}{\omega - \tred{\omega^{(\gl)}_{+}}(k)}  +   \frac{\tilde{v}_{22}(k) \tilde{u}_{12}^{\ast}(k)}{\omega - \tred{\omega^{(\gl)}_{-}}(k)} 
+  \frac{\tilde{v}_{23}(k) \tilde{u}_{13}^{\ast}(k)}{\omega - \tred{\omega^{(\gf)}_{+}}(k)} +  \frac{\tilde{v}_{24}(k) \tilde{u}_{14}^{\ast}(k)}{\omega - \tred{\omega^{(\gf)}_{-}}(k)} 
$$
\begin{equation}
-   \frac{\tilde{u}^{\ast}_{21}(-k) \tilde{v}_{11}(-k)}{\omega + \tred{\omega^{(\gl)}_{-}}(k)} -   \frac{\tilde{u}^{\ast}_{22}(-k) \tilde{v}_{12}(-k)}{\omega + \tred{\omega^{(\gl)}_{+}}(k)} 
-  \frac{\tilde{u}^{\ast}_{23}(-k) \tilde{v}_{13}(-k)}{\omega + \tred{\omega^{(\gf)}_{-}}(k)} -   \frac{\tilde{u}^{\ast}_{24}(-k) \tilde{v}_{14}(-k)}{\omega + \tred{\omega^{(\gf)}_{+}}(k)},  
\end{equation}
$$
G^{-+}_{b_1 a_1} (k,\omega) =  \frac{\tilde{u}_{31}(k) \tilde{u}_{11}^{\ast}(k)}{\omega - \tred{\omega^{(\gl)}_{+}}(k)}  +   \frac{\tilde{u}_{32}(k) \tilde{u}_{12}^{\ast}(k)}{\omega - \tred{\omega^{(\gl)}_{-}}(k)} 
+  \frac{\tilde{u}_{33}(k) \tilde{u}_{13}^{\ast}(k)}{\omega - \tred{\omega^{(\gf)}_{+}}(k)} +  \frac{\tilde{u}_{34}(k) \tilde{u}_{14}^{\ast}(k)}{\omega - \tred{\omega^{(\gf)}_{-}}(k)} 
$$
\begin{equation}
-   \frac{\tilde{v}^{\ast}_{31}(-k) \tilde{v}_{11}(-k)}{\omega + \tred{\omega^{(\gl)}_{-}}(k)} -   \frac{\tilde{v}^{\ast}_{32}(-k) \tilde{v}_{12}(-k)}{\omega + \tred{\omega^{(\gl)}_{+}}(k)} 
-  \frac{\tilde{v}^{\ast}_{33}(-k) \tilde{v}_{13}(-k)}{\omega + \tred{\omega^{(\gf)}_{-}}(k)} -   \frac{\tilde{v}^{\ast}_{34}(-k) \tilde{v}_{14}(-k)}{\omega + \tred{\omega^{(\gf)}_{+}}(k)},  
\end{equation}
$$
G^{++}_{b_1 a_1} (k,\omega) =  \frac{\tilde{v}_{31}(k) \tilde{u}_{11}^{\ast}(k)}{\omega - \tred{\omega^{(\gl)}_{+}}(k)}  +   \frac{\tilde{v}_{32}(k) \tilde{u}_{12}^{\ast}(k)}{\omega - \tred{\omega^{(\gl)}_{-}}(k)} 
+  \frac{\tilde{v}_{33}(k) \tilde{u}_{13}^{\ast}(k)}{\omega - \tred{\omega^{(\gf)}_{+}}(k)} +  \frac{\tilde{v}_{34}(k) \tilde{u}_{14}^{\ast}(k)}{\omega - \tred{\omega^{(\gf)}_{-}}(k)} 
$$
\begin{equation}
-   \frac{\tilde{u}^{\ast}_{31}(-k) \tilde{v}_{11}(-k)}{\omega + \tred{\omega^{(\gl)}_{-}}(k)} -   \frac{\tilde{u}^{\ast}_{32}(-k) \tilde{v}_{12}(-k)}{\omega + \tred{\omega^{(\gl)}_{+}}(k)} 
-  \frac{\tilde{u}^{\ast}_{33}(-k) \tilde{v}_{13}(-k)}{\omega + \tred{\omega^{(\gf)}_{-}}(k)} -   \frac{\tilde{u}^{\ast}_{34}(-k) \tilde{v}_{14}(-k)}{\omega + \tred{\omega^{(\gf)}_{+}}(k)},  
\end{equation}
$$
G^{-+}_{b_2 a_1} (k,\omega) =  \frac{\tilde{u}_{41}(k) \tilde{u}_{11}^{\ast}(k)}{\omega - \tred{\omega^{(\gl)}_{+}}(k)}  +   \frac{\tilde{u}_{42}(k) \tilde{u}_{12}^{\ast}(k)}{\omega - \tred{\omega^{(\gl)}_{-}}(k)} 
+  \frac{\tilde{u}_{43}(k) \tilde{u}_{13}^{\ast}(k)}{\omega - \tred{\omega^{(\gf)}_{+}}(k)} +  \frac{\tilde{u}_{44}(k) \tilde{u}_{14}^{\ast}(k)}{\omega - \tred{\omega^{(\gf)}_{-}}(k)} 
$$
\begin{equation}
-   \frac{\tilde{v}^{\ast}_{41}(-k) \tilde{v}_{11}(-k)}{\omega + \tred{\omega^{(\gl)}_{-}}(k)} -   \frac{\tilde{v}^{\ast}_{42}(-k) \tilde{v}_{12}(-k)}{\omega + \tred{\omega^{(\gl)}_{+}}(k)} 
-  \frac{\tilde{v}^{\ast}_{43}(-k) \tilde{v}_{13}(-k)}{\omega + \tred{\omega^{(\gf)}_{-}}(k)} -   \frac{\tilde{v}^{\ast}_{44}(-k) \tilde{v}_{14}(-k)}{\omega + \tred{\omega^{(\gf)}_{+}}(k)},  
\end{equation}
$$
G^{++}_{b_2 a_1} (k,\omega) =  \frac{\tilde{v}_{41}(k) \tilde{u}_{11}^{\ast}(k)}{\omega - \tred{\omega^{(\gl)}_{+}}(k)}  +   \frac{\tilde{v}_{42}(k) \tilde{u}_{12}^{\ast}(k)}{\omega - \tred{\omega^{(\gl)}_{-}}(k)} 
+  \frac{\tilde{v}_{43}(k) \tilde{u}_{13}^{\ast}(k)}{\omega - \tred{\omega^{(\gf)}_{+}}(k)} +  \frac{\tilde{v}_{44}(k) \tilde{u}_{14}^{\ast}(k)}{\omega - \tred{\omega^{(\gf)}_{-}}(k)} 
$$
\begin{equation}  \label{Gb2a1new}
-   \frac{\tilde{u}^{\ast}_{41}(-k) \tilde{v}_{11}(-k)}{\omega + \tred{\omega^{(\gl)}_{-}}(k)} -   \frac{\tilde{u}^{\ast}_{42}(-k) \tilde{v}_{12}(-k)}{\omega + \tred{\omega^{(\gl)}_{+}}(k)} 
-  \frac{\tilde{u}^{\ast}_{43}(-k) \tilde{v}_{13}(-k)}{\omega + \tred{\omega^{(\gf)}_{-}}(k)} -   \frac{\tilde{u}^{\ast}_{44}(-k) \tilde{v}_{14}(-k)}{\omega + \tred{\omega^{(\gf)}_{+}}(k)}. 
\end{equation}

Comparison of (\ref{Ga1Ga1new}-\ref{Gb2a1new}) with (\ref{Ga1a1old}-\ref{Gb2a1old}) shows that the required $uv$-coefficients are 
\begin{equation}
 \tilde{u}_{11}(k)  = \left[ \frac{- \Delta_{11}\left( k, \tred{\omega^{(\gl)}_{+}}(k) \right)}{2 \tred{\omega^{(\gl)}_{+}}(k) \left\{   \left[ \tred{\omega^{(\gl)}_{+}}(k) \right]^2 - \left[ \tred{\omega^{(\gf)}_{+}}(k) \right]^2  \right\}  \left\{   \left[ \tred{\omega^{(\gl)}_{+}}(k) \right]^2 - \left[ \tred{\omega^{(\gl)}_{-}}(k) \right]^2  \right\}  \left\{   \left[ \tred{\omega^{(\gl)}_{+}}(k) \right]^2 - \left[ \tred{\omega^{(\gf)}_{-}}(k) \right]^2  \right\}}
\right]^{\frac12} e^{i\varphi_1 (k)}, 
\end{equation}
\begin{equation}
 \tilde{u}_{12}(k)  = \left[  \frac{- \Delta_{11}\left( k, \tred{\omega^{(\gl)}_{-}}(k) \right)}{2 \tred{\omega^{(\gl)}_{-}}(k) \left\{   \left[ \tred{\omega^{(\gl)}_{-}}(k) \right]^2 - \left[ \tred{\omega^{(\gf)}_{+}}(k) \right]^2  \right\}  \left\{   \left[ \tred{\omega^{(\gl)}_{-}}(k) \right]^2 - \left[ \tred{\omega^{(\gl)}_{+}}(k) \right]^2  \right\}  \left\{   \left[ \tred{\omega^{(\gl)}_{-}}(k) \right]^2 - \left[ \tred{\omega^{(\gf)}_{-}}(k) \right]^2  \right\}}
\right]^{\frac12} e^{i\varphi_2 (k)}, 
\end{equation}
\begin{equation}
 \tilde{u}_{13}(k)  = \left[ \frac{-\Delta_{11}\left( k, \tred{\omega^{(\gf)}_{+}}(k) \right)}{2 \tred{\omega^{(\gf)}_{+}}(k) \left\{   \left[ \tred{\omega^{(\gf)}_{+}}(k) \right]^2 - \left[ \tred{\omega^{(\gl)}_{+}}(k) \right]^2  \right\}  \left\{   \left[ \tred{\omega^{(\gf)}_{+}}(k) \right]^2 - \left[ \tred{\omega^{(\gl)}_{-}}(k) \right]^2  \right\}  \left\{   \left[ \tred{\omega^{(\gf)}_{+}}(k) \right]^2 - \left[ \tred{\omega^{(\gf)}_{-}}(k) \right]^2  \right\}}
\right]^{\frac12} e^{i\varphi_3 (k)}, 
\end{equation}
\begin{equation}
 \tilde{u}_{14}(k)  =  \left[  \frac{-\Delta_{11}\left( k, \tred{\omega^{(\gf)}_{-}}(k) \right)}{2 \tred{\omega^{(\gf)}_{-}}(k) \left\{   \left[ \tred{\omega^{(\gf)}_{-}}(k) \right]^2 - \left[ \tred{\omega^{(\gl)}_{+}}(k) \right]^2  \right\}  \left\{   \left[ \tred{\omega^{(\gf)}_{-}}(k) \right]^2 - \left[ \tred{\omega^{(\gf)}_{+}}(k) \right]^2  \right\}  \left\{   \left[ \tred{\omega^{(\gf)}_{-}}(k) \right]^2 - \left[ \tred{\omega^{(\gl)}_{-}}(k) \right]^2  \right\}}
\right]^{\frac12} e^{i\varphi_4 (k)}, 
\end{equation}
$$
\tilde{v}_{11}(k) = \frac{\Delta_{12}\left( k, \tred{\omega^{(\gl)}_{+}}(k) \right)}{\left[ - \Delta_{11}\left( k, \tred{\omega^{(\gl)}_{+}}(k) \right) \right]^{\frac12}}  e^{i\varphi_1 (k)}
$$
\begin{equation}
\times  \frac{1}{\left[ 2 \tred{\omega^{(\gl)}_{+}}(k)  \left\{   \left[ \tred{\omega^{(\gl)}_{+}}(k) \right]^2 - \left[ \tred{\omega^{(\gf)}_{+}}(k) \right]^2  \right\}  \left\{   \left[ \tred{\omega^{(\gl)}_{+}}(k) \right]^2 - \left[ \tred{\omega^{(\gl)}_{-}}(k) \right]^2  \right\}  \left\{   \left[ \tred{\omega^{(\gl)}_{+}}(k) \right]^2 - \left[ \tred{\omega^{(\gf)}_{-}}(k) \right]^2  \right\} \right]^{\frac12}}, 
\end{equation} 
$$
\tilde{v}_{12}(k) = \frac{\Delta_{12}\left( k, \tred{\omega^{(\gl)}_{-}}(k) \right)}{\left[ - \Delta_{11}\left( k, \tred{\omega^{(\gl)}_{-}}(k) \right) \right]^{\frac12}}  e^{i\varphi_2 (k)}
$$
\begin{equation}
\times  \frac{1}{\left[ 2 \tred{\omega^{(\gl)}_{-}}(k)  \left\{   \left[ \tred{\omega^{(\gl)}_{-}}(k) \right]^2 - \left[ \tred{\omega^{(\gf)}_{+}}(k) \right]^2  \right\}  \left\{   \left[ \tred{\omega^{(\gl)}_{-}}(k) \right]^2 - \left[ \tred{\omega^{(\gl)}_{+}}(k) \right]^2  \right\}  \left\{   \left[ \tred{\omega^{(\gl)}_{-}}(k) \right]^2 - \left[ \tred{\omega^{(\gf)}_{-}}(k) \right]^2  \right\} \right]^{\frac12}}, 
\end{equation} 
$$
\tilde{v}_{13}(k) = \frac{\Delta_{12}\left( k, \tred{\omega^{(\gf)}_{+}}(k) \right)}{\left[ - \Delta_{11}\left( k, \tred{\omega^{(\gf)}_{+}}(k) \right) \right]^{\frac12}}  e^{i\varphi_3 (k)}
$$
\begin{equation}
\times  \frac{1}{\left[ 2 \tred{\omega^{(\gf)}_{+}}(k)  \left\{   \left[ \tred{\omega^{(\gf)}_{+}}(k) \right]^2 - \left[ \tred{\omega^{(\gl)}_{+}}(k) \right]^2  \right\}  \left\{   \left[ \tred{\omega^{(\gf)}_{+}}(k) \right]^2 - \left[ \tred{\omega^{(\gl)}_{-}}(k) \right]^2  \right\}  \left\{   \left[ \tred{\omega^{(\gf)}_{+}}(k) \right]^2 - \left[ \tred{\omega^{(\gf)}_{-}}(k) \right]^2  \right\} \right]^{\frac12}}, 
\end{equation} 
$$
\tilde{v}_{14}(k) = \frac{\Delta_{12}\left( k, \tred{\omega^{(\gf)}_{-}}(k) \right)}{\left[ - \Delta_{11}\left( k, \tred{\omega^{(\gf)}_{-}}(k) \right) \right]^{\frac12}}  e^{i\varphi_4 (k)}
$$
\begin{equation}
\times  \frac{1}{\left[ 2 \tred{\omega^{(\gf)}_{-}}(k)  \left\{   \left[ \tred{\omega^{(\gf)}_{-}}(k) \right]^2 - \left[ \tred{\omega^{(\gl)}_{+}}(k) \right]^2  \right\}  \left\{   \left[ \tred{\omega^{(\gf)}_{-}}(k) \right]^2 - \left[ \tred{\omega^{(\gl)}_{-}}(k) \right]^2  \right\}  \left\{   \left[ \tred{\omega^{(\gf)}_{-}}(k) \right]^2 - \left[ \tred{\omega^{(\gf)}_{+}}(k) \right]^2  \right\} \right]^{\frac12}}, 
\end{equation} 
$$
\tilde{u}_{21}(k) = - \frac{\Delta_{13}\left( k, \tred{\omega^{(\gl)}_{+}}(k) \right)}{\left[ - \Delta_{11}\left( k, \tred{\omega^{(\gl)}_{+}}(k) \right) \right]^{\frac12}}  e^{i\varphi_1 (k)}
$$
\begin{equation}
\times  \frac{1}{\left[ 2 \tred{\omega^{(\gl)}_{+}}(k)  \left\{   \left[ \tred{\omega^{(\gl)}_{+}}(k) \right]^2 - \left[ \tred{\omega^{(\gf)}_{+}}(k) \right]^2  \right\}  \left\{   \left[ \tred{\omega^{(\gl)}_{+}}(k) \right]^2 - \left[ \tred{\omega^{(\gl)}_{-}}(k) \right]^2  \right\}  \left\{   \left[ \tred{\omega^{(\gl)}_{+}}(k) \right]^2 - \left[ \tred{\omega^{(\gf)}_{-}}(k) \right]^2  \right\} \right]^{\frac12}}, 
\end{equation} 
$$
\tilde{u}_{22}(k) = - \frac{\Delta_{13}\left( k, \tred{\omega^{(\gl)}_{-}}(k) \right)}{\left[ - \Delta_{11}\left( k, \tred{\omega^{(\gl)}_{-}}(k) \right) \right]^{\frac12}}  e^{i\varphi_2 (k)}
$$
\begin{equation}
\times  \frac{1}{\left[ 2 \tred{\omega^{(\gl)}_{-}}(k)  \left\{   \left[ \tred{\omega^{(\gl)}_{-}}(k) \right]^2 - \left[ \tred{\omega^{(\gf)}_{+}}(k) \right]^2  \right\}  \left\{   \left[ \tred{\omega^{(\gl)}_{-}}(k) \right]^2 - \left[ \tred{\omega^{(\gl)}_{+}}(k) \right]^2  \right\}  \left\{   \left[ \tred{\omega^{(\gl)}_{-}}(k) \right]^2 - \left[ \tred{\omega^{(\gf)}_{-}}(k) \right]^2  \right\} \right]^{\frac12}}, 
\end{equation} 
$$
\tilde{u}_{23}(k) = - \frac{\Delta_{13}\left( k, \tred{\omega^{(\gf)}_{+}}(k) \right)}{\left[ - \Delta_{11}\left( k, \tred{\omega^{(\gf)}_{+}}(k) \right) \right]^{\frac12}}  e^{i\varphi_3 (k)}
$$
\begin{equation}
\times  \frac{1}{\left[ 2 \tred{\omega^{(\gf)}_{+}}(k)  \left\{   \left[ \tred{\omega^{(\gf)}_{+}}(k) \right]^2 - \left[ \tred{\omega^{(\gl)}_{+}}(k) \right]^2  \right\}  \left\{   \left[ \tred{\omega^{(\gf)}_{+}}(k) \right]^2 - \left[ \tred{\omega^{(\gl)}_{-}}(k) \right]^2  \right\}  \left\{   \left[ \tred{\omega^{(\gf)}_{+}}(k) \right]^2 - \left[ \tred{\omega^{(\gf)}_{-}}(k) \right]^2  \right\} \right]^{\frac12}}, 
\end{equation} 
$$
\tilde{u}_{24}(k) = - \frac{\Delta_{13}\left( k, \tred{\omega^{(\gf)}_{-}}(k) \right)}{\left[ - \Delta_{11}\left( k, \tred{\omega^{(\gf)}_{-}}(k) \right) \right]^{\frac12}}  e^{i\varphi_4 (k)}
$$
\begin{equation}
\times  \frac{1}{\left[ 2 \tred{\omega^{(\gf)}_{-}}(k)  \left\{   \left[ \tred{\omega^{(\gf)}_{-}}(k) \right]^2 - \left[ \tred{\omega^{(\gl)}_{+}}(k) \right]^2  \right\}  \left\{   \left[ \tred{\omega^{(\gf)}_{-}}(k) \right]^2 - \left[ \tred{\omega^{(\gl)}_{-}}(k) \right]^2  \right\}  \left\{   \left[ \tred{\omega^{(\gf)}_{-}}(k) \right]^2 - \left[ \tred{\omega^{(\gf)}_{+}}(k) \right]^2  \right\} \right]^{\frac12}}. 
\end{equation} 
$$
\tilde{v}_{21}(k) = \frac{\Delta_{14}\left( k, \tred{\omega^{(\gl)}_{+}}(k) \right)}{\left[ - \Delta_{11}\left( k, \tred{\omega^{(\gl)}_{+}}(k) \right) \right]^{\frac12}}  e^{i\varphi_1 (k)}
$$
\begin{equation}
\times  \frac{1}{\left[ 2 \tred{\omega^{(\gl)}_{+}}(k)  \left\{   \left[ \tred{\omega^{(\gl)}_{+}}(k) \right]^2 - \left[ \tred{\omega^{(\gf)}_{+}}(k) \right]^2  \right\}  \left\{   \left[ \tred{\omega^{(\gl)}_{+}}(k) \right]^2 - \left[ \tred{\omega^{(\gl)}_{-}}(k) \right]^2  \right\}  \left\{   \left[ \tred{\omega^{(\gl)}_{+}}(k) \right]^2 - \left[ \tred{\omega^{(\gf)}_{-}}(k) \right]^2  \right\} \right]^{\frac12}}, 
\end{equation} 
$$
\tilde{v}_{22}(k) = \frac{\Delta_{14}\left( k, \tred{\omega^{(\gl)}_{-}}(k) \right)}{\left[ - \Delta_{11}\left( k, \tred{\omega^{(\gl)}_{-}}(k) \right) \right]^{\frac12}}  e^{i\varphi_2 (k)}
$$
\begin{equation}
\times  \frac{1}{\left[ 2 \tred{\omega^{(\gl)}_{-}}(k)  \left\{   \left[ \tred{\omega^{(\gl)}_{-}}(k) \right]^2 - \left[ \tred{\omega^{(\gf)}_{+}}(k) \right]^2  \right\}  \left\{   \left[ \tred{\omega^{(\gl)}_{-}}(k) \right]^2 - \left[ \tred{\omega^{(\gl)}_{+}}(k) \right]^2  \right\}  \left\{   \left[ \tred{\omega^{(\gl)}_{-}}(k) \right]^2 - \left[ \tred{\omega^{(\gf)}_{-}}(k) \right]^2  \right\} \right]^{\frac12}}, 
\end{equation} 
$$
\tilde{v}_{23}(k) = \frac{\Delta_{14}\left( k, \tred{\omega^{(\gf)}_{+}}(k) \right)}{\left[ - \Delta_{11}\left( k, \tred{\omega^{(\gf)}_{+}}(k) \right) \right]^{\frac12}}  e^{i\varphi_3 (k)}
$$
\begin{equation}
\times  \frac{1}{\left[ 2 \tred{\omega^{(\gf)}_{+}}(k)  \left\{   \left[ \tred{\omega^{(\gf)}_{+}}(k) \right]^2 - \left[ \tred{\omega^{(\gl)}_{+}}(k) \right]^2  \right\}  \left\{   \left[ \tred{\omega^{(\gf)}_{+}}(k) \right]^2 - \left[ \tred{\omega^{(\gl)}_{-}}(k) \right]^2  \right\}  \left\{   \left[ \tred{\omega^{(\gf)}_{+}}(k) \right]^2 - \left[ \tred{\omega^{(\gf)}_{-}}(k) \right]^2  \right\} \right]^{\frac12}}, 
\end{equation} 
$$
\tilde{v}_{24}(k) = \frac{\Delta_{14}\left( k, \tred{\omega^{(\gf)}_{-}}(k) \right)}{\left[ - \Delta_{11}\left( k, \tred{\omega^{(\gf)}_{-}}(k) \right) \right]^{\frac12}}  e^{i\varphi_4 (k)}
$$
\begin{equation}
\times  \frac{1}{\left[ 2 \tred{\omega^{(\gf)}_{-}}(k)  \left\{   \left[ \tred{\omega^{(\gf)}_{-}}(k) \right]^2 - \left[ \tred{\omega^{(\gl)}_{+}}(k) \right]^2  \right\}  \left\{   \left[ \tred{\omega^{(\gf)}_{-}}(k) \right]^2 - \left[ \tred{\omega^{(\gl)}_{-}}(k) \right]^2  \right\}  \left\{   \left[ \tred{\omega^{(\gf)}_{-}}(k) \right]^2 - \left[ \tred{\omega^{(\gf)}_{+}}(k) \right]^2  \right\} \right]^{\frac12}}.  
\end{equation} 
$$
\tilde{u}_{31}(k) = - \frac{\Delta_{15}\left( k, \tred{\omega^{(\gl)}_{+}}(k) \right)}{\left[ - \Delta_{11}\left( k, \tred{\omega^{(\gl)}_{+}}(k) \right) \right]^{\frac12}}  e^{i\varphi_1 (k)}
$$
\begin{equation}
\times  \frac{1}{\left[ 2 \tred{\omega^{(\gl)}_{+}}(k)  \left\{   \left[ \tred{\omega^{(\gl)}_{+}}(k) \right]^2 - \left[ \tred{\omega^{(\gf)}_{+}}(k) \right]^2  \right\}  \left\{   \left[ \tred{\omega^{(\gl)}_{+}}(k) \right]^2 - \left[ \tred{\omega^{(\gl)}_{-}}(k) \right]^2  \right\}  \left\{   \left[ \tred{\omega^{(\gl)}_{+}}(k) \right]^2 - \left[ \tred{\omega^{(\gf)}_{-}}(k) \right]^2  \right\} \right]^{\frac12}}, 
\end{equation} 
$$
\tilde{u}_{32}(k) = - \frac{\Delta_{15}\left( k, \tred{\omega^{(\gl)}_{-}}(k) \right)}{\left[ - \Delta_{11}\left( k, \tred{\omega^{(\gl)}_{-}}(k) \right) \right]^{\frac12}}  e^{i\varphi_2 (k)}
$$
\begin{equation}
\times  \frac{1}{\left[ 2 \tred{\omega^{(\gl)}_{-}}(k)  \left\{   \left[ \tred{\omega^{(\gl)}_{-}}(k) \right]^2 - \left[ \tred{\omega^{(\gf)}_{+}}(k) \right]^2  \right\}  \left\{   \left[ \tred{\omega^{(\gl)}_{-}}(k) \right]^2 - \left[ \tred{\omega^{(\gl)}_{+}}(k) \right]^2  \right\}  \left\{   \left[ \tred{\omega^{(\gl)}_{-}}(k) \right]^2 - \left[ \tred{\omega^{(\gf)}_{-}}(k) \right]^2  \right\} \right]^{\frac12}}, 
\end{equation} 
$$
\tilde{u}_{33}(k) = - \frac{\Delta_{15}\left( k, \tred{\omega^{(\gf)}_{+}}(k) \right)}{\left[ - \Delta_{11}\left( k, \tred{\omega^{(\gf)}_{+}}(k) \right) \right]^{\frac12}}  e^{i\varphi_3 (k)}
$$
\begin{equation}
\times  \frac{1}{\left[ 2 \tred{\omega^{(\gf)}_{+}}(k)  \left\{   \left[ \tred{\omega^{(\gf)}_{+}}(k) \right]^2 - \left[ \tred{\omega^{(\gl)}_{+}}(k) \right]^2  \right\}  \left\{   \left[ \tred{\omega^{(\gf)}_{+}}(k) \right]^2 - \left[ \tred{\omega^{(\gl)}_{-}}(k) \right]^2  \right\}  \left\{   \left[ \tred{\omega^{(\gf)}_{+}}(k) \right]^2 - \left[ \tred{\omega^{(\gf)}_{-}}(k) \right]^2  \right\} \right]^{\frac12}}, 
\end{equation} 
$$
\tilde{u}_{34}(k) = - \frac{\Delta_{15}\left( k, \tred{\omega^{(\gf)}_{-}}(k) \right)}{\left[ - \Delta_{11}\left( k, \tred{\omega^{(\gf)}_{-}}(k) \right) \right]^{\frac12}}  e^{i\varphi_4 (k)}
$$
\begin{equation}
\times  \frac{1}{\left[ 2 \tred{\omega^{(\gf)}_{-}}(k)  \left\{   \left[ \tred{\omega^{(\gf)}_{-}}(k) \right]^2 - \left[ \tred{\omega^{(\gl)}_{+}}(k) \right]^2  \right\}  \left\{   \left[ \tred{\omega^{(\gf)}_{-}}(k) \right]^2 - \left[ \tred{\omega^{(\gl)}_{-}}(k) \right]^2  \right\}  \left\{   \left[ \tred{\omega^{(\gf)}_{-}}(k) \right]^2 - \left[ \tred{\omega^{(\gf)}_{+}}(k) \right]^2  \right\} \right]^{\frac12}}. 
\end{equation} 
$$
\tilde{v}_{31}(k) = \frac{\Delta_{16}\left( k, \tred{\omega^{(\gl)}_{+}}(k) \right)}{\left[ - \Delta_{11}\left( k, \tred{\omega^{(\gl)}_{+}}(k) \right) \right]^{\frac12}}  e^{i\varphi_1 (k)}
$$
\begin{equation}
\times  \frac{1}{\left[ 2 \tred{\omega^{(\gl)}_{+}}(k)  \left\{   \left[ \tred{\omega^{(\gl)}_{+}}(k) \right]^2 - \left[ \tred{\omega^{(\gf)}_{+}}(k) \right]^2  \right\}  \left\{   \left[ \tred{\omega^{(\gl)}_{+}}(k) \right]^2 - \left[ \tred{\omega^{(\gl)}_{-}}(k) \right]^2  \right\}  \left\{   \left[ \tred{\omega^{(\gl)}_{+}}(k) \right]^2 - \left[ \tred{\omega^{(\gf)}_{-}}(k) \right]^2  \right\} \right]^{\frac12}}, 
\end{equation} 
$$
\tilde{v}_{32}(k) = \frac{\Delta_{16}\left( k, \tred{\omega^{(\gl)}_{-}}(k) \right)}{\left[ - \Delta_{11}\left( k, \tred{\omega^{(\gl)}_{-}}(k) \right) \right]^{\frac12}}  e^{i\varphi_2 (k)}
$$
\begin{equation}
\times  \frac{1}{\left[ 2 \tred{\omega^{(\gl)}_{-}}(k)  \left\{   \left[ \tred{\omega^{(\gl)}_{-}}(k) \right]^2 - \left[ \tred{\omega^{(\gf)}_{+}}(k) \right]^2  \right\}  \left\{   \left[ \tred{\omega^{(\gl)}_{-}}(k) \right]^2 - \left[ \tred{\omega^{(\gl)}_{+}}(k) \right]^2  \right\}  \left\{   \left[ \tred{\omega^{(\gl)}_{-}}(k) \right]^2 - \left[ \tred{\omega^{(\gf)}_{-}}(k) \right]^2  \right\} \right]^{\frac12}}, 
\end{equation} 
$$
\tilde{v}_{33}(k) = \frac{\Delta_{16}\left( k, \tred{\omega^{(\gf)}_{+}}(k) \right)}{\left[ - \Delta_{11}\left( k, \tred{\omega^{(\gf)}_{+}}(k) \right) \right]^{\frac12}}  e^{i\varphi_3 (k)}
$$
\begin{equation}
\times  \frac{1}{\left[ 2 \tred{\omega^{(\gf)}_{+}}(k)  \left\{   \left[ \tred{\omega^{(\gf)}_{+}}(k) \right]^2 - \left[ \tred{\omega^{(\gl)}_{+}}(k) \right]^2  \right\}  \left\{   \left[ \tred{\omega^{(\gf)}_{+}}(k) \right]^2 - \left[ \tred{\omega^{(\gl)}_{-}}(k) \right]^2  \right\}  \left\{   \left[ \tred{\omega^{(\gf)}_{+}}(k) \right]^2 - \left[ \tred{\omega^{(\gf)}_{-}}(k) \right]^2  \right\} \right]^{\frac12}}, 
\end{equation} 
$$
\tilde{v}_{34}(k) = \frac{\Delta_{16}\left( k, \tred{\omega^{(\gf)}_{-}}(k) \right)}{\left[ - \Delta_{11}\left( k, \tred{\omega^{(\gf)}_{-}}(k) \right) \right]^{\frac12}}  e^{i\varphi_4 (k)}
$$
\begin{equation}
\times  \frac{1}{\left[ 2 \tred{\omega^{(\gf)}_{-}}(k)  \left\{   \left[ \tred{\omega^{(\gf)}_{-}}(k) \right]^2 - \left[ \tred{\omega^{(\gl)}_{+}}(k) \right]^2  \right\}  \left\{   \left[ \tred{\omega^{(\gf)}_{-}}(k) \right]^2 - \left[ \tred{\omega^{(\gl)}_{-}}(k) \right]^2  \right\}  \left\{   \left[ \tred{\omega^{(\gf)}_{-}}(k) \right]^2 - \left[ \tred{\omega^{(\gf)}_{+}}(k) \right]^2  \right\} \right]^{\frac12}}.  
\end{equation} 
$$
\tilde{u}_{41}(k) = - \frac{\Delta_{17}\left( k, \tred{\omega^{(\gl)}_{+}}(k) \right)}{\left[ - \Delta_{11}\left( k, \tred{\omega^{(\gl)}_{+}}(k) \right) \right]^{\frac12}}  e^{i\varphi_1 (k)}
$$
\begin{equation}
\times  \frac{1}{\left[ 2 \tred{\omega^{(\gl)}_{+}}(k)  \left\{   \left[ \tred{\omega^{(\gl)}_{+}}(k) \right]^2 - \left[ \tred{\omega^{(\gf)}_{+}}(k) \right]^2  \right\}  \left\{   \left[ \tred{\omega^{(\gl)}_{+}}(k) \right]^2 - \left[ \tred{\omega^{(\gl)}_{-}}(k) \right]^2  \right\}  \left\{   \left[ \tred{\omega^{(\gl)}_{+}}(k) \right]^2 - \left[ \tred{\omega^{(\gf)}_{-}}(k) \right]^2  \right\} \right]^{\frac12}}, 
\end{equation} 
$$
\tilde{u}_{42}(k) = - \frac{\Delta_{17}\left( k, \tred{\omega^{(\gl)}_{-}}(k) \right)}{\left[ - \Delta_{11}\left( k, \tred{\omega^{(\gl)}_{-}}(k) \right) \right]^{\frac12}}  e^{i\varphi_2 (k)}
$$
\begin{equation}
\times  \frac{1}{\left[ 2 \tred{\omega^{(\gl)}_{-}}(k)  \left\{   \left[ \tred{\omega^{(\gl)}_{-}}(k) \right]^2 - \left[ \tred{\omega^{(\gf)}_{+}}(k) \right]^2  \right\}  \left\{   \left[ \tred{\omega^{(\gl)}_{-}}(k) \right]^2 - \left[ \tred{\omega^{(\gl)}_{+}}(k) \right]^2  \right\}  \left\{   \left[ \tred{\omega^{(\gl)}_{-}}(k) \right]^2 - \left[ \tred{\omega^{(\gf)}_{-}}(k) \right]^2  \right\} \right]^{\frac12}}, 
\end{equation} 
$$
\tilde{u}_{43}(k) = - \frac{\Delta_{17}\left( k, \tred{\omega^{(\gf)}_{+}}(k) \right)}{\left[ - \Delta_{11}\left( k, \tred{\omega^{(\gf)}_{+}}(k) \right) \right]^{\frac12}}  e^{i\varphi_3 (k)}
$$
\begin{equation}
\times  \frac{1}{\left[ 2 \tred{\omega^{(\gf)}_{+}}(k)  \left\{   \left[ \tred{\omega^{(\gf)}_{+}}(k) \right]^2 - \left[ \tred{\omega^{(\gl)}_{+}}(k) \right]^2  \right\}  \left\{   \left[ \tred{\omega^{(\gf)}_{+}}(k) \right]^2 - \left[ \tred{\omega^{(\gl)}_{-}}(k) \right]^2  \right\}  \left\{   \left[ \tred{\omega^{(\gf)}_{+}}(k) \right]^2 - \left[ \tred{\omega^{(\gf)}_{-}}(k) \right]^2  \right\} \right]^{\frac12}}, 
\end{equation} 
$$
\tilde{u}_{44}(k) = - \frac{\Delta_{17}\left( k, \tred{\omega^{(\gf)}_{-}}(k) \right)}{\left[ - \Delta_{11}\left( k, \tred{\omega^{(\gf)}_{-}}(k) \right) \right]^{\frac12}}  e^{i\varphi_4 (k)}
$$
\begin{equation}
\times  \frac{1}{\left[ 2 \tred{\omega^{(\gf)}_{-}}(k)  \left\{   \left[ \tred{\omega^{(\gf)}_{-}}(k) \right]^2 - \left[ \tred{\omega^{(\gl)}_{+}}(k) \right]^2  \right\}  \left\{   \left[ \tred{\omega^{(\gf)}_{-}}(k) \right]^2 - \left[ \tred{\omega^{(\gl)}_{-}}(k) \right]^2  \right\}  \left\{   \left[ \tred{\omega^{(\gf)}_{-}}(k) \right]^2 - \left[ \tred{\omega^{(\gf)}_{+}}(k) \right]^2  \right\} \right]^{\frac12}}. 
\end{equation} 
$$
\tilde{v}_{41}(k) = \frac{\Delta_{18}\left( k, \tred{\omega^{(\gl)}_{+}}(k) \right)}{\left[ - \Delta_{11}\left( k, \tred{\omega^{(\gl)}_{+}}(k) \right) \right]^{\frac12}}  e^{i\varphi_1 (k)}
$$
\begin{equation}
\times  \frac{1}{\left[ 2 \tred{\omega^{(\gl)}_{+}}(k)  \left\{   \left[ \tred{\omega^{(\gl)}_{+}}(k) \right]^2 - \left[ \tred{\omega^{(\gf)}_{+}}(k) \right]^2  \right\}  \left\{   \left[ \tred{\omega^{(\gl)}_{+}}(k) \right]^2 - \left[ \tred{\omega^{(\gl)}_{-}}(k) \right]^2  \right\}  \left\{   \left[ \tred{\omega^{(\gl)}_{+}}(k) \right]^2 - \left[ \tred{\omega^{(\gf)}_{-}}(k) \right]^2  \right\} \right]^{\frac12}}, 
\end{equation} 
$$
\tilde{v}_{42}(k) = \frac{\Delta_{18}\left( k, \tred{\omega^{(\gl)}_{-}}(k) \right)}{\left[ - \Delta_{11}\left( k, \tred{\omega^{(\gl)}_{-}}(k) \right) \right]^{\frac12}}  e^{i\varphi_2 (k)}
$$
\begin{equation}
\times  \frac{1}{\left[ 2 \tred{\omega^{(\gl)}_{-}}(k)  \left\{   \left[ \tred{\omega^{(\gl)}_{-}}(k) \right]^2 - \left[ \tred{\omega^{(\gf)}_{+}}(k) \right]^2  \right\}  \left\{   \left[ \tred{\omega^{(\gl)}_{-}}(k) \right]^2 - \left[ \tred{\omega^{(\gl)}_{+}}(k) \right]^2  \right\}  \left\{   \left[ \tred{\omega^{(\gl)}_{-}}(k) \right]^2 - \left[ \tred{\omega^{(\gf)}_{-}}(k) \right]^2  \right\} \right]^{\frac12}}, 
\end{equation}
$$
\tilde{v}_{43}(k) = \frac{\Delta_{18}\left( k, \tred{\omega^{(\gf)}_{+}}(k) \right)}{\left[ - \Delta_{11}\left( k, \tred{\omega^{(\gf)}_{+}}(k) \right) \right]^{\frac12}}  e^{i\varphi_3 (k)}
$$
\begin{equation}
\times  \frac{1}{\left[ 2 \tred{\omega^{(\gf)}_{+}}(k)  \left\{   \left[ \tred{\omega^{(\gf)}_{+}}(k) \right]^2 - \left[ \tred{\omega^{(\gl)}_{+}}(k) \right]^2  \right\}  \left\{   \left[ \tred{\omega^{(\gf)}_{+}}(k) \right]^2 - \left[ \tred{\omega^{(\gl)}_{-}}(k) \right]^2  \right\}  \left\{   \left[ \tred{\omega^{(\gf)}_{+}}(k) \right]^2 - \left[ \tred{\omega^{(\gf)}_{-}}(k) \right]^2  \right\} \right]^{\frac12}}, 
\end{equation} 
$$
\tilde{v}_{44}(k) = \frac{\Delta_{18}\left( k, \tred{\omega^{(\gf)}_{-}}(k) \right)}{\left[ - \Delta_{11}\left( k, \tred{\omega^{(\gf)}_{-}}(k) \right) \right]^{\frac12}}  e^{i\varphi_4 (k)}
$$
\begin{equation}
\times  \frac{1}{\left[ 2 \tred{\omega^{(\gf)}_{-}}(k)  \left\{   \left[ \tred{\omega^{(\gf)}_{-}}(k) \right]^2 - \left[ \tred{\omega^{(\gl)}_{+}}(k) \right]^2  \right\}  \left\{   \left[ \tred{\omega^{(\gf)}_{-}}(k) \right]^2 - \left[ \tred{\omega^{(\gl)}_{-}}(k) \right]^2  \right\}  \left\{   \left[ \tred{\omega^{(\gf)}_{-}}(k) \right]^2 - \left[ \tred{\omega^{(\gf)}_{+}}(k) \right]^2  \right\} \right]^{\frac12}}.  
\end{equation} 

Similar to the analysis for the longitudinal modes, the phases $\varphi_a$ ($a=1,\ldots,4$)  are chosen zero. 

\end{widetext}

\section{High-frequency expansion}
\label{app:hfe}

To calculate the Floquet Hamiltonian in the high-frequency limit, the high-frequency expansion (HFE)  proves to be an effective method.  

Within this approach the effective Hamiltonian can be expressed as the series introduced in Appendix~D of the main article, whose first terms are given by\cite{Goldman2014,Mikami2016,Sato2025} 
\begin{align}
\hat{H}^{(0)} & = \hat{H}_0,  \\ 
\hat{H}^{(1)} & = - \sum_{m=1}^{\infty} \frac{[\hat{H}_{+m}, \hat{H}_{-m}]}{m\hbar\omega} ,
\end{align}
\begin{align}
\hat{H}^{(2)} & = \sum_{\substack{m=-\infty \\ (m\neq 0)}}^{\infty}
\frac{[[\hat{H}_{-m}, \hat{H}_0], \hat{H}_m]}{2m^2(\hbar\omega)^2} \\ 
 & + \sum_{\substack{m=-\infty \\ (m\neq 0)}}^{\infty}
\sum_{\substack{n=-\infty \\ (n\neq 0,m)}}^{\infty}
\frac{[[\hat{H}_{-m}, \hat{H}_{m-n}], \hat{H}_n]}{3mn(\hbar\omega)^2},
\end{align}

\begin{widetext}
\begin{align}
\hat{H}^{(3)} &=
\sum_{\substack{m=-\infty \\ (m\neq 0)}}^{\infty}
\frac{[[[\hat{H}_{-m}, \hat{H}_0], \hat{H}_0], \hat{H}_m]}{2m^3(\hbar\omega)^3}
+ \sum_{\substack{m=-\infty \\ (m\neq 0)}}^{\infty}
\sum_{\substack{n=-\infty \\ (n\neq 0,m)}}^{\infty}
\frac{[[[\hat{H}_{-m}, \hat{H}_0], \hat{H}_{m-n}], \hat{H}_n]}{3m^2n(\hbar\omega)^3}  \\[4pt]
&\quad
+ \sum_{\substack{m=-\infty \\ (m\neq 0)}}^{\infty}
\sum_{\substack{n=-\infty \\ (n\neq 0,m)}}^{\infty}
\frac{[[[\hat{H}_{-m}, \hat{H}_{m-n}], \hat{H}_0], \hat{H}_n]}{4mn^2(\hbar\omega)^3}
- \sum_{\substack{m=-\infty \\ (m\neq 0)}}^{\infty}
\sum_{\substack{n=-\infty \\ (n\neq 0)}}^{\infty}
\frac{[[[\hat{H}_{-m}, \hat{H}_m], \hat{H}_{-n}], \hat{H}_n]}{12mn^2(\hbar\omega)^3} \notag \\[4pt]
&\quad
+ \sum_{\substack{m=-\infty \\ (m\neq 0)}}^{\infty}
\sum_{\substack{n=-\infty \\ (n\neq 0,m)}}^{\infty}
\frac{[[\hat{H}_{-m}, \hat{H}_0], [\hat{H}_{m-n}, \hat{H}_n]]}{12m^2n(\hbar\omega)^3}
+ \sum_{\substack{m=-\infty \\ (m\neq 0)}}^{\infty}
\sum_{\substack{n=-\infty \\ (n\neq 0)}}^{\infty}
\sum_{\substack{\ell=-\infty \\ (\ell\neq 0,m,n)}}^{\infty}
\frac{[[[\hat{H}_{-m}, \hat{H}_{m-\ell}], \hat{H}_{\ell-n}], \hat{H}_n]}{6\ell mn(\hbar\omega)^3} \notag \\[4pt]
&\quad
+ \sum_{\substack{m=-\infty \\ (m\neq 0)}}^{\infty}
\sum_{\substack{n=-\infty \\ (n\neq 0,m)}}^{\infty}
\sum_{\substack{\ell=-\infty \\ (\ell\neq 0,m-n)}}^{\infty}
\frac{[[[\hat{H}_{-m}, \hat{H}_{m-n-\ell}], \hat{H}_\ell], \hat{H}_n]}{24\ell mn(\hbar\omega)^3}
+ \sum_{\substack{m=-\infty \\ (m\neq 0)}}^{\infty}
\sum_{\substack{n=-\infty \\ (n\neq 0,m)}}^{\infty}
\sum_{\substack{\ell=-\infty \\ (\ell\neq 0,m,n)}}^{\infty}
\frac{[[\hat{H}_{-m}, \hat{H}_{m-\ell}], [\hat{H}_{\ell-n}, \hat{H}_n]]}{24\ell mn(\hbar\omega)^3} . 
\end{align}
\end{widetext}

\end{document}